\documentclass[review]{elsarticle}
\usepackage{amssymb}
\usepackage{mathtools}
\usepackage{hyperref}
\usepackage[monochrome]{xcolor}
\usepackage{xspace}
\usepackage{soul}

\newcommand{\mc}{$m_\mathrm{c}$\xspace}
\newcommand{\pt}{$p_\mathrm{T}$\xspace}
\newcommand{\kt}{$k_\mathrm{T}$\xspace}
\newcommand{\pty}{$p_\mathrm{T}^{\gamma}$\xspace}
\newcommand{\etay}{$\eta^{\gamma}$\xspace}
\newcommand{\yy}{$y^{\gamma}$\xspace}
\newcommand{\xf}{$x_\mathrm{F}$\xspace}
\newcommand{\xfq}{$x_\mathrm{F}^\mathrm{Q}$\xspace}
\newcommand{\mev}{MeV/c\xspace}
\newcommand{\gev}{GeV/c\xspace}
\newcommand{\tev}{TeV/c\xspace}
\newcommand{\gevs}{(GeV/c)$^2$\xspace}

\biboptions{numbers,sort&compress}
\journal{Progress in Particle and Nuclear Physics}

\begin{document}
\begin{frontmatter}

\title{The Physics of Heavy Quark Distributions in Hadrons: Collider Tests}

\author[slac]{S.J.~Brodsky}
\author[jinr]{V.A.~Bednyakov}
\author[jinr]{G.I.~Lykasov\corref{correspondingauthor}}
\cortext[correspondingauthor]{Corresponding author}
\ead{lykasov@jinr.ru}
\author[comenius]{J.~Smiesko}
\author[comenius]{S.~Tokar}

\address[slac]{SLAC National Accelerator Laboratory,
               Stanford University, Menlo Park, CA 94025, United States}
\address[jinr]{Joint Institute for Nuclear Research,
               Dubna 141980, Moscow region, Russia}
\address[comenius]{Comenius University in Bratislava,
                   Faculty of Mathematics, Physics and Informatics,
                   Mlynska Dolina, 842 48 Bratislava, Slovakia}

\begin{abstract}
We present a review of the current understanding of the heavy quark
distributions in the nucleon and their impact on collider physics. The origin
of strange, charm and bottom quark pairs at high light-front (LF) momentum
fractions in hadron wavefunctions~---~the ``intrinsic'' quarks, is reviewed.
The determination of heavy-quark parton distribution functions (PDFs) is
particularly significant for the analysis of hard processes at LHC energies.
We show that a careful study of the inclusive production of open charm and
the production of $\gamma$/$Z$/$W$ particles, accompanied by the heavy jets
at large transverse momenta can give essential information on the intrinsic
heavy quark (IQ) distributions. We also focus on the theoretical predictions
concerning other observables which are very sensitive to the intrinsic charm
contribution to PDFs including Higgs production at high \xf\ and novel fixed
target measurements which can be tested at the LHC\@.
\end{abstract}

\begin{keyword}
Quarks \sep\ gluons \sep\ charm \sep\ QCD \sep\ PDF\@.
\end{keyword}
\end{frontmatter}

\newpage
\tableofcontents{}
\newpage

\newpage


\section{Introduction}
\label{sec:intro}
\subsection{Motivation of this review}

The knowledge of the  heavy quark distributions of the proton can provide
fundamental information on nucleon structure. Since many hard processes within
the Standard Model (SM) and beyond, such as the production of heavy jets and
Higgs boson production, etc., are sensitive to the heavy quark content of the
nucleon, heavy quark distributions play an increasingly significant role in the
physics program of the Large Hadron Collider (LHC). The parton distribution
functions (PDFs) of the strange, charm, and bottom quarks and their evolution
are essential inputs for the calculation of these processes within perturbative
quantum chromodynamics (pQCD). The heavy quark PDFs reflect both the initial
conditions at momentum transfers below a factorization scale $Q^2 < Q^2_0$ as
dictated by the non-perturbative QCD color-confining dynamics of the proton
plus the effects of pQCD which generates heavy quarks from the evolution of the
light quark and gluon PDFs. A global QCD analysis allows one to extract the
PDFs from the comparison of hard-scattering data based on the factorization
properties of QCD\@.

In fact, QCD predicts two separate and distinct contributions to the heavy
quark distributions $q(x, Q^2)$ of the nucleons at low and high $x$. Here ${x =
{k^+ \over P^+}} = {k^0 + k^3 \over P^0 + P^3}$ is the frame-independent
light-front (LF) momentum fraction carried by the heavy quark in a hadron with
momentum $P^\mu$. In the case of deep inelastic lepton-proton scattering, the
LF momentum fraction variable $x$ in the proton structure functions can be
identified with the Bjorken variable $x_\mathrm{Bj} = {Q^2 \over 2 p \cdot q}$.
At small $x$, heavy-quark pairs are dominantly produced via gluon-splitting
subprocess $g \to Q \bar{Q}$. The \textcolor{purple}{existence} of the heavy
quarks in \textcolor{purple}{the} nucleon from this standard contribution is a
result of the QCD evolution of the light quark and gluon PDFs. The gluon
splitting contribution to the heavy-quark degrees of freedom are perturbatively
calculable using the Dokshitzer-Gribov-Lipatov-Altarelli-Parisi (DGLAP)
evolution equation~\cite{Gribov:1972ri,Altarelli:1977zs,Dokshitzer:1977sg}. To
first approximation, the heavy quark distribution falls as $(1 - x)$ times the
gluon distribution.

However, \textcolor{purple}{the} QCD also predicts additional Fock state
contributions to \textcolor{purple}{the} proton structure at high $x$, such as
$|uud Q\bar{Q}\rangle$\textcolor{purple}{,} where the heavy quark pair is
multiply connected to two or more valence quarks of the proton. This
contribution depends on non-perturbative distribution of valence quarks in the
hadron, and is maximal at minimal off-shellness; i.e., when the constituents
all have the same rapidity $y_i$ and thus $x_i \propto \sqrt{m_i^2 +
\vec{k}_{\mathrm{T} i}^2}$. The heavy quark contributions to the nucleon's PDF
thus peak at large $x$. Since they depend on the correlations determined by the
valence quark distributions, these heavy quark contributions are
\emph{intrinsic} contributions to the hadron's fundamental structure.
Furthermore, since all of the intrinsic quarks in the $|uud Q \bar{Q}\rangle$
Fock state have similar rapidities they can reinteract, leading to significant
$Q$ vs. $\bar{Q}$ asymmetries. In contrast, the contribution to the heavy quark
PDFs arising from gluon splitting are symmetric in $Q$ vs. $\bar{Q}$. Since
they only depend on the gluon distribution, the contributions generated by
DGLAP evolution can be considered as \emph{extrinsic} contributions.

The PDFs at a fundamental level are computed from the squares of the hadrons'
light-front wavefunctions, the frame-independent eigensolutions of the QCD
Light-Front Hamiltonian. The intrinsic contributions are associated with
amplitudes such as $gg \to Q \bar{Q} \to gg$ in the
\textcolor{purple}{self-energy} of the proton, the analogs of
\textcolor{purple}{light-by-light} scattering $\gamma \gamma \to \ell \bar{\ell}
\to \gamma \gamma$ in QED, i.e., twist-6 contributions proportional to the
gluon field strength \textcolor{purple}{of the order 3 in} the operator product
expansion (OPE). Thus the OPE provides a first-principle derivation for the
existence of intrinsic heavy quarks.

Unlike the conventional $\log m^2_\mathrm{Q}$ dependence of the low $x$
extrinsic gluon-splitting contribution, the probabilities for the intrinsic
heavy quark Fock states at high $x$ scale as $1 \over m_\mathrm{Q}^2$ in
non-Abelian QCD\@. In contrast the probability for a higher Fock state in an
atom such as $|e^+ e^- \ell \bar \ell\rangle$ in positronium scales as $1 \over
m_\ell^4$ in Abelian QED, corresponding to the twist-8 Euler-Heisenberg
light-by-light insertion. Detailed derivations based on the OPE have been given
in~\cite{Brodsky:1984nx,Franz:2000ee}.

When a proton collides with another proton or nucleus, the off-shell intrinsic
heavy quark Fock state fluctuations such as $|uudc\bar{c}\rangle$ are
materialized and can produce open or hidden charm states at high momentum
fraction. For example, the comoving $udc$ quarks in a Fock state such as
$|uudc\bar{c}\rangle$ can coalesce to produce a $\Lambda_\mathrm{c}(udc)$
baryon with a high Feynman momentum fraction $x_\mathrm{F} = x_\mathrm{c} +
x_\mathrm{u} + x_\mathrm{d}$ or produce a $J/\psi$ with $x_\mathrm{F} = c
\bar{c}$. Such high \xf\ heavy hadron events have been observed and measured
with substantial cross sections at the ISR proton-proton collider at fixed
target experiments such as NA3 at CERN and SELEX at FermiLab. The
$\Lambda_\mathrm{b}(udb)$ baryon was first observed at the ISR in forward $pp
\to \Lambda_\mathrm{b} X$ reactions at high \xf.

The hypothesis of the \emph{intrinsic} quark components in the proton was
suggested in~\cite{Brodsky:1980pb} to explain the large cross-section for the
forward open charm production in $pp$ collision at ISR
energies~\cite{Drijard:1978gv,Giboni:1979rm,Lockman:1979aj,Drijard:1979vd}.
The magnitude of the ISR cross sections suggests that the intrinsic charm
probability in the proton is approximately of order 1~\%; however, theoretical
and experimental uncertainties make it difficult to make an accurate estimate.

Intrinsic charm is also important for charm production in cosmic ray
experiments that measure charm production from high energy experiments
interacting in the earth's atmosphere. It also is important for estimating the
high energy flux of neutrinos observed in the IceCube experiment. In fact one
finds~\cite{Laha:2016dri} that the prompt neutrino flux arising from charm
hadroproduction by protons interacting in the earth's atmosphere which is due
to  intrinsic charm is comparable to the extrinsic contribution if one
normalizes the intrinsic charm differential cross sections to the ISR and the
LEBC-MPS collaboration data.

Most measurements of $c$ and $b$-jet production in deep inelastic lepton-proton
scattering are consistent with the extrinsic gluon-splitting perturbative
origin of heavy flavor quarks~\cite{Aktas:2005iw} alone. However, these
experiments are not sensitive to the heavy quark distributions at large $x$
region, where the theory predicts a non zero contribution of the so called
\emph{intrinsic} heavy-quark components in the proton wave
function~\cite{Brodsky:1980pb,Brodsky:1981se,Brodsky:1984nx}, as is suggested
in Refs.~\cite{Brodsky:2004er,Brodsky:2015fna, Pumplin:2005yf}.

The first direct experimental indication for the \emph{intrinsic} heavy quarks
in a nucleon was observed at the EMC measurement of the charm structure
function at large $x$~\cite{Aubert:1982tt}. The measurement of the charm
structure function at high $x_\mathrm{Bj}$ by the EMC experiment at CERN using
deep inelastic muon-nucleus scattering showed a significant contribution to the
proton structure function at large $x_\mathrm{Bj}$~\cite{Aubert:1982tt}. In
fact, the charm structure function $c(x, Q)$ measured by the EMC collaboration
was approximately 30 times higher than expected from gluon splitting and at
$x_\mathrm{Bj} = 0.42$ and $Q^2 = 75$~\gevs. As shown in
Refs.~\cite{Ball:2016neh,Ball:2014uwa,Dulat:2015mca}, the inclusion of both the
HERA data and the EMC data to the global NNPDF analysis allows one to do a
model independent determination of the charm content of the proton. The results
of Ref.~\cite{Ball:2016neh} suggest the perturbative origin of the charm PDF at
$x < 0.1$, which vanishes at $\mu \simeq 1.5$~\gev, but it also indicate
presence of a large $x$ \emph{intrinsic} component picked at $x \sim 0.5$ and
carried $0.7 \pm 0.3$~\% of the nucleon momentum with the 68~\%~CL at low scale
$\mu_0 = 1.65$~\gev.

There have been attempts to describe the data~\cite{Adamovich:1993kc,
Alves:1993mp,Aitala:1996hf} on inclusive spectra of the open charm production
in soft $pp$ collisions and the asymmetry between the $D^-$ and $D^+$
production in $\pi^- p$ collisions within the string
model~\cite{Kaidalov:1982xg,Capella:1992yb} without the assumption on the
\emph{intrinsic} charm in a nucleon. However, the description of data is very
sensitive to the fragmentation functions of $c$-quarks to charmed hadrons, the
knowledge of which up to now is not sufficient, see, e.g.
Refs.~\cite{Kaidalov:1985jg, Shabelski:1995ei,Lykasov:1999fj}. Moreover, the
experimental data on the open charm production, for example,
$\Lambda_\mathrm{c}^+$ in $pp$ collision at the ISR energies and large
$x$\textcolor{purple}{,} have large uncertainty~\cite{Barlag:1990hg,Bari:1991in}.

The existence of the \emph{intrinsic} heavy quark components can be observed
not only in the forward kinematic region, but also in hard $pp$ processes of
photon or vector boson production in association with the heavy flavor jets.
The first hint on this was observed at the Tevatron ($\sqrt{s} = 1.96$~TeV) in
the production of prompt photons accompanied by heavy jets ($c$ or
$b$)~\cite{Abazov:2009de,Abazov:2012ea,D0:2012gw,Aaltonen:2009wc,
Aaltonen:2013ama}. The basic underlying hard subprocesses are $gc \to \gamma c$
and $gb \to \gamma b$. The measurements of $pp \to b\gamma X$ at high \pt\ are
consistent with standard analyses where the $b$ quark PDF arises from gluon
splitting $g \to b \bar{b}$. However, the $pp \to c \gamma X$ data show a
significant excess, indicating that the charm PDF has significant support at
large $x$. This is consistent with the fact that the ratio of intrinsic bottom
to charm probabilities in the proton scales as $m_\mathrm{c}^2/
m_\mathrm{b}^2$~\cite{Polyakov:1998rb}.

It was observed that the ratio between the experimental spectrum of the prompt
photon accompanied by the $c$-jets and the theoretical calculation, which used
the PDF without the IC, increases with \pty\ up to factor of about 3 at
$p_\mathrm{T}^\gamma \simeq 110$~\gev. The inclusion of the IC contribution
obtained within the BHPS model~\cite{Brodsky:1980pb,Brodsky:1981se} allows one
to reduce this ratio up to 1.5~\cite{Stavreva:2009vi}. \textcolor{purple}{This}
stimulated us to investigate the hard processes of the production of prompt
photons or gauge vector bosons accompanied by heavy flavor jets ($c$ or $b$) in
$pp$ collisions at the LHC energies~\cite{Bednyakov:2013zta,
Beauchemin:2014rya}. It was shown that at large transverse momenta of $\gamma$
or $Z/W$ ($p_\mathrm{T} > 100$~\gev) and their rapidities $y > 1.5$ the
\emph{intrinsic} charm can contribute to the $p_\mathrm{T}$-spectra of these
particles. Therefore, these processes can be used as a laboratory to search for
the IC contribution. It was shown~\cite{Lipatov:2016feu} that a possible
observation of the IC contribution looks very promising at the LHC in processes
like $pp \to \gamma/Z/W + c/b + X$. The main goal of this review is to show,
how the \emph{intrinsic} heavy quark components in a nucleon can be tested at
the LHC\@.

In addition to the processes discussed above the more exotic observables, for
example, the diffractive and inclusive Higgs boson production in $pp$
collisions at LHC energies can give an information on the non-zero contribution
of the \emph{intrinsic} heavy flavor components to the proton
PDF~\cite{Brodsky:2006wb,Brodsky:2007yz}. In the \xf-distribution of the
diffractive Higgs boson production in $pp$ collision a peak is predicted at
$x_\mathrm{F} \sim 0.9$, which is due to the \emph{intrinsic charm} (IC)
contribution~\cite{Brodsky:2006wb}. In~\cite{Brodsky:2007yz} a similar
enhancement in the \xf-spectrum of the inclusive Higgs boson production is
predicted at $x_\mathrm{F} \sim$ 0.8~--~0.9 due to the \emph{intrinsic bottom}
(IB) contribution to the proton PDF\@. Recently a proposal to observe the IC
signal at the high luminosity fixed-target experiment using the LHC beam
(AFTER@LHC)~\cite{Brodsky:2012vg} was suggested, see
review~\cite{Brodsky:2015fna} and references therein. All these predictions
concern mainly the forward production of heavy hadrons at large longitudinal
fraction of their momentum $x_\mathrm{F} > 0.5$, therefore, in principle, the
\emph{intrinsic} heavy quark signal in such observables could be visible.
However, a high experimental accuracy is needed to observe the enhancement in
the heavy hadron spectra due to the IC or IB contribution to the proton PDF\@.

\subsection{Outline of this Review}

The review consists of 7 sections. In Section~\ref{sec:ex_in_comp} we present a
more detailed overview of the concepts of extrinsic and intrinsic quark
components in a nucleon and their distinction. In
Section~\ref{sec:heavy_quarks} we show why the \emph{intrinsic} quark
components can be visible in the hard $pp$ processes at high energies.
Section~\ref{sec:is} is devoted to the \emph{intrinsic strangeness} (IS)
content of the nucleon and search for it in hard $pp$ collisions at high
energies. Section~\ref{sec:glob_analy} is a short overview of the global
analysis of PDFs with intrinsic charm. Section~\ref{sec:ic_hard} is devoted to
search for the \emph{intrinsic} heavy quark in processes at collider energies,
namely, in the inclusive production of open charm or Higgs boson and production
of prompt photons \textcolor{purple}{$\gamma$} or gauge vector bosons $W/Z$
accompanied by $c(b)$-jets or open charm in $pp$ collision at LHC energies.
There we discuss a possible test of the IC contribution in proton at LHC\@.
Finally, in Section~\ref{sec:conclusion} we present the conclusion.

\section{Extrinsic and intrinsic quark components in nucleon}
\label{sec:ex_in_comp}

By definition, the PDF $f_\mathrm{a}(x, \mu)$ is a function of the proton
momentum fraction $x$ carried by parton $a$ (quark $q$ or gluon $g$) at the QCD
momentum transfer scale $\mu$. For small values of $\mu$, corresponding to the
long distance scales greater than $1/\mu_0$, the PDF cannot be calculated from
the first principles of QCD (although some progress in this direction has been
recently achieved within the lattice methods~\cite{Negele:2004iu,
Schroers:2005rm}). \textcolor{purple}{If the} PDF $f_\mathrm{a}(x, \mu)$,
\textcolor{purple}{is known} at a scale $\mu > \mu_0$, \textcolor{purple}{one} can
\textcolor{purple} {calculate it} at any other scale $\mu$ by solving the
perturbative QCD evolution equations
(DGLAP)~\cite{Gribov:1972ri,Altarelli:1977zs,
Dokshitzer:1977sg}\textcolor{purple}{;} usually, the value of the starting scale
$\mu_0$ is chosen about a few \gev. As the input of evolution
functions $f_\mathrm{a}(x, \mu_0)$ can be taken at any scale the functions can
be found empirically from some ``QCD global analysis''~\cite{Pumplin:2002vw,
Stump:2003yu,Thorne:2004ci} of a large variety of data, typically at $\mu >
\mu_0$.

In general, almost all $pp$ processes that took place at the LHC energies,
including the Higgs boson production, are sensitive to the charm
$f_\mathrm{c}(x, \mu)$ or bottom $f_\mathrm{b}(x, \mu)$ PDFs.  Heavy quark
become visible in proton at scales $\mu_\mathrm{c}$ and $\mu_\mathrm{b}$ and
their contents increases with increasing $Q^2$-scale through the gluon
splitting in the DGLAP $Q^2$ evolution~\cite{Gribov:1972ri,Altarelli:1977zs,
Dokshitzer:1977sg}. Direct measurement of the open charm and open bottom
production in the deep inelastic processes (DIS) confirms the QCD picture of
heavy quarks dynamics in proton~\cite{Aktas:2005iw}.

\begin{figure}[h!]
\centering
\includegraphics[width=.65\textwidth]{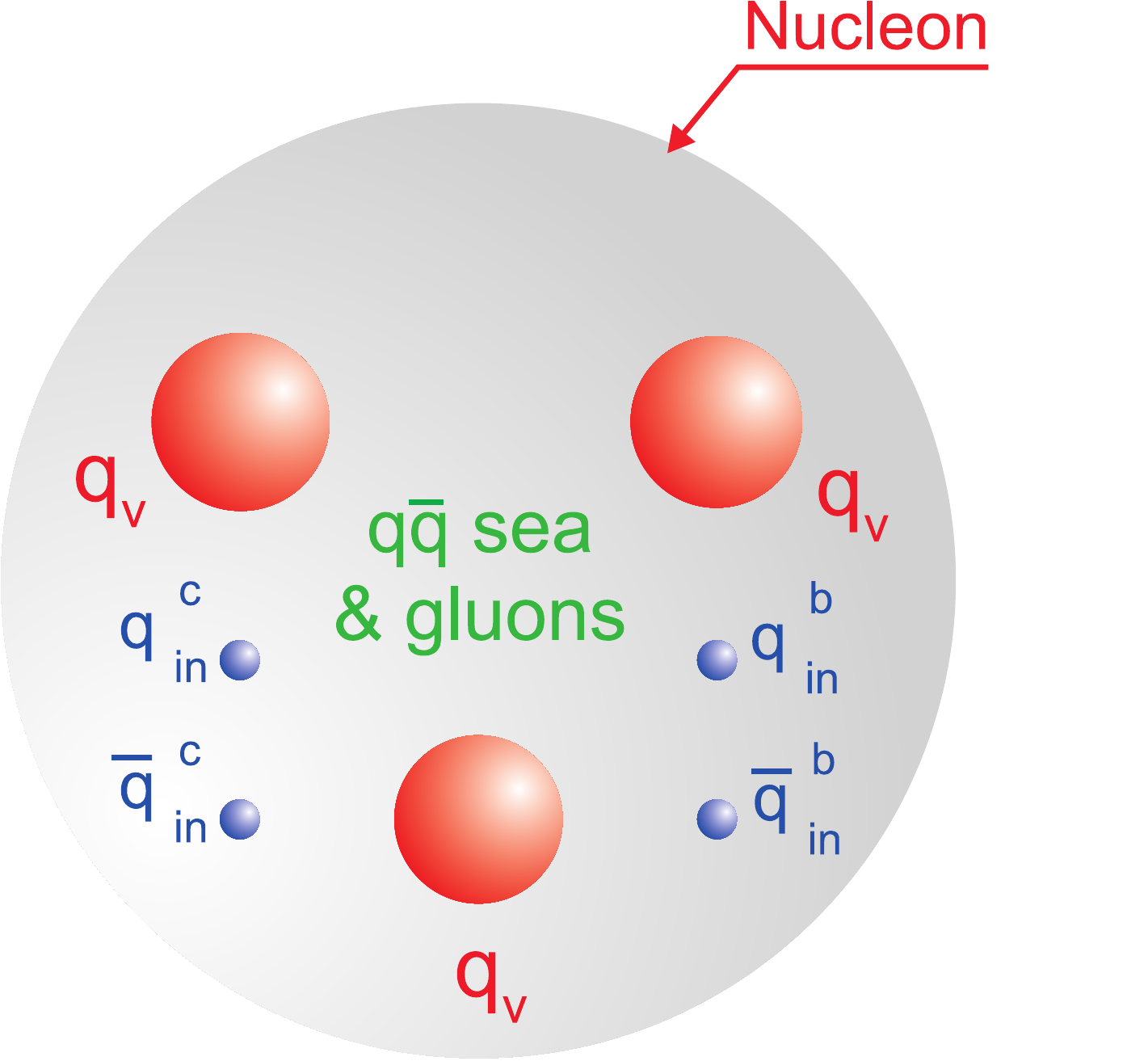}
\caption{Schematic presentation of a proton during the $pp$ collision during
the $pp$ collision at a value of $Q^2$, when the lifetime of the
\emph{intrinsic} Fock-state is much larger than the interaction time.}
\label{fig:nucleon}
\end{figure}

As was assumed in~\cite{Brodsky:1980pb,Brodsky:1981se}, there are
\emph{extrinsic} and \emph{intrinsic} contributions to the quark-gluon
structure of the proton. The \emph{extrinsic} (or perturbative sea) quarks and
gluons are generated on a short time scale associated with a large transverse
momentum processes as a result of gluon dynamics. Their distribution functions
satisfy the standard QCD evolution equations~\cite{Gribov:1972ri,
Altarelli:1977zs,Dokshitzer:1977sg}. The \emph{extrinsic} quark contribution
$q_\mathrm{ex}(x)$ to PDFs is mostly significant at low $x$ and decreases when
the quark momentum fraction grows. It depends logarithmically on the heavy
quark mass $M_\mathrm{Q}$, while the \emph{intrinsic} quark contribution
$q_\mathrm{in}(x)$, the residual part of the proton structure not coming from
the perturbative gluon splitting, is almost zero at low $x$ and dominates
compared to the \emph{extrinsic} one at large $x > 0.1$. \textcolor{purple}{It
depends} on the heavy quark mass as $1/M_\mathrm{Q}^2$. Initially
in~\cite{Brodsky:1980pb, Brodsky:1981se} authors have proposed existence of the
5-quark state in a proton, which consists of valence $uud$ quarks and the
charm-anticharm pair $c\bar{c}$, i.e., $|uudc\bar{c}\rangle$. This model is
called as the BHPS model, in the frame of which it was assumed that $c(x) =
\bar{c}(x)$. Later it was shown within the light-cone meson-baryon fluctuation
model~\cite{Brodsky:1996hc} that the \emph{intrinsic} quarks and antiquarks
cannot be identical in contrast to the \emph{extrinsic} quarks and antiquarks,
which are necessarily CP symmetric because they are produced from the gluon
splitting. The distributions of $q_\mathrm{ex}(x)$ in comparison with
$q_\mathrm{in}(x)$ exhibits a series of distinctions~\cite{Brodsky:1980pb,
Brodsky:1981se,Brodsky:2015fna} explained in the following. The
\emph{extrinsic} heavy quarks are generated by gluon splitting, therefore their
PDFs are softer than those of the parent gluon by a factor of $(1 - x)$. In
contrast, the \emph{intrinsic} heavy quark state in proton
$|uudq_\mathrm{in}\bar{q}_\mathrm{in}\rangle$ is dominated at high $x$, when it
is minimally off shell and the \emph{intrinsic} quarks distributions correspond
to those of constituent quarks. The resulting momentum and spin distributions
of \emph{intrinsic} $q_\mathrm{in}$ and $\bar{q}_\mathrm{in}$ can be
$q_\mathrm{in}(x) \neq \bar{q}_\mathrm{in}(x)$. One can argue that the proton
quantum numbers are identical with those of $uud$ and so the intrinsic PDFs of
$q_\mathrm{in}$ and $\bar{q}_\mathrm{in}$ should be identical but with opposite
spins. Fig.~\ref{fig:nucleon} shows a schematic view of a nucleon, which
consists of three valence quarks $q_\mathrm{v}$, the quark-antiquark pairs can
be created by a diagram similar to the cut self-energy graph within the QED\@.
In principle, these pairs can be both the light quark-antiquark pair
\textcolor{purple}{$q\bar{q}$} and heavy quark-antiquark $Q\bar{Q}$ pair.
However, the distribution of the \emph{intrinsic} heavy quark $Q\bar{Q}$
components dominates compared to the \emph{intrinsic} light $q\bar{q}$
components at large $x > 0.1$, as will be discussed later, see
also~\cite{Brodsky:2015fna,Pumplin:2007wg}. Therefore, we will focus mainly on
the properties of heavy flavor \emph{intrinsic} $Q\bar{Q}$ pairs in a nucleon.

\begin{figure}[h!]
\centering
\includegraphics[width=.65\textwidth]{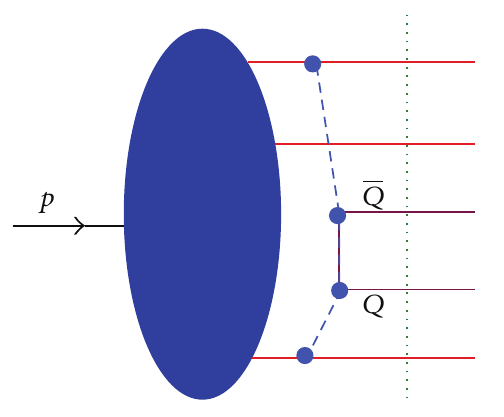}
\caption{Schematic graph of the $Q\bar{Q}$ pair creation in a nucleon.}
\label{fig:QQbar}
\end{figure}

According to Fig.~\ref{fig:QQbar}, if the gluon-gluon scattering box diagram,
$gg \to Q\bar{Q} \to gg$, is inserted into the proton self-energy graph, the
cut of this amplitude ge\-ne\-ra\-tes five-quark Fock states of the proton
\textcolor{purple}{$|uudQ\bar{Q}\rangle$}~\cite{Brodsky:2015fna}.

It was shown in~\cite{Brodsky:1981se} that the existence of \emph{intrinsic}
heavy quark pairs $c\bar{c}$ and $b\bar{b}$ within the proton state could be
due \textcolor{purple}{to the diagrams} where \textcolor{purple}{heavy} quarks are
multiply connected to \textcolor{purple}{valence} quarks. On this basis, within
the MIT bag model~\cite{Donoghue:1977qp}, the probability to find the
five-quark component $|uudc\bar{c}\rangle$ bound within the nucleon bag was
estimated to be about 1~--~2~\%. Later some other models were also
developed~\cite{Paiva:1996dd,Melnitchouk:1997ig,Steffens:1999hx,Pumplin:2005yf,
Pumplin:2007wg,Hobbs:2013bia}. One of them considered a quasi-two-body state
$\bar{D}^0(u\bar{c})\,\bar{\Lambda}_\mathrm{c}^+(udc)$ in the
proton~\cite{Pumplin:2005yf}. In~\cite{Pumplin:2005yf,Pumplin:2007wg,
Nadolsky:2008zw} the probability to find the intrinsic charm (IC) in the proton
(the weight of the relevant Fock state in the proton) was assumed to be
1~--~3.5~\%. The probability of the intrinsic bottom (IB) in the proton is
suppressed by the factor $m^2_\mathrm{c}/m^2_\mathrm{b} \simeq
0.1$~\cite{Polyakov:1998rb}, where $m_\mathrm{c}$ and $m_\mathrm{b}$ are the
masses of the charmed and bottom quarks.

The probability distribution for the 5-quark state ($|uudc\bar{c}\rangle$) in
the light-cone description of the proton was first calculated
in~\cite{Brodsky:1980pb}. The general form for this distribution calculated
within the light-cone dynamics in the so-called BHPS model~\cite{Brodsky:1980pb,
Brodsky:1981se} can be written as~\cite{Pumplin:2005yf}
\begin{equation}
\mathrm{d}P = N \prod_{j = 1}^5 \frac{\mathrm{d}x_j}{x_j} \delta
\left( 1 - \sum_{j = 1}^5 x_j \right) \prod_{j = 1}^5
\mathrm{d}^2 p_{\mathrm{T}j} \delta^{(2)} \left( \sum_{j = 1}^5 p_{\mathrm{T}j}
\right) \frac{F^2(s)}{{(s - m_\mathrm{N}^2)}^2},
\label{def:B}
\end{equation}
where
\begin{equation}
s = \sum_{j = 1}^{5} \frac{p_{\mathrm{T}j}^2 + m_j^2}{x_j}
\end{equation}
and $x_j$ is the longitudinal momentum fraction of the parton, $m_j$ is its
mass and $m_\mathrm{N}$ is the nucleon mass. The form factor $F^2$
characterizes the dynamics of the bound state. It suppresses the contributions
at large values of $p_{\mathrm{T}j}$ and small $x_j$ to make the integrated
probability $P$ converge.

LFWFs (light front wave functions) are defined at a fixed LF (light front) time
$\tau = t + z/c$. They are arbitrarily off-shell in $P^{-} = P^{0} - P^{3}$ and
thus invariant mass $W$.

The intrinsic heavy quark LF Fock states such as $|uud Q \bar{Q}\rangle$ arise
from standard QCD --- any diagram where the $Q \bar{Q}$ is multi-connected to
the valence Fock states.

The simplest IQ Fock states correspond to the insertion of
gluon-gluon scattering $gg \to Q\bar{Q} \to gg$ in the hadron self-energy.
Its cut diagram is presented in Fig.~\ref{fig:QQbar}.

The LFWFs are maximal at minimal off-shellness; i.e.,
minimum invariant mass. This occurs when all of the constituents have the same
rapidity and thus $x_i \propto \sqrt{m^2_i + k^2_{\mathrm{T}i}}$. This is why
the heavy partons have the highest LF $x_i$.

Even in QED one has intrinsic heavy lepton states such as $|e^+ e^- \mu^{+}
\mu^{-}\rangle$ LF Fock states. In the Abelian case the intrinsic heavy lepton
probability falls as $1 \over \cal{M}^\mathrm{4}$\textcolor{purple}{, where
$\cal{M}$ is the effective mass of this system.} In contrast, the probability
falls as $1 \over \cal{M}^\mathrm{2}$ in non-Abelian QCD\@. Thus one has a
direct way to validate QCD\@.

One could actually measure the $x$ of distributions \textcolor{purple}{of}
such Fock states in positronium by colliding and dissociating relativistic
positronium atoms in a gas jet target: $[e^+ e^- ] Z \to e^+ e^- \mu^+ \mu^-
Z$.

This is the analog of the Ashery experiment $\pi A \to q
\bar q A$ which measured the LFWF of the pion.

Since the quarks in $|uud Q\bar{A}\rangle$ tend to have the
same rapidity, they will continue to reinteract. This causes the
$s(x)\bar{s}(x)$ asymmetry in the proton PDF\@.

IQ is rigorous --- it is  a first-principle prediction of
QCD\@. It can be analysed using the OPE as shown by M.~Polyakov et
al.~\cite{Polyakov:1998rb} and earlier by paper from S.J.~Brodsky et
al.~\cite{Brodsky:1984nx}.

The production cross-section for $\gamma^* p \to c \bar{c}$
vanishes as a power of ${W^2 - W^2_\mathrm{threshold}}$. This is why one cannot
reliably use SLAC data to test IC, as noted by Gardner and Brodsky. Very few
events satisfy this cut~\cite{Brodsky:2015uwa}.

The IQ states will be produced at small rapidity $\Delta y$
relative to the SMOG nuclear target in the LHCb fixed target
experiment~\cite{Ammar:1986nk}. This provides an amazing testing ground for
IQ\@.

One can thus validate the ISR experiments which measured
$\Lambda_\mathrm{c}$ and $\Lambda_\mathrm{b}$ at high \xf~\cite{Barlag:1990hg,
Bari:1991in}. This leading particle phenomena was also observed by
SELEX~\cite{Russ:1998rr} in $pA \to \Lambda_\mathrm{c} X$.

A dramatic consequence of IQ\@: Higgs boson production at high $x_\mathrm{F} >
0.8$ at the ISR\@. \textcolor{purple}{This is supported by the measurement of
$J/\psi$ by Badier et al. (NA3)} and even double $J/\psi$ hadroproduction at
high \xf~\cite{Brodsky:2007yz,Brodsky:2006wb,Brodsky:2014hia}.

Neglecting all the quark masses in comparison with the
nucleon mass and the charm quark mass $m_\mathrm{c}$, and
integrating (\ref{def:B}) over $\mathrm{d}x_1 \dots \mathrm{d}x_4$ and putting
$F^2 = 1$ in Eq.~(\ref{def:B}) one can get the following probability to find
the intrinsic charm with momentum fraction $x\equiv x_5$ in the
nucleon~\cite{Blumlein:2015qcn}
\begin{equation}
P(x) = \frac{Nx^2}{6 {(1 - cx)}^5}
\Big( \phi_1(x) + \phi_2(x)\big[\ln(x) - \ln[1 - c(1 - x)x]\big] \Big),
\label{def:fcBlueml}
\end{equation}
where $x = x_5$, $c = m^2_\mathrm{N}/m^2_\mathrm{c}$,
\begin{equation}
\phi_1(x) = (1 - x) (1 - cx) \bigg[1 + x\Big[10 + x - c (1 - x)
\Big(x \big(10 - c(1 - x)\big) + 2\Big)\Big]\bigg]\,,
\label{def:phixa}
\end{equation}
and
\begin{equation}
\phi_2(x) = 6x\big[1 + x\big(1 - c(1 - x)\big)\big][1 - c(1 - x)x].
\label{def:phixb}
\end{equation}
Here $N$ is found from the normalization equation:
\begin{equation}
\int\limits_0^1 P(x)\mathrm{d}x = w,
\label{def:normal}
\end{equation}
where $w$ is the integral fraction of the intrinsic charm. Setting
$c \to 0$ leads to the BHPS result~\cite{Brodsky:1980pb,
Pumplin:2005yf,Blumlein:2015qcn}:
\begin{equation}
P(x) = 600wx^2 \Big[6x(1 + x)\ln x + (1 - x)(1 + 10x + x^2)\Big],
\label{def:fcPumpl}
\end{equation}
Equation~(\ref{def:fcPumpl}) was first derived by Brodsky, Hoyer, Peterson and
Sakai~\cite{Brodsky:1980pb} and, usually, this approach is called as the BHPS
model. In principle, the form factor $F^2$ can be less than 1. It is due to the
suppression of the high mass configurations in Eq.~(\ref{def:fcPumpl}). For
example, it can be chosen as the \emph{exponential suppression} factor
\cite{Pumplin:2005yf}:
\begin{equation}
F^2 = \exp \left( \frac{-(s - m_\mathrm{N}^2)}{\Lambda^2} \right).
\label{def:exponf}
\end{equation}
or the \emph{power-low suppression} factor:
\begin{equation}
F^2 = \frac{1}{{\left(s + \Lambda^2\right)}^{n}}\,.
\label{def:powerf}
\end{equation}
As is shown in~\cite{Pumplin:2005yf}, the results for $P(x)$ using the
exponential or power-low suppression factor with $n = 4$ are rather similar to
the BHPS model, but are somewhat smaller at $x > 0.5$. Values $n \leq 2$ are
unphysical because they result in the total probability
divergence~\cite{Pumplin:2005yf} and the use
Eq.~(\ref{def:powerf}) at $n = 3$ leads to a dependence $F^2\sim
1/m_\mathrm{c}^2$ that is similar to the result of~\cite{Franz:2000ee}.

Another way to suppress the high-mass Fock space components can be made on the
basis of quasi-two-body states~\cite{Brodsky:1996hc,Pumplin:2005yf}. In
particular, the relevant 5-quark Fock configurations can be grouped as
$(udc)(u\bar{c})$ considered as the off-shell two-body state $\bar{D}^0
\Lambda_\mathrm{c}^+$. In this case the factor $F^2$ is chosen, for example,
in~\cite{Pumplin:2005yf} as power-low suppression and can be different for
$\bar{D}^0$ and $\Lambda_\mathrm{c}^+$. It leads to the asymmetry between the
$c$ and \textcolor{purple}{$\bar{c}$-distribution}, which was considered first
in~\cite{Brodsky:1996hc}. The observation of a $c(x) - \bar{c}(x)$ difference
can definitively prove a non-perturbative charm quark component since the
$c\bar{c}$ pairs produced by gluon splitting are symmetric, according to the
NLO calculations, whereas the NNLO calculations can give different shapes for
$c(x)$ and $\bar{c}(x)$, see for example,~\cite{Pumplin:2005yf}
and~\cite{Catani:2004nc}. The charm and anti-charm distributions obtained
within the cloud-model and can be parametrized by the following form at $\mu^2
= m_\mathrm{c}^2$~\cite{Pumplin:2007wg}:
\begin{equation}
c(x) = N x^{1.897} {(1 - x)}^{6.095}; \quad
\bar{c}(x) = \bar{N} x^{2.511} {(1 - x)}^{4.929},
\label{def:c_barc}
\end{equation}
where $N/\bar{N}$ is determined by the quark number sum rule
\cite{Pumplin:2007wg}:
\begin{equation}
  \int\limits_0^1 \left[c(x) - \bar{c}(x)\right] \mathrm{d}x = 0,
\label{def:sr}
\end{equation}
and the normalization factor is related to the IC probability. In fact,
the asymmetry between $\bar{\Lambda}_\mathrm{c}^-$ and $\Lambda_\mathrm{c}^+$
baryons was observed by the SELEX Collaboration~\cite{Garcia:2001xj} in
$\Sigma^- p$ collision at the beam momentum about 600~\gev. This asymmetry was
about 50~\% at $x_\mathrm{F} \simeq 0.3$. However, in $\pi^- p$ and $pp$
collisions the asymmetry was practically invisible because
of the very poor statistics.

\textcolor{purple}{For the strange-antistrange pairs asymmetry similar to the
charm-anticharm one was predicted in~\cite{Brodsky:1996hc}}
and~\cite{Steffens:1999hx} within the light-cone five meson-baryon fluctuation
model assuming hypothesis about the \emph{intrinsic strangeness} in a nucleon.
The difference between $c(x)$ and $\bar{c}(x)$ was studied, for example,
in~\cite{Brodsky:1996hc}. Later we will discuss a possible  verification of
this quark-antiquark asymmetry of the nucleon sea at the LHC\@.
 
Alternatively, the relevant 5-quark configurations can be grouped as
$(uud)(c\bar{c})$ or the off-shell proton and
$J/\Psi$-meson~\cite{Pumplin:2005yf}. However, this quark combination does not
result in the $c-\bar{c}$ asymmetry.

In contrast to the light-cone approach of heavy flavor Fock quark components in
a nucleon, the alternative purely phenomenological scenario was
suggested~\cite{Pumplin:2007wg}. In this model the charm distribution in a
nucleon is \emph{sea like}, i.e., similar to that of the light-flavor sea
quarks $c(x) = \bar{c}(x) = \bar{d}(x) + \bar{u}(x)$ at $\mu_0^2 =
m_\mathrm{c}^2$.

The initial non-perturbative forms for $c(x)$ and $\bar{c}(x)$ at $\mu_0$
specified above are usually used as inputs to the general-mass perturbative
DGLAP evolution. The \emph{global analysis} of distributions of the
\emph{extrinsic} and \emph{intrinsic} heavy-flavor and light-flavor quark
distributions and gluons was performed in~\cite{Pumplin:2007wg} on basis of the
BHPS, the meson-cloud model, \emph{sea like} scenario and the PDF of type
CTEQ6.5, which does not contain the IC contribution. Therefore, we will
not redraw the figures from~\cite{Pumplin:2007wg} and~\cite{Pumplin:2005yf},
but let us discuss the main properties of these parton distributions as a
function of $x$ at different values of \textcolor{purple}{scale} $\mu$. The
BHPS model for the $x$ distribution for intrinsic heavy flavor quarks was
derived within the light-cone approach. The derivation uses some additional
simplifying approximations, which can be modified. The shape of the intrinsic
quark distribution versus $x$ is due to its dependence on the energy propagator
$1/(s - m_\mathrm{N}^2)$ in Eq.~(\ref{def:B}). Using the DGLAP
$\mu^2$-evolution for parton distributions it was shown that the intrinsic
charm provides the dominant contribution to $c(x)$ and $\bar{c}(x)$ at any
$\mu$, if the form of the IC distribution is given by the BHPS model with the
non-zero IC probability $w$. The comparison of the charm distribution to other
flavor distributions presented, for example, in~\cite{Pumplin:2007wg} showed
that the inclusion of the IC contribution with $w = 1$ and 3.5~\% leads to an
enhancement, which is larger than the values of light sea quark distributions
($u_\mathrm{sea} = \bar{u}_\mathrm{sea}$, $d_\mathrm{sea} =
\bar{d}_\mathrm{sea}$, $s_\mathrm{sea} = \bar{s}_\mathrm{sea}$) at $x > 0.4$ at
any QCD scale $\mu$. The intrinsic charm contribution $c_\mathrm{in}(x)$
dominates compared to the extrinsic one $c_\mathrm{ex}(x)$ at $x > 0.1$. The
shape of the charm quark distribution $c(x) = c_\mathrm{ex}(x) +
c_\mathrm{in}(x)$ is similar to the valence quark distribution, but is smaller
by a factor of about 10 in the whole region of $0.003 < x < 1$ at low $\mu$ and
in the region of $x > 0.1$ at large $\mu$.  The results for the IC distribution
using the BHPS model and the $D_0 \Lambda_\mathrm{c}^+$ meson-cloud model are
close~\cite{Pumplin:2007wg}. The \emph{sea like} scenario results
in no any enhancement for the charm quark distribution at $x > 0.1$, as it is
within the BHPS and the meson-cloud models. However, the \emph{sea like} model
gives for $c(x)$ an excess about a factor of 1.5~--~2 compared to the
$c_\mathrm{ex}(x)$ at $0.003 < x < 0.1$ and low $\mu$ and there is no excess at
large $\mu \geq 100$~\gev.

\subsection{Intrinsic charm density in a proton as a function of IC
probability $w$} 

According to~\cite{Brodsky:2015fna,Rostami:2015iva,Catani:1990eg}, the
intrinsic charm distribution at the starting scale $\mu_0^2$ as a function of
$x$ can be presented in the following approximated form
\textcolor{purple}{similar to Eq.~\ref{def:fcPumpl}}:
\begin{eqnarray}
c_\mathrm{int}(x, \mu_0^2) = c_0 w x^2
\Big[(1 - x) (1 + 10x + x^2) + 6x(1 + x)\ln(x) \Big],
\label{def:fcPumplw}
\end{eqnarray}
where $w$ is the probability to find the Fock state $|uudc\bar{c}\rangle$ in
the proton, $c_0$ is the normalization constant and the masses of the light
quarks and the nucleon are negligible compared to the charm
quark mass. The inclusion of the non-zero nucleon mass leads to a more
complicated analytic form~\cite{Pumplin:2007wg,Blumlein:2015qcn}. According to
the BHPS model~\cite{Brodsky:1980pb,Brodsky:1981se}, the charm density in a
proton is the sum of the \emph{extrinsic} and \emph{intrinsic} charm densities,
\begin{eqnarray}
xc(x, \mu_0^2) = xc_\mathrm{ext}(x, \mu_0^2) + xc_\mathrm{int}(x, \mu_0^2).
\label{def:cdens_start}
\end{eqnarray}
The \emph{extrinsic}, or ordinary quarks and gluons are generated on a
short-time scale associated with the large-transverse-momentum processes. Their
distribution functions satisfy the standard QCD evolution equations.
Contrariwise, the \emph{intrinsic} quarks and gluons can be associated with a
bound-state hadron dynamics and one believes that they have a non-perturbative
origin. It was argued~\cite{Brodsky:1981se} that existence of \emph{intrinsic}
heavy quark pairs $c\bar{c}$ and $b\bar{b}$ within the proton state can be due
to the gluon-exchange and vacuum-polarization graphs presented in
Fig.~\ref{fig:QQbar}.

\begin{figure}[h!]
\centering
\includegraphics[width=.65\textwidth,angle=270]{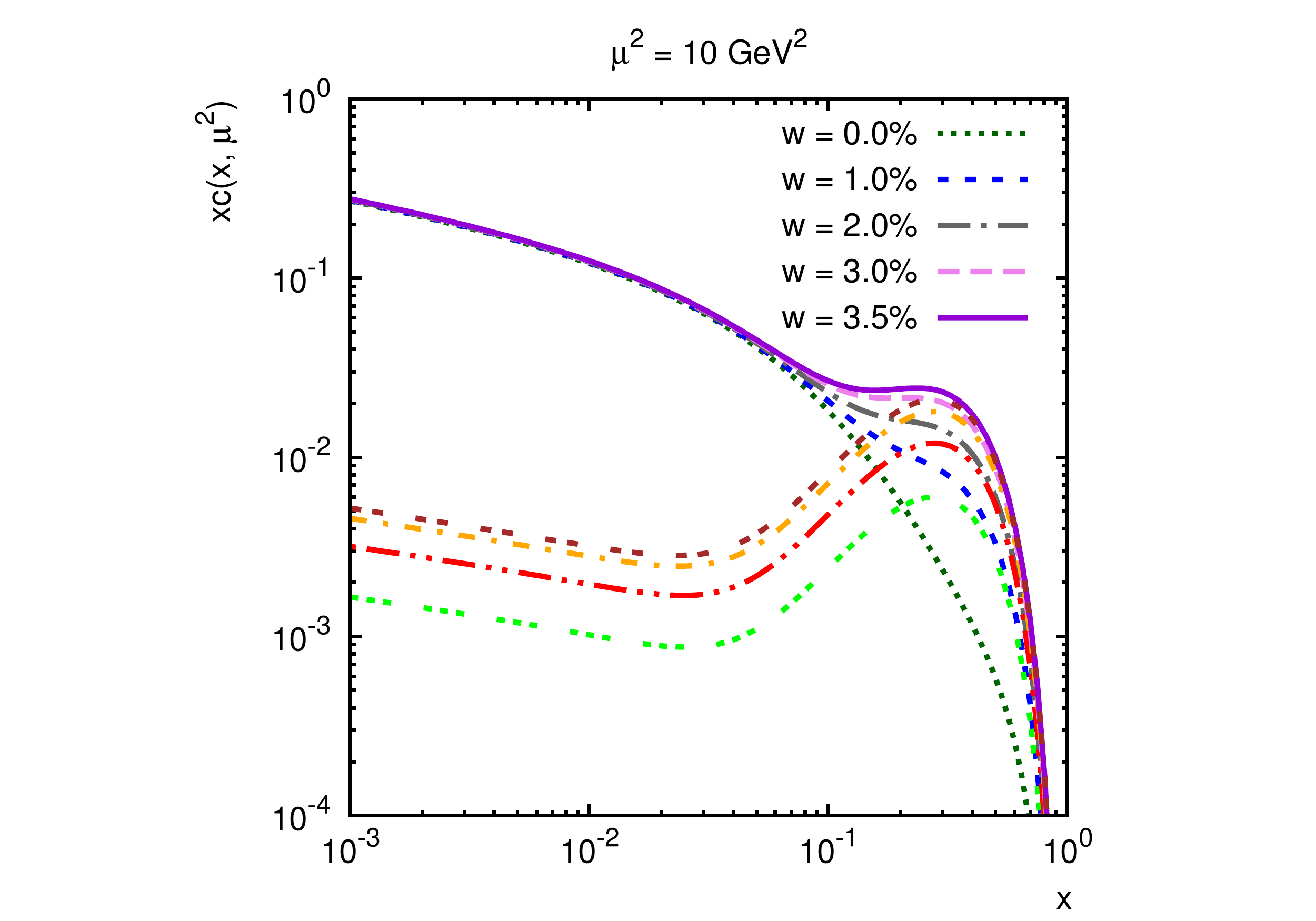}
\includegraphics[width=.65\textwidth,angle=270]{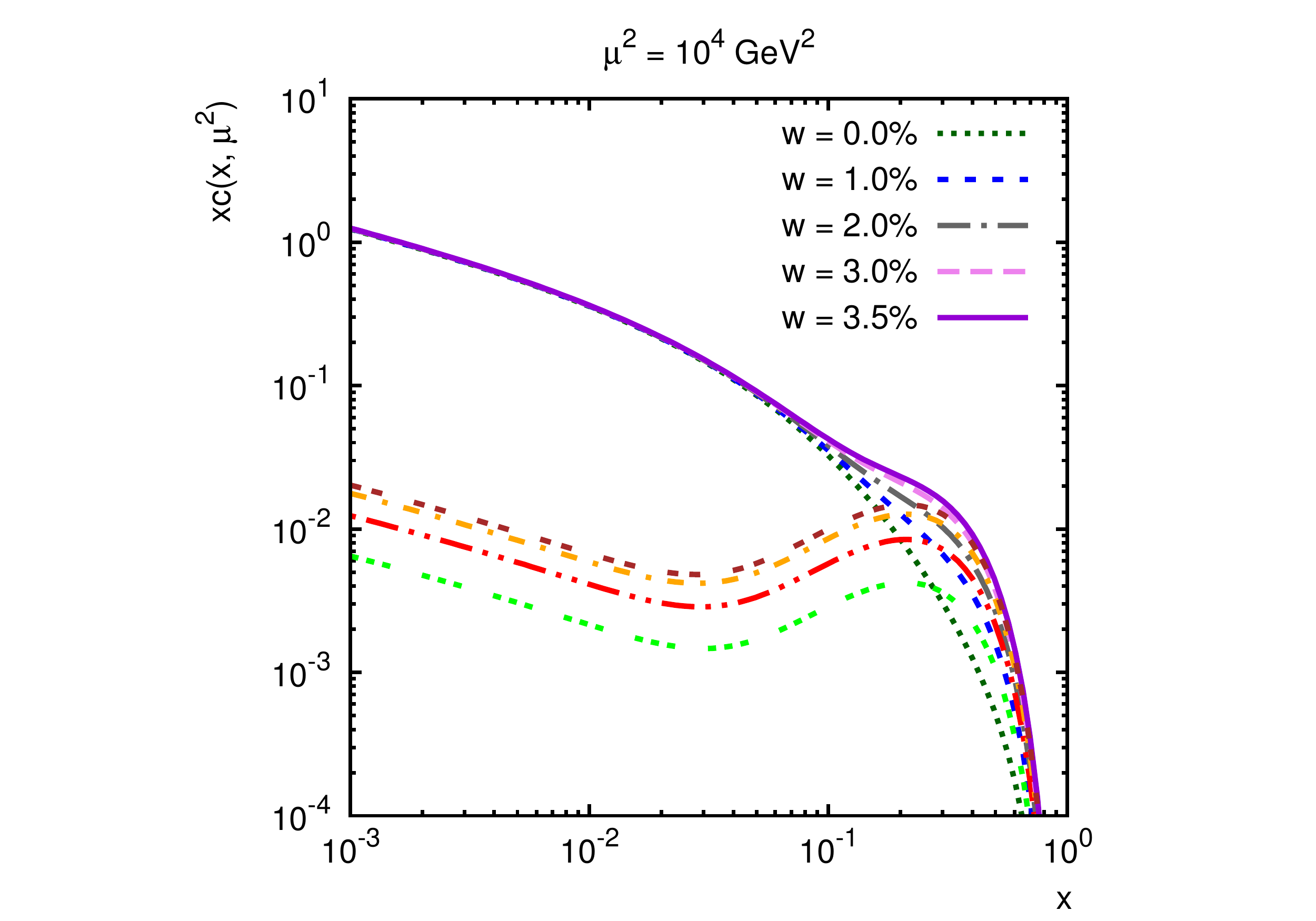}
\caption{The total charmed quark density $xc(x,\mu^2)$ as a function of $x$ at
different values of $w$ at $\mu^2 = 10$~\gevs\ (top) and $\mu^2 =
10^4$~\gevs\ (bottom). The triple-dashed line is the IC contribution at $w =
1$~\%, the dashed-double-dotted line corresponds to the IC at $w = 2$~\%, the
dashed-dotted curve is the IC at $w = 3$~\% and the double-dashed line
corresponds to the IC at $w = 3.5$~\%.}
\label{fig:xc}
\end{figure}

The charm density $xc(x, \mu^2)$ at an arbitrary scale $\mu^2$ is calculated
using the Dokshitzer-Gribov-Lipatov-Altarelli-Parisi (DGLAP)
equations~\cite{Gribov:1972ri,Altarelli:1977zs,Dokshitzer:1977sg}.  Such
calculations were done by the CTEQ~\cite{Nadolsky:2008zw} and CT14
groups~\cite{Dulat:2015mca} at some fixed values of the IC probability $w$.
Namely, the CTEQ group used $w = 1$~\% and $w = 3.5$~\%, and CT14 used $w =
1$~\% and $w = 2$~\%. To calculate the charm density function $xc(x, \mu^2)$ at
any reasonable value of $w$, we use the following
approximation~\textcolor{purple}{Eq.~(\ref{def:xcint})}. In the general case,
there is some mixing between two parts of
\textcolor{purple}{Eq.~(\ref{def:cdens_start})} during the DGLAP evolution. It
can be seen from comparison of our calculations of charmed quark densities
presented in Fig.~\ref{fig:xc}, where this mixing was included within the
CTEQ~\cite{Nadolsky:2008zw} set, and Fig.~2 of~\cite{Rostami:2015iva}, when the
mixing between two parts of the charm density was neglected.  Our results on
the total charm density $xc(x,\mu^2)$ are in good agreement with the
calculations of~\cite{Rostami:2015iva} at the whole kinematical region of $x$
because at $x < 0.1$ the IC contribution $xc_\mathrm{int}$ is much smaller than
the \emph{extrinsic} one $xc_\mathrm{ext}$. However, such mixing is
negligible~\cite{Rostami:2015iva}, especially at large $\mu^2$ and $x$.
Therefore, one can apply the DGLAP evolution separately to the first part
$xc_\mathrm{ext}(x, \mu_0^2)$ and the second part $xc_\mathrm{int}(x,\mu_0^2)$
of Eq.~(\ref{def:cdens_start}).

Taking into account that the IC probability $w$ enters into
Eq.~(\ref{def:cdens_start}) as a constant in front of the function dependent on
$x$ and $\mu^2$, one can suggest a simple relation at any ${w \leq w_{\max}}$:
\begin{eqnarray}
xc_\mathrm{int}(x, \mu^2) = \frac{w}{w_{\max}} xc_\mathrm{int}(x, \mu^2)
\Bigg|_{w = w_{\max}}.
\label{def:xcint}
\end{eqnarray}
Actually, Eq.~\ref{def:xcint} is the linear interpolation between two charm
densities at the scale $\mu^2$, obtained at $w = w_{\max}$ and $w = 0$. Later
we adopt the charm distribution function from the CTEQ66M
set~\cite{Nadolsky:2008zw}. We assume $w_{\max} = 3.5$~\% everywhere, which
corresponds to the CTEQ66c1 set~\cite{Nadolsky:2008zw}.

The charmed quark densities at different $w$ and $\mu^2$ are shown in
Fig.~\ref{fig:xc}.

Our results are in good agreement with the calculations~\cite{Rostami:2015iva}
but are obtained in the more straightforward manner. Additionally, we performed
the three-point interpolation of the charmed quark distributions (over $w = 0$,
$w = 1$~\% and $w = 3.5$~\%, which correspond to the CTEQ66M, CTEQ66c0 and
CTEQ66c1 sets, respectively). These results differ from the ones based
on~(\ref{def:xcint}) by no more than $0.5$~\%, thus giving us the confidence in
our starting point. \textcolor{purple}{The comparison to CT14 with IC 1~\% and
2~\% was also performed and it is in very good agreement with results obtained
using Eq.~(\ref{def:xcint}). We conclude that the sensitivity of resulting
distributions to choice of intrinsic charm PDF (CTEQ66 or CT14) is very small.}

Below we apply the charmed quark density obtained by (\ref{def:cdens_start})
and (\ref{def:xcint}) to calculate the total and differential cross-sections of
associated prompt photon \textcolor{purple}{$\gamma$} or $Z$ boson and heavy
flavor jet production, $\gamma(Z) + Q$, at the LHC conditions. The suggested
procedure to calculate $xc_\mathrm{int}(x, \mu^2)$ at any $w \leq w_{\max}$
allows us to reduce significantly the time for the calculation of these
observables.

As a rule, the gluons and sea quarks play the key role in hard processes of
open charm hadroproduction. Simultaneously, due to the non-perturbative
\emph{intrinsic} heavy quark components one can expect some excess of these
heavy quark PDFs over the ordinary sea quark PDFs at $x > 0.1$. Therefore the
existence of this intrinsic charm component can lead to some enhancement in the
inclusive spectra of open charm hadrons, in particular $D$-mesons, produced at
the LHC in $pp$-collisions at large pseudo-rapidities $\eta$ and large
transverse momenta \pt~\cite{Lykasov:2012hf}. Furthermore, as we know
from~\cite{Brodsky:1980pb,Brodsky:1981se,Donoghue:1977qp,Pumplin:2005yf,
Pumplin:2007wg,Nadolsky:2008zw} photons produced in association with heavy
quarks $Q(\equiv c,b)$ in the final state of $pp$-collisions provide valuable
information about the parton distributions in the proton~\cite{Polyakov:1998rb,
Pumplin:2005yf,Pumplin:2007wg,Nadolsky:2008zw,Goncalves:2008sw,Lykasov:2012hf,
Peng:2012rn,Airapetian:2008qf,Litvine:1999sv,Abazov:2009de,D0:2012gw,
Aaltonen:2009wc,Abazov:2012ea,Bednyakov:2013zta,Lipatov:2012rg,Lipatov:2005wk,
Vogt:2000sk,Navarra:1995rq,Melnitchouk:1997ig}.

In this paper, having in mind these considerations we will first discuss
where the above-mentioned heavy flavor Fock states in the proton could be
searched for at the LHC energies. Following this we analyze in detail, and give
predictions for, the LHC semi-inclusive $pp$-production of prompt photons
accompanied by $c$-jets including the \emph{intrinsic} charm component in the
PDF\@.

\section{Intrinsic heavy quarks and hard $pp$ collisions}
\label{sec:heavy_quarks}

\subsection{Where can one look for the intrinsic heavy quarks?}
\label{subsec:where_IQ}

It is known that in the open charm/\textcolor{purple}{bottom} $pp$-production at large momentum
transfer the hard QCD interactions of two sea quarks, two gluons and a gluon
with a sea quark play the main role. \textcolor{purple}{The cross section for
hard inclusive hadronic reactions $pp \to h X$ can be
factorized~\cite{Collins:1989gx} as the product of structure and fragmentation
functions convoluted with the sum of contributing $2 \to 2$ quark and gluon
subprocess $i + j \to i^\prime + j^\prime$ cross sections. This result can be
presented in the following general form~\cite{Feynman:1978dt} (see
also~\cite{Bednyakov:2011hj})}:
\begin{equation}
E \frac{\mathrm{d}\sigma}{\mathrm{d}^3p} =
\sum_{i,j}\! \int\! \mathrm{d}^2 k_{\mathrm{T}i}\! \int\!
\mathrm{d}^2k_{\mathrm{T}j}\!\!\! \int\limits_{x_i^{\min}}^1\!\!\!
\mathrm{d}x_i\!\!\! \int\limits_{x_j^{\min}}^1\!\!\!\mathrm{d}x_j
f_i(x_i, k_{\mathrm{T}i}) f_j(x_j, k_{\mathrm{T}j})
\frac{\mathrm{d}\sigma_{ij}(\hat{s},\hat{t})}{\mathrm{d}\hat{t}}
\frac{D_{i,j}^\mathrm{h}(z_\mathrm{h})}{\pi z_\mathrm{h}}.
\label{def:rho_c}
\end{equation}
Here $k_{i,j}$ and $k_{i,j}^\prime$ are the four-momenta of the partons $i$ or
$j$ before and after the elastic parton-parton scattering, respectively;
$k_{\mathrm{T}i}$, $k_{\mathrm{T}j}$ are the transverse momenta of the partons
$i$ and $j$; $z$ is the fraction of the hadron momentum from the parton
momentum; $f_{i,j}$ is the PDF\@; and $D_{i,j}$ is the fragmentation function
(FF) of the parton $i$ or $j$ into a hadron $h$.

When the transverse momenta of the partons are neglected in comparison with the
longitudinal momenta, the variables $\hat{s}$, $\hat{t}$, $\hat{u}$ and
$z_\mathrm{h}$ can be presented in the following
forms~\textcolor{purple}{\cite{Collins:1989gx}}:
\begin{eqnarray}
\hat{s} = x_i x_j s, \quad
\hat{t} = x_i \frac{t}{z_\mathrm{h}}, \quad
\hat{u} = x_j \frac{u}{z_\mathrm{h}}, \quad
z_\mathrm{h} = \frac{x_1}{x_i} + \frac{x_2}{x_j},
\label{def:stuzh}
\end{eqnarray}
where
\begin{equation}
x_1 = -\frac{u}{s} = \frac{x_\mathrm{T}}{2} \cot\frac{\theta}{2}, \quad
x_2 = -\frac{t}{s} = \frac{x_\mathrm{T}}{2}\tan\frac{\theta}{2}, \quad
x_\mathrm{T} = 2\frac{\sqrt{t u}}{s} = 2 \frac{p_\mathrm{T}}{\sqrt{s}}.
\end{equation}
Here as usual, $s = {(p_1 + p_2)}^2$, $t = {(p_1 - p_1^\prime)}^2$, $u = {(p_2 -
p_1^\prime)}^2$, and $p_1$, $p_2$, $p_1^\prime$ are the 4-momenta of the
colliding protons and the produced hadron $h$, respectively; $\theta$ is the
scattering angle for the hadron $h$ in the $pp$ c.m.s. The lower limits of the
integration in (\ref{def:rho_c}) are
\begin{equation}
x_i^{\min} = \frac{x_\mathrm{T} \cot\displaystyle\frac{\theta}{2}}
{2 - x_\mathrm{T} \tan\displaystyle\frac{\theta}{2}}, \qquad
x_j^{\min} = \frac{x_i x_\mathrm{T} \tan\displaystyle\frac{\theta}{2}}
{2x_i - x_\mathrm{T} \cot\displaystyle\frac{\theta}{2}}.
\label{def:xijmn}
\end{equation}
Actually, the parton distribution functions $f_i(x_i, k_{iT})$ also depend on
the four-momentum transfer squared $Q^2$ that is related to the Mandelstam
variables $\hat{s}$, $\hat{t}$, $\hat{u}$ for the elastic parton-parton
scattering~\textcolor{purple}{\cite{Collins:1989gx,Feynman:1978dt}}
\begin{equation}
Q^2 = \frac{2\hat{s}\hat{t}\hat{u}}{\hat{s}^2 + \hat{t}^2 + \hat{u}^2}
\label{def:Qsqr}
\end{equation}
One can see that the Feynman variable \xf\ of the produced hadron, can be
expressed via the variables \pt\ and $\eta$, or $\theta$ the hadron scattering
angle in the $pp$ c.m.s,
\begin{equation}
x_\mathrm{F} \equiv \frac{2p_\mathrm{z}}{\sqrt{s}} =
\frac{2p_\mathrm{T}}{\sqrt{s}} \frac{1}{\tan\theta} =
\frac{2p_\mathrm{T}}{\sqrt{s}}\sinh\eta.
\label{def:xFptteta}
\end{equation}
At small scattering angles of the produced hadron this formula becomes
\begin{eqnarray}
x_\mathrm{F} \sim \frac{2p_\mathrm{T}}{\sqrt{s}}\frac{1}{\theta}.
\label{def:xFtetapt}
\end{eqnarray}
It is clear that for fixed \pt\ an outgoing hadron must possess a very small
$\theta$ or very large $\eta$ in order to have large \xf\ (to follow forward, or
backward direction).

In the fragmentation region (of large \xf) the Feynman variable \xf\ of the
produced hadron is related to the variable $x$ of the intrinsic charm quark in
the proton, and  according to the longitudinal momentum conservation law, the
$x_\mathrm{F} \simeq x$ (and $x_\mathrm{F} < x$). Therefore, the visible excess
of the inclusive spectrum, for example, of $K$-mesons can be due to the
enhancement of the IS distribution (see Fig.~\ref{fig:IS}) at $x > 0.1$.

The lower limits of the integration in (\ref{def:rho_c}) can be presented in
the following form:
\begin{equation}
x_i^{\min} = \frac{x_\mathrm{R} + x_\mathrm{F}}{2 - (x_\mathrm{R} - x_\mathrm{F})}, \qquad
x_j^{\min} = \frac{x_i(x_\mathrm{R} - x_\mathrm{F})}{2x_i - (x_\mathrm{R} + x_\mathrm{F})},
\label{def:xijmin}
\end{equation}
where $x_\mathrm{R} = 2p/\sqrt{s}$. One can see from (\ref{def:xijmin}) that,
at least, one of the low limits $x_i^{\min}$ of the integral (\ref{def:rho_c})
must be $\geq x_\mathrm{F}$. Thus if $x_\mathrm{F} \geq 0.1$, then $x_i^{\min}
> 0.1$, where the ordinary (extrinsic) charm distribution is completely
negligible in comparison with the intrinsic charm distribution. Therefore, at
$x_\mathrm{F} \geq 0.1$, or equivalently at the charm momentum fraction
$x_\mathrm{c} > 0.1$ the intrinsic charm distribution intensifies the charm PDF
contribution into charm hadroproduction substantially (see Fig.~\ref{fig:xc}).
As a result, the spectrum of the open charm hadroproduction can be increased in
a certain region of \pt\ and $\eta$ (which corresponds to $x_\mathrm{F} \geq
0.1$ in accordance to~\ref{def:xijmin}). We stress that this excess (or even
the very possibility to observe relevant events in this region) is due to the
non-zero contribution of IC component at $x_\mathrm{c} > x_\mathrm{F} > 0.1$
(where non-IC component completely vanishes).

This possibility was demonstrated for the $D$-meson production at the LHC
in~\cite{Lykasov:2012hf}. It was shown that the \pt\ spectrum of $D$-mesons is
enhanced at pseudo-rapidities of $3 < \eta < 5.5$ and 10~\gev\ $< p_\mathrm{T} <
25$~\gev\ due to the IC contribution, which was included using the CTEQ66c
PDF~\cite{Nadolsky:2008zw}. For example, due to the IC PDF, with probability
about 3.5~\%, the \pt-spectrum increases by a factor of 2 at $\eta = 4.5$. A
similar effect was predicted in~\cite{Kniehl:2012ti}.

One expects a similar enhancement in the experimental spectra of the open
bottom production due to the (hidden) intrinsic bottom (IB) in the proton,
which could have a distribution very similar to the one given in
\textcolor{purple}{(\ref{def:fcPumplw})}. However, the probability
$w_\mathrm{IB}$ to find the Fock state with the IB contribution
$|uudb\bar{b}\rangle$ in the proton is about 10 times smaller than the IC
probability $w_\mathrm{IC}$ due to relation $w_\mathrm{IB}/w_\mathrm{IC} \sim
m_\mathrm{c}^2/m_\mathrm{b}^2$~\cite{Brodsky:1981se,Polyakov:1998rb}.

The IC ``signal'' can be studied not only in the inclusive open (forward) charm
hadroproduction at the LHC, but also in some other processes, such as
production of real prompt photons $\gamma$ or virtual ones $\gamma^*$, or
$Z^0$-bosons (decaying into dileptons) accompanied by $c$-jets in the
kinematics available to the ATLAS and CMS experiments. The contributions of the
heavy quark states in the proton could be investigated also in the $c(b)$-jet
production accompanied by the vector bosons $W^\pm$, $Z^0$. Similar kinematics
given by (\ref{def:xijmin}) and (\ref{def:xFptteta}) can also be applied to
these hard processes.

In the next section we analyze in detail the hard process of the
open strangeness production in $pp$ collisions including the
intrinsic strangeness (IS) contribution in the proton.

\section{Intrinsic strangeness in proton}
\label{sec:is}

\begin{figure}[h!]
\centering
\includegraphics[width=.65\textwidth]{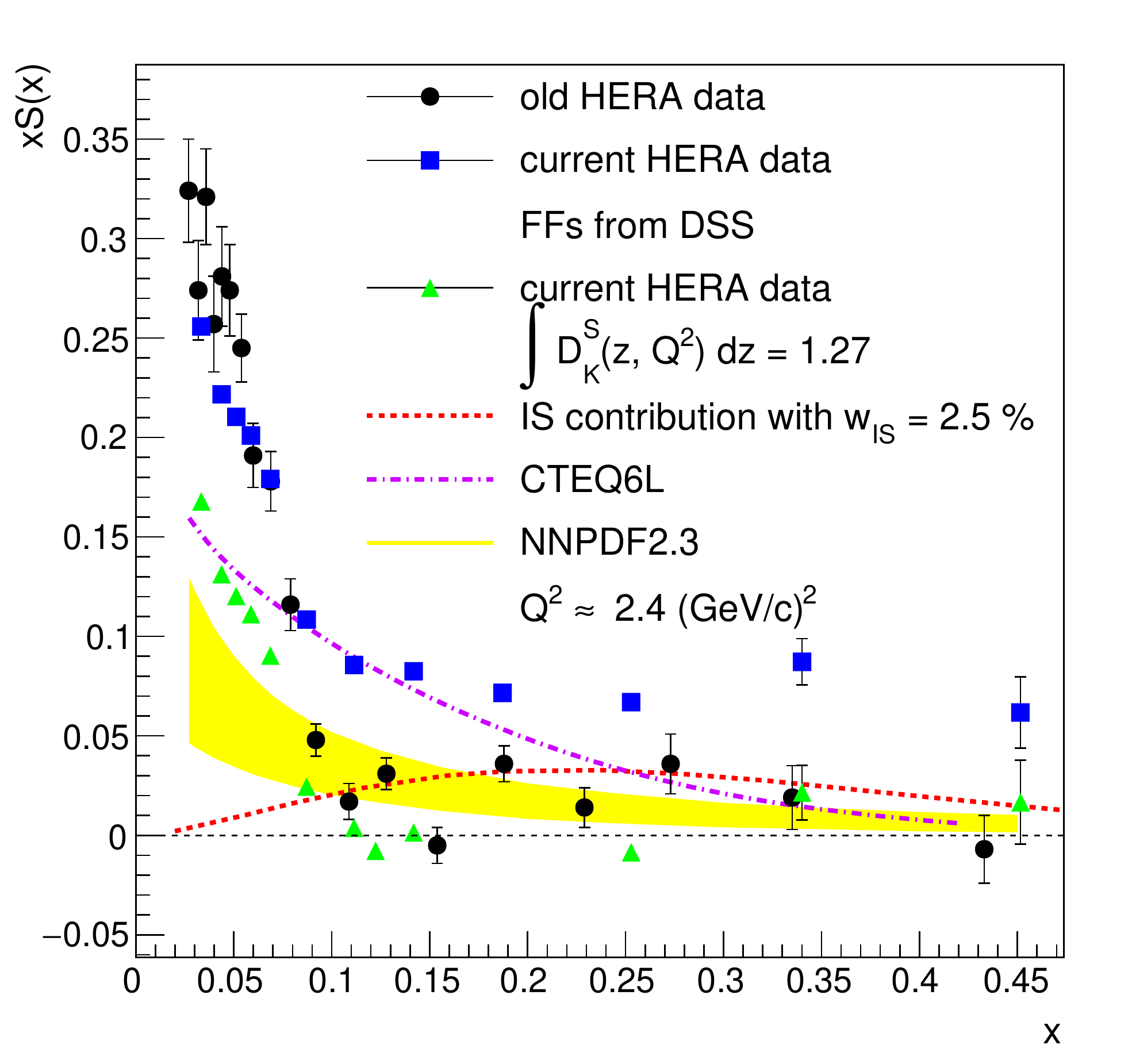}
\caption{The distributions of strange quarks $xS(x) = x(s(x) + \bar{s}(x))$ in
the proton at ${Q^2 = 2.4}$~\gevs, the black points are the old HERMES
data~\cite{Airapetian:2008qf}; the blue squares are the new
data~\cite{Airapetian:2013zaw}, when the FFs are taken
from~\cite{deFlorian:2007aj} (DSS); the green triangles are the new data
assuming he normalization given by Eq.~(\ref{def:normFF}); the dashed curve is
the contribution of the intrinsic strangeness (IS) in the proton with the
probability 2.5~\%; the yellow band corresponds to NNPDF2.3 set
\cite{Ball:2014uwa}.}
\label{fig:IS}
\end{figure}

The BHPS model can be applied also for the search for the intrinsic strangeness
(IS), see, for example,~\cite{Peng:2012rn}, where the early HERMES
data~\cite{Airapetian:2008qf} on the strange quark distribution $xS(x, Q_0^2) =
x[s(x, Q_0^2) + \bar{s}(x, Q_0^2)]$ at $x > 0.1$ and ${Q_0^2 = 2.5}$~\gevs\ has
described rather satisfactorily using the IS contribution in a form about
2.5~\%. Unfortunately later the new HERMES data
appeared~\cite{Airapetian:2013zaw}. The HERMES information on $xS(x, Q_0^2)$
were extracted from the data on the multiplicities of charged $K$-mesons
produced in the deep inelastic $ep$ scattering. The extraction of the
polarization-averaged strange quark distribution from the HERMES data had an
uncertainty related to the fragmentation functions (FFs). In Fig.~\ref{fig:IS}
the old HERMES data~\cite{Airapetian:2008qf} (circles) and the new ones
(squares and triangles) are presented. The square points in Fig.~\ref{fig:IS}
correspond to the FFs taken from~\cite{deFlorian:2007aj} (DSS) and the
triangles correspond to the following assumption:
\begin{equation}
\int D_\mathrm{S}^\mathrm{K}(z, Q_0^2) \mathrm{d}z = 1.27,
\label{def:normFF}
\end{equation}
where $D_\mathrm{S}^\mathrm{K}(z, Q_0^2)$ is the FF of $S$ to $K\equiv K^+ +
K^-$ (the sum of $K^+$ and $K^-$ mesons).

The dashed-dotted line in Fig.~\ref{fig:IS} corresponds to the PDF of type
CTEQ6L~\cite{Pumplin:2002vw}, the dash curve is the IS contribution with its
probability $w_\mathrm{IS} = 2.5$~\% calculated within the BHPS
model~\cite{Brodsky:1980pb,Brodsky:1981se}. One can see from Fig.~\ref{fig:IS}
a large uncertainty of the HERMES data. Therefore, unfortunately the HERMES data
do not allow to extract a reliable information on the IS contribution to the
PDF of strange quarks.

Let us analyze now how the possible existence of the intrinsic strangeness in
the proton can be visible in $pp$ collisions. For example, consider the
$K^-$-meson production in the process $pp \to K^- + X$. Considering the
intrinsic strangeness in the proton~\cite{Lykasov:2013rva} we calculated the
inclusive spectrum $\mathrm{d}\sigma/\mathrm{d}^3p$ of such mesons within the
hard scattering model (Eq.~(\ref{def:rho_c})), which describes satisfactorily
the HERA and HERMES data on the DIS\@. The FF and the parton cross-sections
were taken from~\cite{Albino:2008fy,Mangano:2010zza}, respectively, as
mentioned above.

We also emphasize that the intrinsic $s(x)$ and $\bar{s}(x)$ are expected to be
different in the proton since the comoving valence and strange quarks in the
$|uuds\bar{s}\rangle$ Fock state can repeatedly interact. This can also be
understood by the duality of this Fock state with meson-nucleon fluctuations
such as the $K^+(\bar{s}u) \Lambda(uds)$ state. This
duality~\cite{Brodsky:1996hc} also predicts very different $s(x)$ and
$\bar{s}(x)$ spin distributions in a polarized proton.

\begin{figure}[h!]
\centering
\includegraphics[width=.65\textwidth]{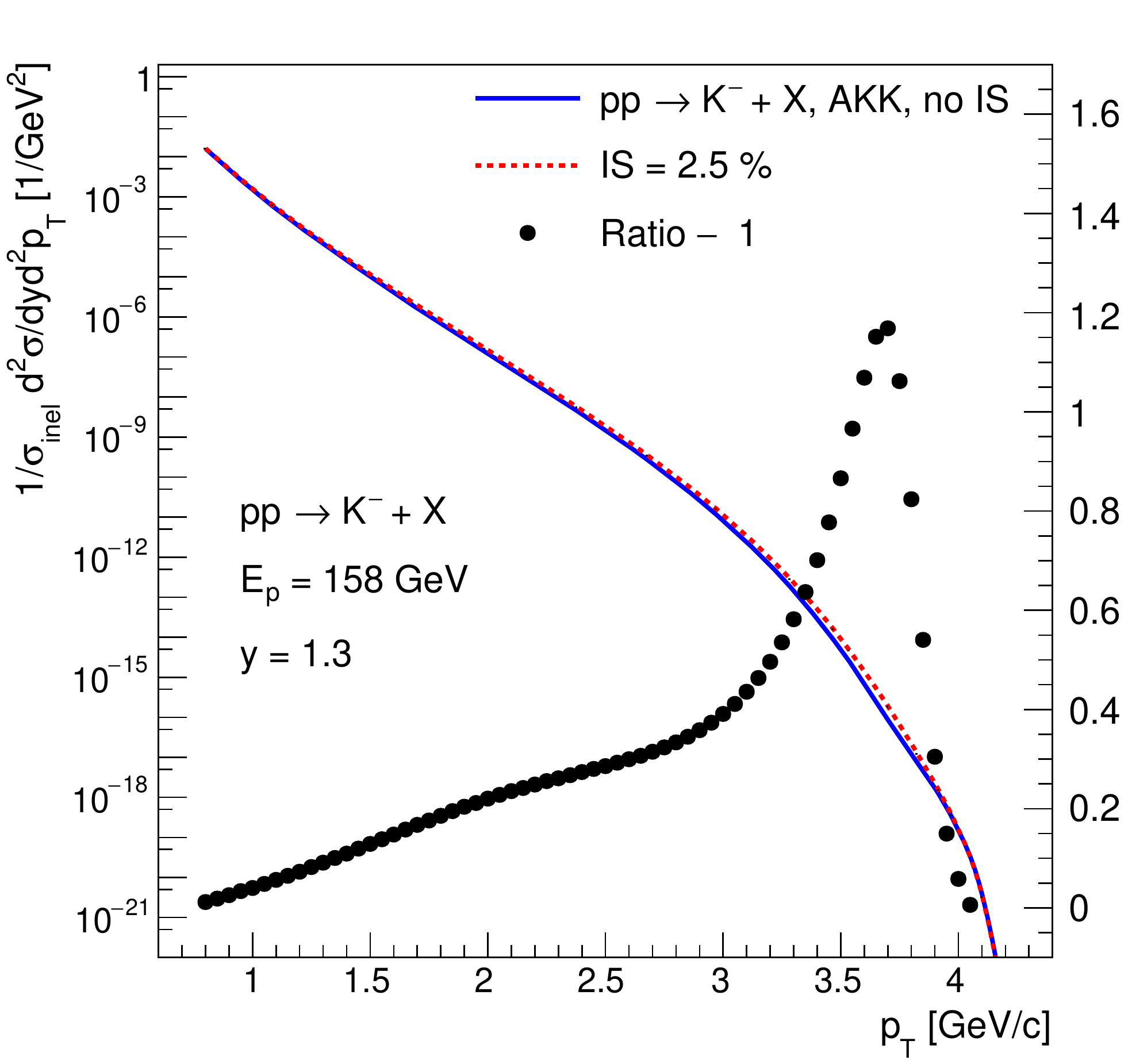}
\caption{The $K^-$-meson distributions (with and without intrinsic strangeness
contribution) over the transverse momentum $p_\mathrm{T}$ for $pp \to
K^- + X$ at the initial energy $E = 158$~\gev, the rapidity $y = 1.3$ and
$p_\mathrm{T} \geq 0.8$~\gev.}
\label{fig:K_meson13}
\end{figure}

\begin{figure}[h!]
\centering
\includegraphics[width=.65\textwidth]{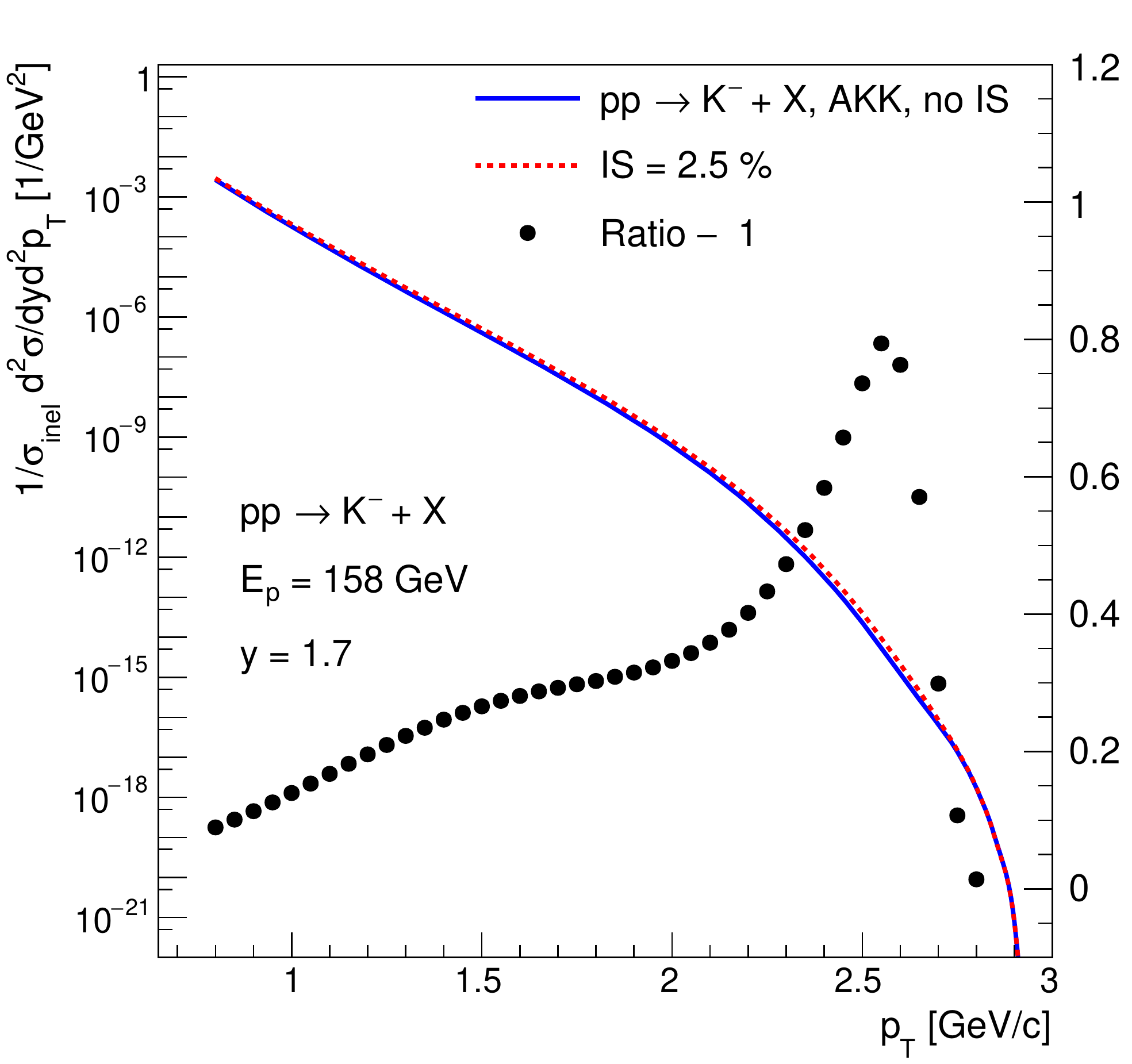}
\caption{The $K^-$-meson distributions (with and without intrinsic strangeness
contribution) over the transverse momentum $p_\mathrm{T}$ for $pp \to
K^- + X$ at the initial energy $E = 158$~\gev, the rapidity $|y| = 1.7$ and
$p_\mathrm{T} \geq 0.8$~\gev.}
\label{fig:K_meson17}
\end{figure}

In Figs.~\ref{fig:K_meson13},~\ref{fig:K_meson17} the inclusive
$p_\mathrm{T}$-spectra of $K^-$-mesons produced in $pp$ collision at the
initial energy $E_p = 158$~\gev\ are presented at the rapidity $|y| = 1.3$
(Fig.~\ref{fig:K_meson13}) and $|y| = 1.7$ (Fig.~\ref{fig:K_meson17}). The solid
lines in  Figs.~\ref{fig:K_meson13},~\ref{fig:K_meson17} correspond to our
calculation ignoring the intrinsic strangeness (IS) in the proton and the
dashed curves correspond to the calculation including the IS with the
probability about 2.5~\%, according to~\cite{Airapetian:2008qf}. The
\textcolor{purple}{dots} show the ratio of our calculation with the IS and
without the IS minus 1. One can see from
Figs.~\ref{fig:K_meson13},~\ref{fig:K_meson17} (right axis) that the IS signal
can be above 200~\% at $|y| = 1.3$, $p_\mathrm{T} = 3.6$~--~3.7~\gev\ and
slightly smaller, than 200~\% at $|y| = 1.7$, $p_\mathrm{T} \simeq 2.5$~\gev.
Actually, this is our prediction for the NA61 experiment that is now under way
at CERN\@.

\section{Global analysis of PDFs with intrinsic charm}
\label{sec:glob_analy}

The charm content of the proton can be studied by the global analysis of PDFs,
the charm quark distributions are generated perturbatively by pair radiation
off gluons and light quarks, vanishing at a scale about the charm mass~\mc, see
details in~\cite{Ball:2014uwa,Ball:2016neh,Rottoli:2016lsg, Brodsky:2015fna}.
In contrast to this, it \textcolor{purple}{was} found in~\cite{Ball:2016neh}
\textcolor{purple}{that} the fitted charm PDF vanishes within uncertainties at
a scale $Q \sim 1.5$~\gev\ at $x\leq 0.1$ independently of the~\mc\ value.
However, \textcolor{purple}{It was also} shown~\cite{Ball:2016neh} that, at $x
\geq 0.1$ and low scales the charm PDF does not vanish and rather has an
intrinsic component, very weakly scale dependent and almost independent of the
\mc\ value, carrying about 1~\% of the total proton momentum. The uncertainties
in all other PDFs are slightly increased by including the IC charm, while the
dependence of these PDFs on \mc\ is significantly reduced~\cite{Ball:2016neh}.
As was shown, the uncertainties in the fitted charm PDF are reduced if the EMC
charm structure function data set is included. The main application of the
results obtained in~\cite{Ball:2014uwa,Ball:2016neh,Rottoli:2016lsg} for the
LHC phenomenology is the increase stability respect to \mc\ persists at high
scales.

The charm PDF can have a non-vanishing intrinsic component of the
non-perturbative origin, see, for example,~\cite{Brodsky:2015fna}. On the other
hand, if one assumes that the charm PDF is purely perturbative in origin,
\textcolor{purple}{it vanishes} below its production threshold.
\textcolor{purple}{The question arises as to the value of this threshold} related
to the charm pole mass, which is not known very precisely. Even if the charm is
entirely perturbative and its production threshold is known,
\textcolor{purple}{there is} a problem \textcolor{purple}{of getting} the accurate
predictions because the \textcolor{purple}{cross-section} of the massive charm
\textcolor{purple}{is} known within the low perturbative order. These difficulties
are solved, for example, within the NNPDF3.0 set~\cite{Ball:2014uwa}. Within
this set the charm PDF is parametrized on the basis of light quark and gluon
PDFs, i.e., with an independent neural network with 37 free parameters.

\section{Intrinsic heavy quark signal in processes at collider energies}
\label{sec:ic_hard}

\subsection{Inclusive production of charmed meson}
\label{subsec:charm_meson}

It was shown that the IC could result in a sizable contribution to the forward
charmed meson production~\cite{Goncalves:2008sw}. Furthermore the IC ``signal''
can constitute almost 100~\% of the inclusive spectrum of $D$-mesons produced
at high pseudo-rapidities $\eta$ and large transverse momenta \pt\ in $pp$
collisions at LHC energies~\cite{Lykasov:2012hf}.

If the distributions of the intrinsic charm or bottom in the proton are hard
enough and are similar in the shape to the valence quark distributions (have
the valence-like form), then the production of the charmed (bottom) mesons or
charmed (bottom) baryons in the fragmentation region should be similar to the
production of pions or nucleons. However, the yield of this production depends
on the probability to find the intrinsic charm or bottom in the proton, but
this yield appears to be rather small. The PDF which
included the IC contribution in the proton have already been used in the
perturbative QCD calculations
in~\cite{Pumplin:2005yf,Pumplin:2007wg,Nadolsky:2008zw}.

The possible existence of the intrinsic charm in the proton can lead to some
enhancement in the inclusive spectra of the open charm hadrons, in particular
$D$-mesons, produced at the LHC in $pp$-collisions at high pseudo-rapidities
$\eta$ and large transverse momenta \pt~\cite{Lykasov:2012hf}.

\begin{figure}[h!]
\centering
\includegraphics[width=.8\textwidth]{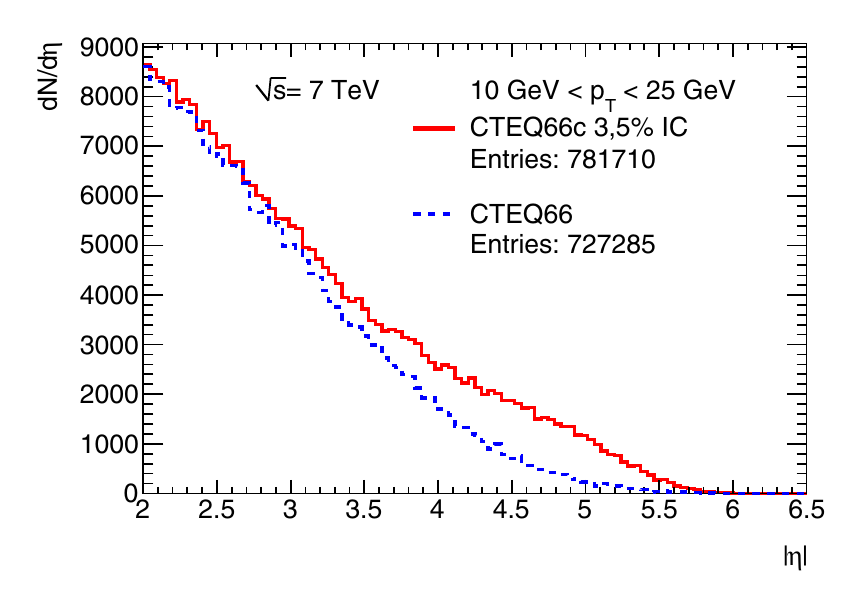}
\caption{The $D_0 + \bar{D̄}_0$ distributions over the pseudorapidity $\eta$ in
${pp \to (D_0 + \bar{D̄}_0)X}$ at ${\sqrt{s} = 7}$~\tev\ and $10 \leq
 p_\mathrm{T} \leq 25$~\gev~\cite{Lykasov:2012hf}.}
\label{fig:D_meson}
\end{figure}

The inclusion of the intrinsic bottom or/and charm in the proton can increase
the yield of the relevant heavy flavour baryons by a factor of 3 to 10. In
particular, we considered a possibility of measuring the reaction $pp
\to \Lambda_\mathrm{c}^+ X \to \Lambda^0 \pi^+ X \to n
\pi^0 \pi^+ X$ using the ATLAS, more specifically, one of its forward detectors
the ZDC\@. This measurement can provide information on the intrinsic charm in
the proton, the probability of which is estimated to be a factor of 10 higher
than the one for the intrinsic bottom in the proton. Finally, it is worth
noticing that any reliable non-observation of this enhancement in the
experiments at the LHC can severely constrain the intrinsic heavy quark
hypothesis.

Our calculations of the charmed meson production in $pp$ collisions were done
within the MC generator PYTHIA8 including the IC contribution with the
probability about 3.5~\% to the PDF are presented in
Fig.~\ref{fig:D_meson}~\cite{Lykasov:2012hf}. It is the distribution of the
single $D$-mesons ($D_0 + {\bar D}_0$) produced in the $pp$ collision at
${\sqrt{s} = 7}$~\tev\ as a function of their pseudo-rapidity $\eta$. We found
that the contribution of the intrinsic charm in the proton could be studied in
the production of $D$-mesons in $pp$ collisions at the LHC\@. The IC
contribution for the single $D^0$-meson production can be sizable, it is about
100~\% at large rapidities $3 \leq |y| \leq 4.5$ and large transverse momenta
$10 \leq p_\mathrm{T} \leq 25$~\gev. As it is shown in~\cite{Lykasov:2012hf},
for the double $D_0$ production this contribution is not larger than 30~\% at
$p_\mathrm{T} \geq 5$~\gev\ and $3 \leq |y| \leq 4.5$.  These IC contributions
for the single and double $D$-meson production were obtained with the
probability of the intrinsic charm taking to be $w_{\mathrm{c}\bar{\mathrm{c}}}
= 3.5 $~\%~\cite{Nadolsky:2008zw}, and they will decrease by a factor of 3 when
$w_{\mathrm{c}\bar{\mathrm{c}}} \simeq 1 $~\%. Therefore, this value can be
verified experimentally at LHCb.

The predictions presented in Fig.~\ref{fig:D_meson}~\cite{Lykasov:2012hf} could
be verified at the LHCb experiment in the kinematic region mentioned above to
observe a possible signal for the intrinsic charm. The intrinsic \textcolor{purple}{bottom} in the
proton is suppressed by a factor of 10~\cite{Polyakov:1998rb}, therefore its
signal in the inclusive spectra of $B$-mesons will probably be very weak.

The IC contributions could be also observed at the SMOG
experiment~\cite{Aaij:2014ida} of the LHCb in the production of the open charm,
for example, $D$-mesons or charmed baryons $\Lambda_\mathrm{c}$. One can use a
gas jet target in a LHC detector~---~such as the SMOG target available at
LHCb~---~to study novel intrinsic heavy quark physics phenomena in $pA$
collisions. Remarkably, heavy-quark hadrons such as the $\Lambda_\mathrm{b}$,
double-charm baryons, and exotic hadrons such as tetraquarks and pentaquarks
containing heavy quarks will be produced at small target rapidities~---~nearly
at rest in the nuclear target rest frame~---~and thus can be easily observed.
We note that the intrinsic heavy quark Fock states of a proton in the nuclear
target, such as $|uudQ\bar{Q}\rangle$, have high light-front momentum fraction
$x_\mathrm{Q}$. The collision materializes the far off-shell light-front wave
functions of Fock states, such as $|uudb\bar{b}\rangle$. The coalescence of the
heavy quarks with the comoving light quarks corresponds to the production of a
heavy hadron such as a $\Lambda_\mathrm{b}(udb)$ at \emph{small rapidity}
$|y_{\Lambda_\mathrm{b}}| \simeq \ln{x_b}$, relative to the rapidity of the
nucleon in the target.

\subsection{Inclusive production of Higgs boson and intrinsic
charm and bottom in proton}

\begin{figure}[h!]
\centering
\includegraphics[width=0.65\textwidth]{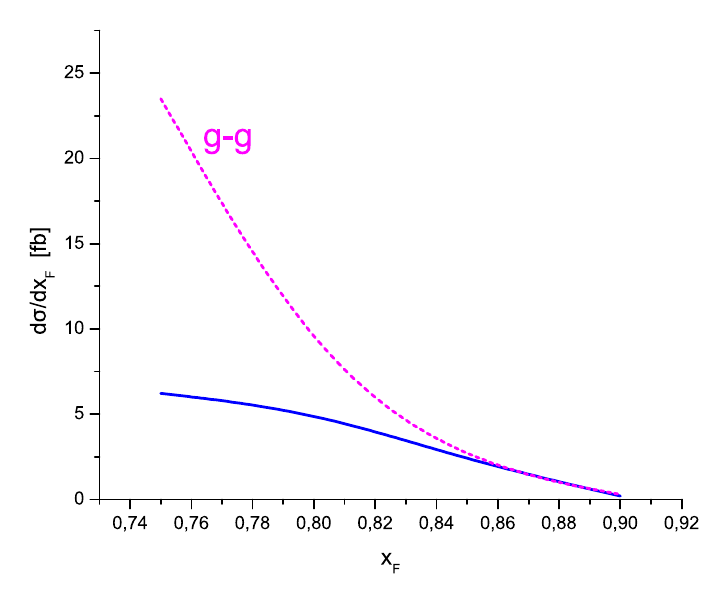}
\includegraphics[width=0.65\textwidth]{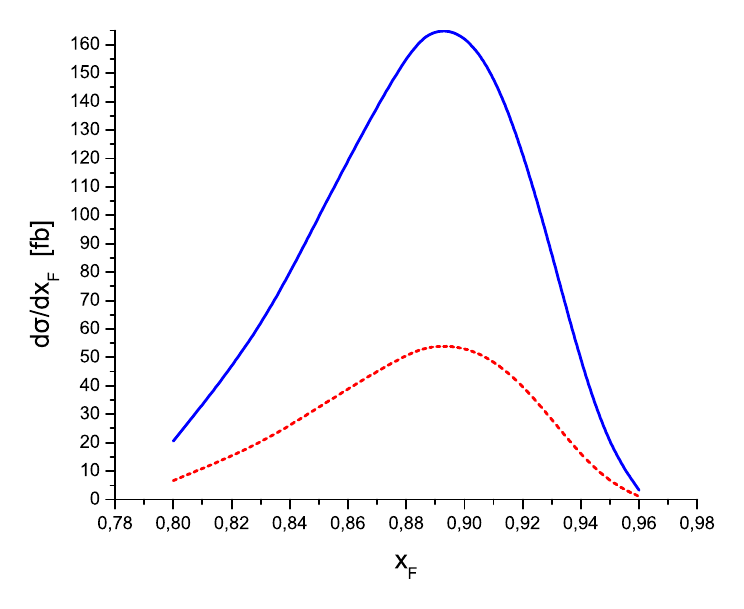}
\caption{The \xf-distribution of the Higgs boson produced in $pp$ collision at
the LHC energy $\sqrt{s} = 14$~\tev\ due to the non-perturbative intrinsic
charm with the probability about 1~\% (solid line). The dashed line corresponds
to the Higgs boson production from the gluon-gluon fusion
(\textcolor{purple}{top}). The same distribution due to the non-perturbative
intrinsic bottom (IB) at the LHC energy $\sqrt{s}= 14$~\tev\ (solid line) and
the TEVATRON energy $\sqrt{s}=$ 2 TeV (dashed line,
\textcolor{purple}{bottom})~\cite{Brodsky:2007yz}.}
\label{fig:higgs}
\end{figure}

The interesting predictions on the possible signal of the intrinsic heavy quark
(IQ) contributions to the inclusive \xf-spectrum of the Higgs bosons produced
at the TEVATRON and LHC energies are presented in~\cite{Brodsky:2007yz}. In
Fig.~\ref{fig:higgs} the contributions of the IC and IB to the \xf-distribution
of the Higgs boson produced in $pp$ collisions at the collider energies LHC and
TEVATRON are presented.

One can see from Fig.~\ref{fig:higgs} that the \xf-distribution for the
inclusive Higgs boson production coming from the IB contribution is much larger
than the one coming from the IC\@. It is due to the fact that the Higgs-$Q$
coupling is proportional to the quark mass $m_\mathrm{Q}$, therefore the
Higgs-$b$ coupling constant is much larger than the coupling constant
Higgs-$c$. Fig.~\ref{fig:higgs} also shows that the cross section
$\mathrm{d}\sigma/\mathrm{d}x_\mathrm{F}$ for the inclusive production of the
Standard Model Higgs boson coming from the IB is of order 150~fb at the LHC
energy $\sqrt{s} = 14$~\tev, peaking in the region of $x_\mathrm{F} \simeq
0.9$. Therefore, the signal of the IB in the differential cross section of the
Higgs boson produced in $pp$ collision at high \xf\ can be tested at the LHC\@.
As is shown in~\cite{Brodsky:2007yz}, it is much larger than the IC signal.

\subsection{Production of prompt photon and $c$ or $b$-jet in hard $pp$
collisions}
\label{subsec:prompt_photon}

\begin{figure}[h!]
\centering
\includegraphics[width=.65\textwidth]{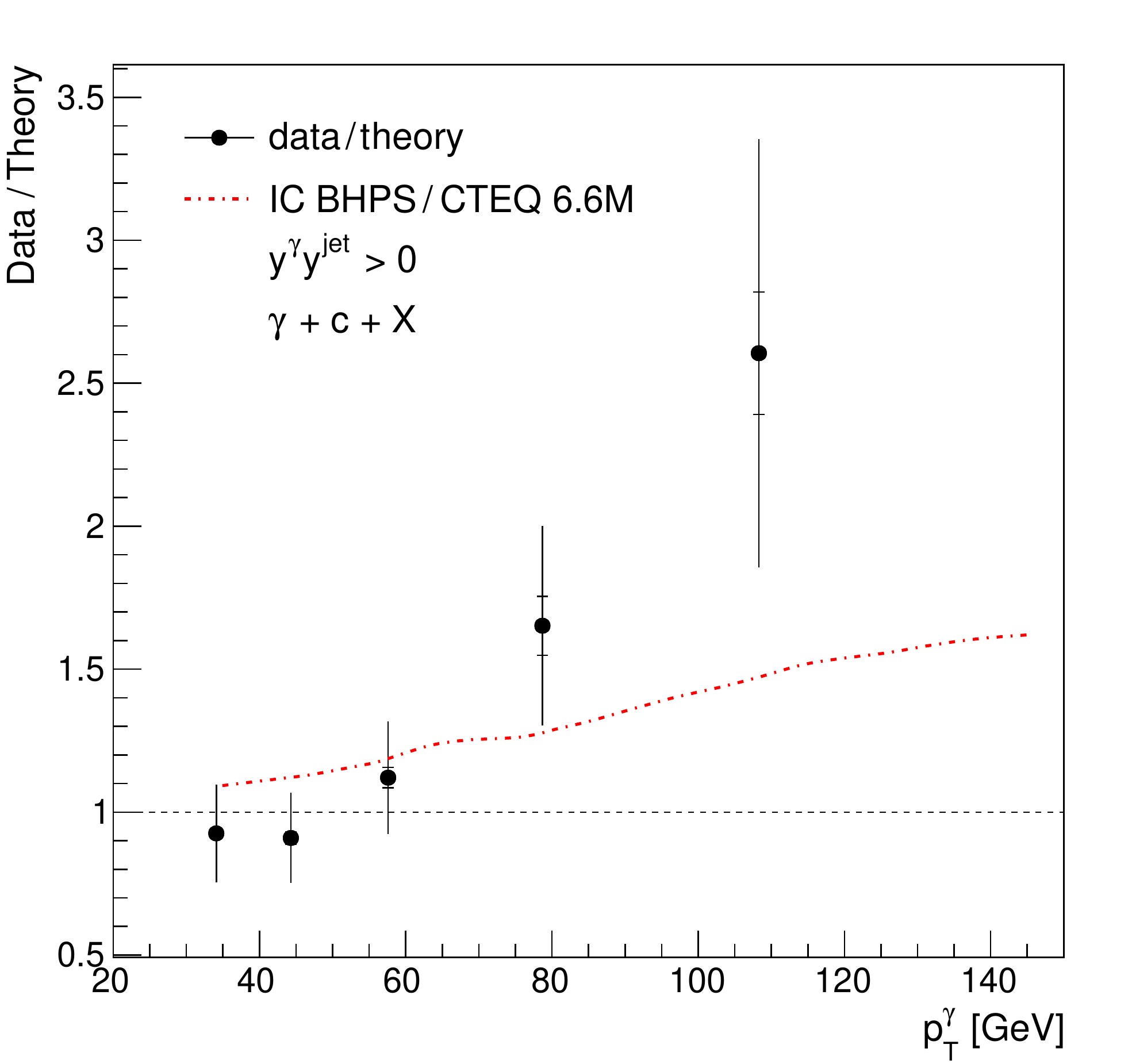}
\includegraphics[width=.65\textwidth]{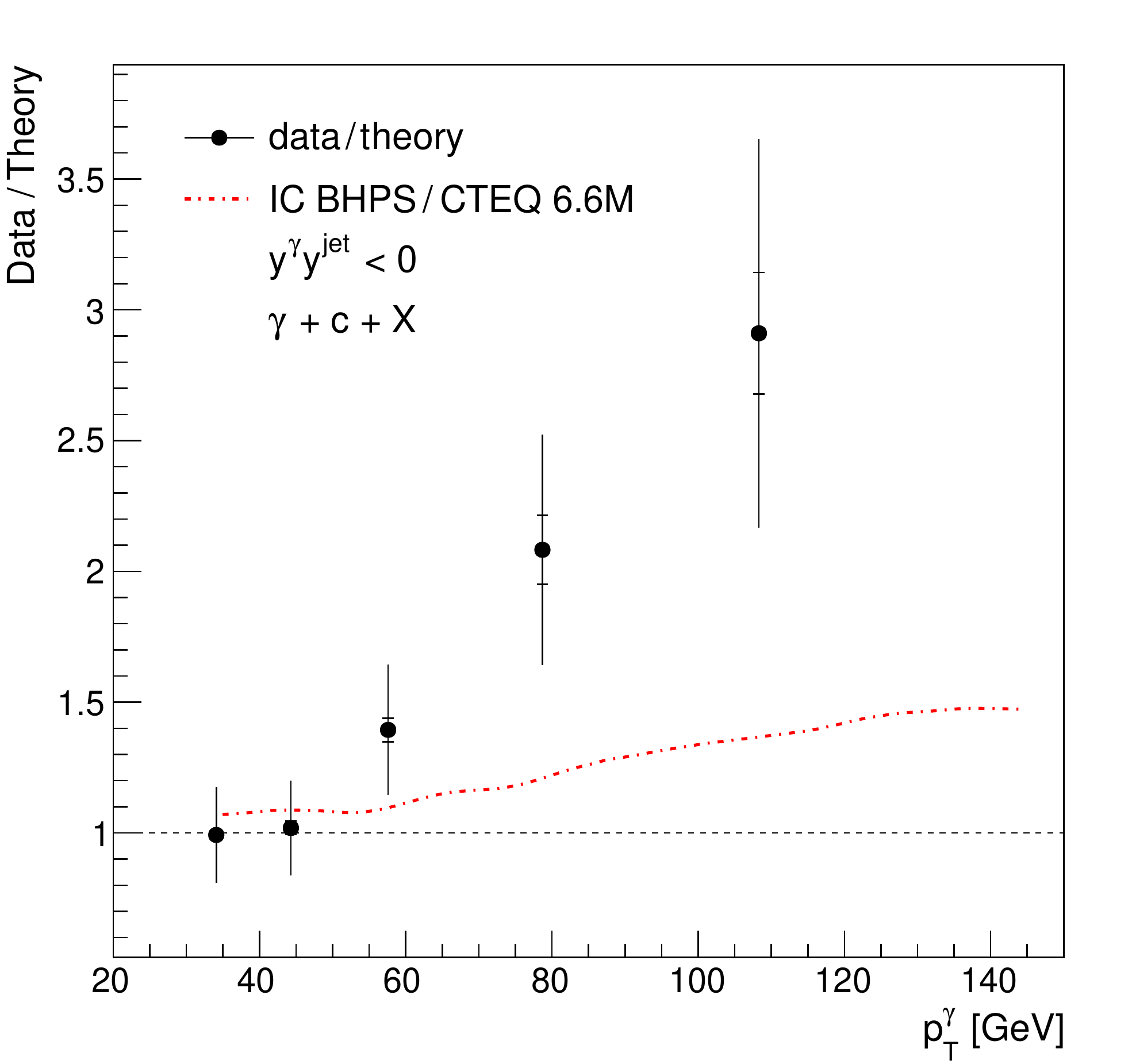}
\caption{The data-to-theory ratio~\cite{Abazov:2009de} for the processes
$p\bar{p} \to \gamma + c + X$, when $y^\gamma y^\mathrm{jet} > 0$ (top)
and the same ratio, when $y^\gamma y^\mathrm{jet} < 0$ (bottom) at $\sqrt{s} =
1.96$~\tev. The dash-dotted line is the calculation of this ratio using the
BHPS IC model with the IC probability about 3.5~\%.}
\label{Fig_D0_pTsp}
\end{figure}

The investigation of prompt photon and $c(b)$-jet production
in $p\bar{p}$ collisions at $\sqrt{s} = 1.96$~\tev\ was carried out at the
TEVATRON~\cite{Abazov:2009de,D0:2012gw,Aaltonen:2009wc,Abazov:2012ea}. In
particular, it was observed that the ratio of the experimental spectrum of the
prompt photons accompanied by the $c$-jets to the relevant theoretical
expectation (based on the conventional PDF, which ignored the IC contribution)
increases up to factor of about 3, when \pty\ becomes above 110~\gev.
Furthermore, taking into account the CTEQ66c PDF, which includes the IC
contribution obtained within the BHPS
model~\cite{Brodsky:1980pb,Brodsky:1981se} one can reduce this ratio up to
1.5~\cite{Stavreva:2009vi}. For the \mbox{$\gamma$ + $b$-jets}
$p\bar{p}$-production no enhancement in the \pty-spectrum was observed at the
beginning of the experiment~\cite{Abazov:2009de,Aaltonen:2009wc}. However, in
2012 the D\O\ collaboration has confirmed observation of such an
enhancement~\cite{Abazov:2012ea}. It is illustrated in Fig.~\ref{Fig_D0_pTsp}.

\begin{figure}[h]
\begin{center}
\includegraphics[width=.65\textwidth]{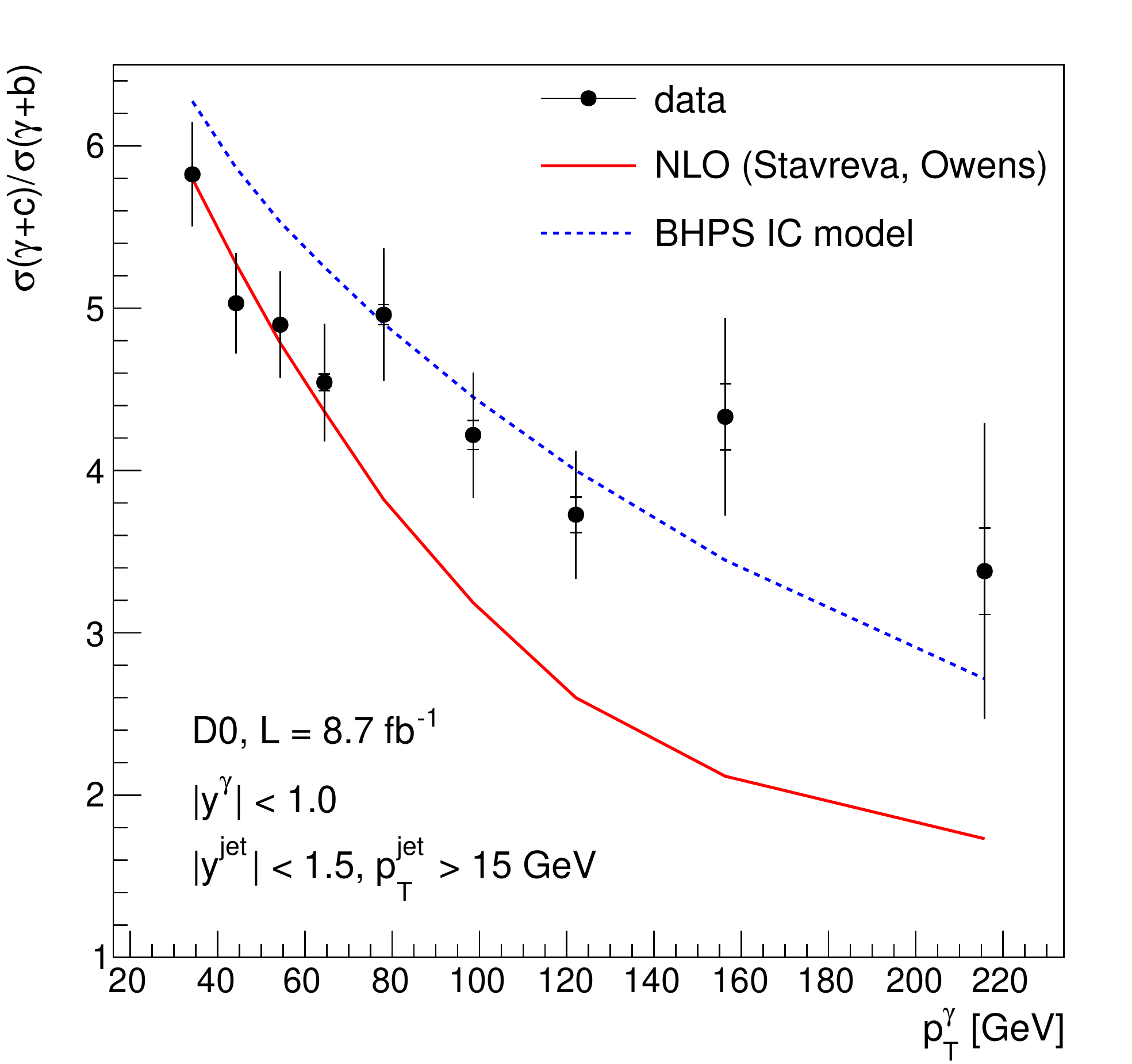}
\caption{The ratio $\sigma(\gamma + c) / \sigma(\gamma + b)$ as a function of
the photon transverse momentum \pty\ in $p\bar{p} \to \gamma + c(b) + X$
process at $\sqrt s = 1.96$~\tev, see~\cite{D0:2012gw}.}
\label{fig2}
\end{center}
\end{figure}

The dash-dotted lines in Fig.~\ref{Fig_D0_pTsp} (bottom) show the ratio between
the NLO calculations of \pty-spectrum obtained within the BHPS model (using the
CTEQ66c with the IC probability about 3.5~\%) and the spectrum using the CTEQ6M
without the IC contribution. In addition to that the IC signal was visible in
the ratio between the differential cross-sections of the photon and $c$-jet
production in $p\bar{p}$ collision, $\gamma + c$, and $\gamma + b$ production,
see Fig.~\ref{fig2}. This figure shows that, according to the pQCD
calculations~\cite{Stavreva:2009vi}, in the absence of the IC contribution this
ratio decreases (solid line), as \pty\ grows, while the TEVATRON data show its
flat behavior at large $p_\mathrm{T} \ge 100$~\gev. As for the prompt photon
production accompanied by the $b$-jet in $p\bar{p}$ annihilation, the
TEVATRON data do not show any signal of the intrinsic $b$
contribution, see Fig.~\ref{Fig_D0_pTsp} (top)~\cite{Abazov:2009de}. It can be
due to very small intrinsic \textcolor{purple}{bottom} probability in a proton, as mentioned above.

This intriguing observation stimulates our interest to look for a similar
IC signal in $pp \to \gamma + c(b) + X$ processes
at LHC energies, see~\cite{Bednyakov:2013zta,Lipatov:2016feu}.

\begin{figure}[h]
\centering
\includegraphics[width=.7\textwidth]{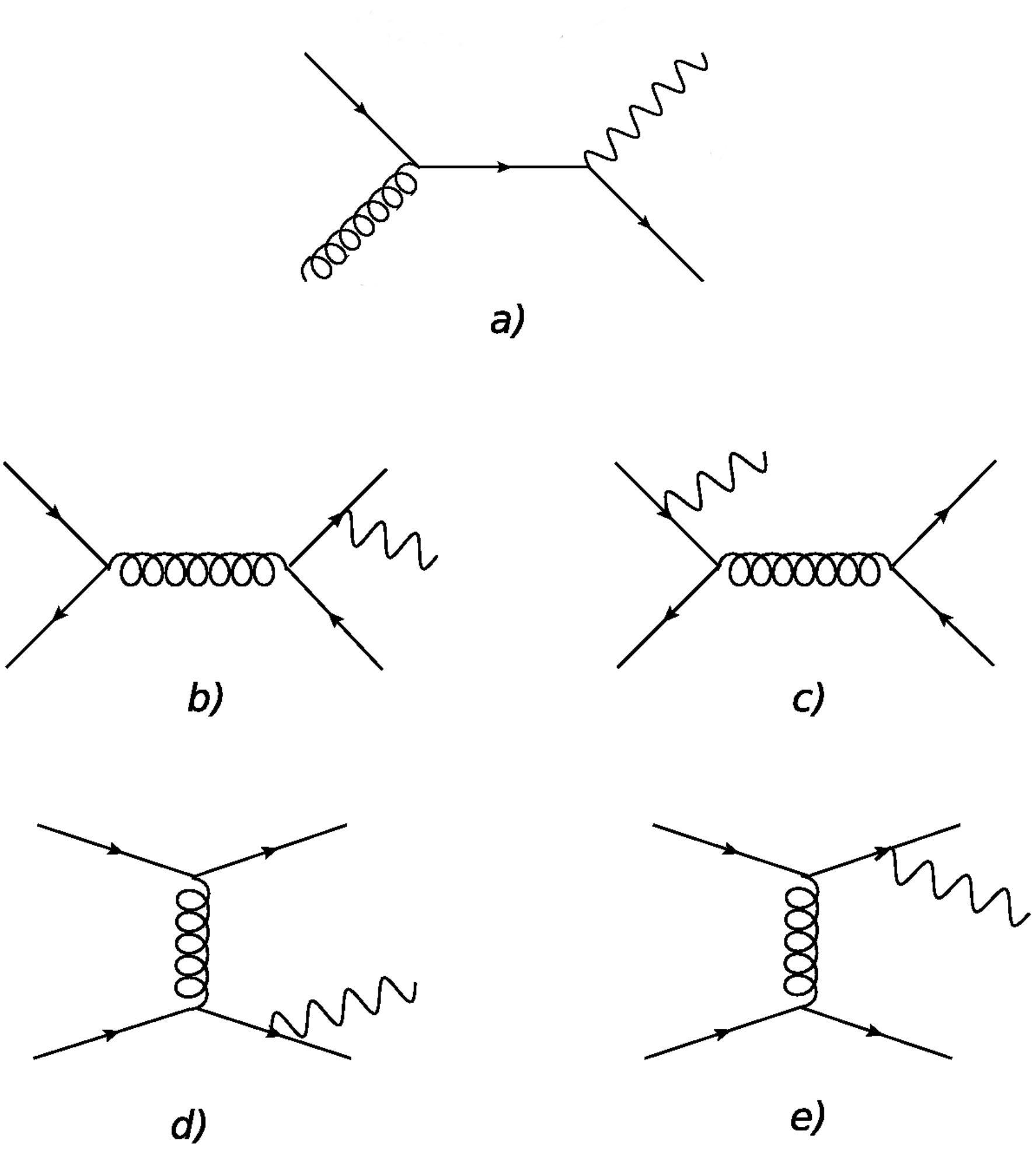}
\caption{The $\cal{O}(\alpha \alpha_\textrm{s})$~(a) and $\cal{O}(\alpha
\alpha_\textrm{s}^\mathrm{2})$~(b)~--~(e) contributions to the $\gamma(Z) + Q$
production.}
\label{fig1}
\end{figure}

The Examples of Feynman diagrams corresponding to $gg \to
\gamma(Z)Q\bar{Q}$ (a), $q\bar{q} \to \gamma(Z)Q\bar{Q}$ (b,c) and $qQ
\to \gamma(Z)qQ$ (d,e) subprocesses are presented in Fig.~\ref{fig1}.

The diagrams within the NLO QCD are more complicated than Fig.~\ref{fig1}.

Let us illustrate qualitatively the kinematical regions where the IC component
can contribute significantly to the spectrum of prompt photons produced
together with a $c$-jet in $pp$ collisions at the LHC\@. For simplicity we
consider only the contribution to the reaction $pp \to \gamma(Z) + Q +
X$ of the diagrams given in Fig.~\ref{fig1} (a). According
to~(\ref{def:xFptteta})
at certain values of the transverse momentum of the photon, \pty, and its
pseudo-rapidity, \etay, (or rapidity \yy) the momentum fraction of $\gamma$
can be $x_\mathrm{F}^\gamma > 0.1$, therefore the fraction of the initial
$c$-quark must
also be above 0.1, where the IC contribution in the proton is enhanced (see
Fig.~\ref{fig:xc}).
Therefore, one can expect some non-zero IC signal in the \pty\ spectrum of the
reaction $pp \to \gamma + c + X$ in this certain region of \pty\ and
\yy~\cite{Bednyakov:2013zta}. A similar IC effect can be visible in the
production of the $Z$ or $W$-boson accompanied by $c$ or $b$-jets in $pp$
collisions. In~\cite{Lipatov:2016feu,Beauchemin:2014rya} the theoretical
predictions about the possible observation of the IC signal in the \pt-spectra
of $Z$ or $W$-bosons accompanied by the $c$ or $b$-jets respectively are
presented. These processes will be considered later. First, let us discuss a
possible search for the IC signal in the production of prompt photons
accompanied by $c$-jet in $pp$ collision at LHC energies.

\begin{figure}[h]
\centering
\includegraphics[width=0.60\textwidth]{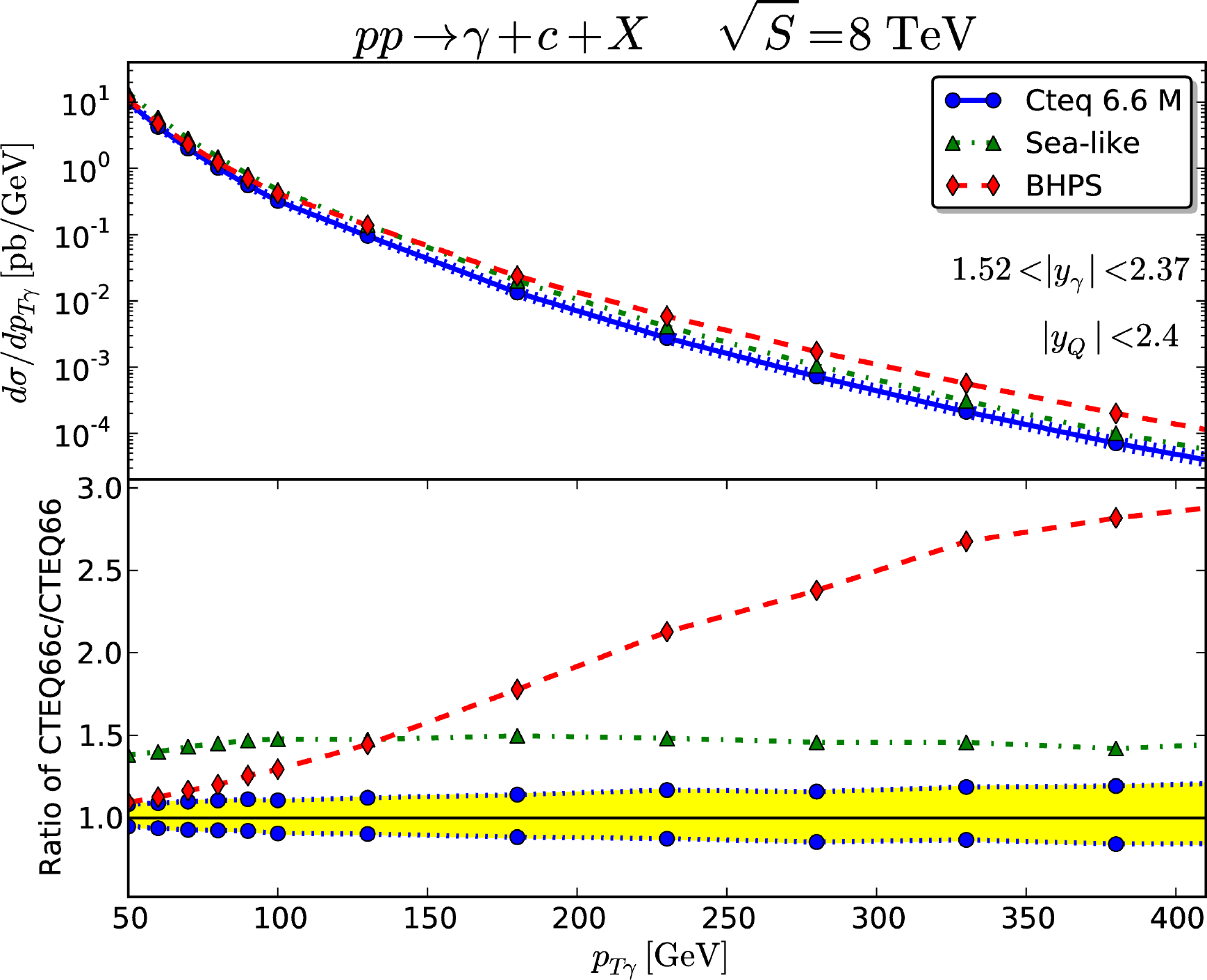}
\caption{The $\mathrm{d}\sigma/\mathrm{d}p_\mathrm{T}^\gamma$ distribution
versus the transverse momentum of the photon for the process $pp \to
\gamma + c + X$ at $\sqrt{s} = 8$\tev\ using CTEQ6.6M (solid blue line), BHPS
CTEQ6c2 (dashed red line) and sea-like CTEQ6c4 (dash-dotted green line), for
forward photon rapidity $1.52 < |y_\gamma| < 2.37$. The ratio of these spectra
with respect to the CTEQ6.6M (solid blue line) distributions (bottom).  The
calculation was done within the NLO QCD approximation.}
\label{Fig_Tzvt2}
\end{figure}

In Fig.~\ref{Fig_Tzvt2} the differential cross-section
$\mathrm{d}\sigma/\mathrm{d}p_\mathrm{T}^\gamma$ calculated at NLO within the
collinear QCD massless quark approximation as described
in~\cite{Stavreva:2009vi} is presented as a function of the transverse momentum
of the prompt photon~\cite{Bednyakov:2013zta}. The following requirements are
applied: $p_\mathrm{T}^\gamma > 45$~\gev, $p_\mathrm{T}^\mathrm{c} > 20$~\gev\
with the $c$-jet pseudorapidity in the interval $|y_\mathrm{c}| \leq 2.4$ and
the photon pseudorapidity in the interval $1.52 < |y_\gamma| < 2.37$ (forward
region).  The solid line represents the differential cross-section calculated
with the radiatively generated charm PDF (CTEQ66), the dash-dotted line uses as
input the sea-like PDF (CTEQ66c4) and the dashed line the BHPS PDF (CTEQ66c2).
In the lower half of Fig.~\ref{Fig_Tzvt2} the above distributions normalized to
the distribution acquired using the CTEQ66 PDF and $\mu_\mathrm{r} =
\mu_\mathrm{f} = \mu_\mathrm{F} = p_\mathrm{T}^\gamma$, are presented. The
shaded yellow region, represents the scale dependence. Clearly the difference
between the spectrum using the BHPS IC PDF and the one using the radiatively
generated PDF increases as \pty\ grows.

Therefore, Fig.~\ref{Fig_Tzvt2} shows that the IC signal could be visible at
the LHC energies with both the ATLAS and  CMS detector in the process $pp
\to \gamma + c + X$ when $p_\mathrm{T}^\gamma \simeq 150$~\gev.  In the
region described above the IC signal dominates over the all non-intrinsic charm
background with significance at a level of a factor of 2 (in fact 170~\%).

\subsection{Production of $\gamma(Z) + c(b)$-jet within the \kt-factorization
and the MCFM}
\label{VII.III}

The \kt-factorization approach~\cite{Campbell:2002tg,Levin:1991ry,
Gribov:1984tu} is based on the small-$x$ Balitsky-Fadin-Kuraev-Lipatov
(BFKL)~\cite{Kuraev:1976ge,Kuraev:1977fs,Balitsky:1978ic} gluon dynamics and
provides solid theoretical ground for the effects of the initial gluon
radiation and the intrinsic parton transverse momentum.\footnote{A detailed
description of the \kt-factorization approach can be found, for example, in
reviews~\cite{Andersson:2002cf,Andersen:2003xj,Andersen:2006pg}}
Our main motivation to use the \kt-fac\-to\-ri\-za\-tion
formalism here is that its predictions for the associated $\gamma + Q$
production better agree with the TEVATRON data compared to the NLO pQCD
(see~\cite{Abazov:2012ea, Aaltonen:2013ama}). The consideration is mainly based
on the $\cal{O}(\alpha \alpha_\mathrm{s})$ off-shell (depending on the
transverse momenta of initial quarks and gluons) quark-gluon Compton-like
scattering subprocess, see Fig.~\ref{fig1}~(a). Within this approach the
transverse momentum dependent (TMD) parton densities include many high order
corrections, while the partonic amplitudes are calculated within the leading
order (LO) of QCD\@. The off-shell quark-gluon Compton scattering amplitude is
calculated within the reggeized parton approach~\cite{Lipatov:2000se,
Bogdan:2006af,Hentschinski:2011tz,Hentschinski:2011xg,Chachamis:2012gh} based
on the effective action formalism~\cite{Lipatov:1995pn,Lipatov:1996ts}, which
ensures the gauge invariance of the obtained amplitudes despite the off-shell
initial quarks and gluons.\footnote{Here we use the expressions derived
earlier~\cite{Lipatov:2016wgr}.} The TMD parton densities are calculated using
the Kimber-Martin-Ryskin (KMR) approach, currently developed within the
NLO~\cite{Martin:2009ii}. This approach is the formalism to construct the TMD
quark and gluon densities from the known conventional parton distributions.
The key assumption is that the \kt\ dependence appears at the last evolution
step, so that the DGLAP evolution can be used up to this step.  Numerically,
for the input we used parton densities derived in Section~2.  Other details of
these calculations are explained in~\cite{Lipatov:2016wgr}.

To improve the \kt-factorization predictions at high transverse momenta, we
take into account some ${\cal O}(\alpha \alpha_\mathrm{s}^2)$ contributions,
namely ${q\bar{q} \to VQ\bar{Q}}$ and ${qQ \to VqQ}$ ones,
where $V$ denotes the photon or the $Z$ boson, see Fig.~\ref{fig1}~((b)~--~(e)).
These contributions are significant at large $x$ and therefore can be
calculated in the usual collinear QCD factorization scheme.  Thus, we rely on
the combination of two techniques that are most suitable.

Let us present the results of our calculations. First of all we describe our
numerical input. Following to~\cite{Martin:2009ii}, we set the charmed and
\textcolor{purple}{bottom} quark masses $m_\mathrm{c} = 1.4$~\gev, $m_\mathrm{b} = 4.75$~\gev, the
$Z$-boson mass $m_\mathrm{Z} = 91.1876$~\gev, and $\sin^2 \theta_\mathrm{W} =
0.23122$. The chosen factorization and renormalization scales are
$\mu_\mathrm{R} = \mu_\mathrm{F} = \xi p_\mathrm{T}$ or $\mu_\mathrm{R} =
\mu_\mathrm{F} = \xi m_\mathrm{T}$, where \pt\ is the produced photon transverse
momentum and $m_\mathrm{T}$ is the $Z$ boson transverse mass. As usual, we vary
the non-physical parameter $\xi$ between $1/2$ and $2$ about the default value
$\xi = 1$ in order to estimate the scale uncertainties of our calculations. We
employ the two-loop formula for the strong coupling constant with active quark
flavors $n_\mathrm{f} = 5$ at $\Lambda_\mathrm{QCD} = 226.2$~\mev\ and use the
running QED coupling constant over a wide region of transverse momenta.  The
multidimensional integration in the \kt-factorization calculations was
performed by means of the Monte Carlo technique, using the VEGAS
routine~\cite{Agashe:2014kda}.

In our calculations we also follow the conclusion obtained in our
papers~\cite{Bednyakov:2013zta,Beauchemin:2014rya} that the IC signal in the
hard processes discussed here can be detected at ATLAS or CMS of the LHC in the
forward rapidity region $1.5 < |\eta| < 2.4$ and  $p_\mathrm{T} > 50$~\gev.
Additionally, we require $|\eta(Q)| < 2.4$ and $p_\mathrm{T}(Q) > 25$~\gev,
where $\eta(Q)$ and $p_\mathrm{T}(Q)$ are the pseudo-rapidity and transverse
momentum of the heavy quark jet in a final state, as was done
in~\cite{Bednyakov:2013zta,Beauchemin:2014rya}.

\begin{figure}[h!]
\centering
\includegraphics[width=.65\textwidth,angle=270]{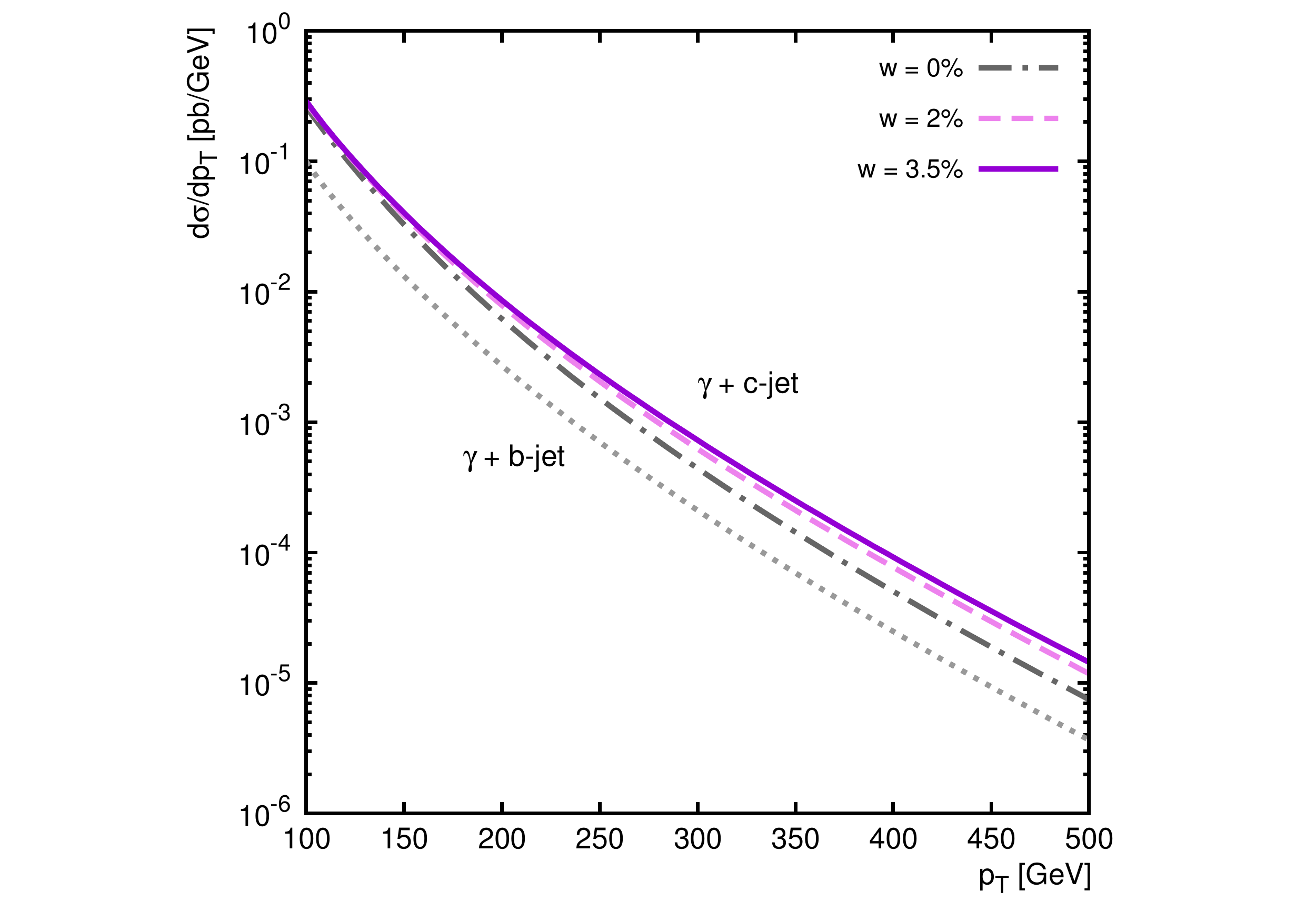}
\includegraphics[width=.65\textwidth,angle=270]{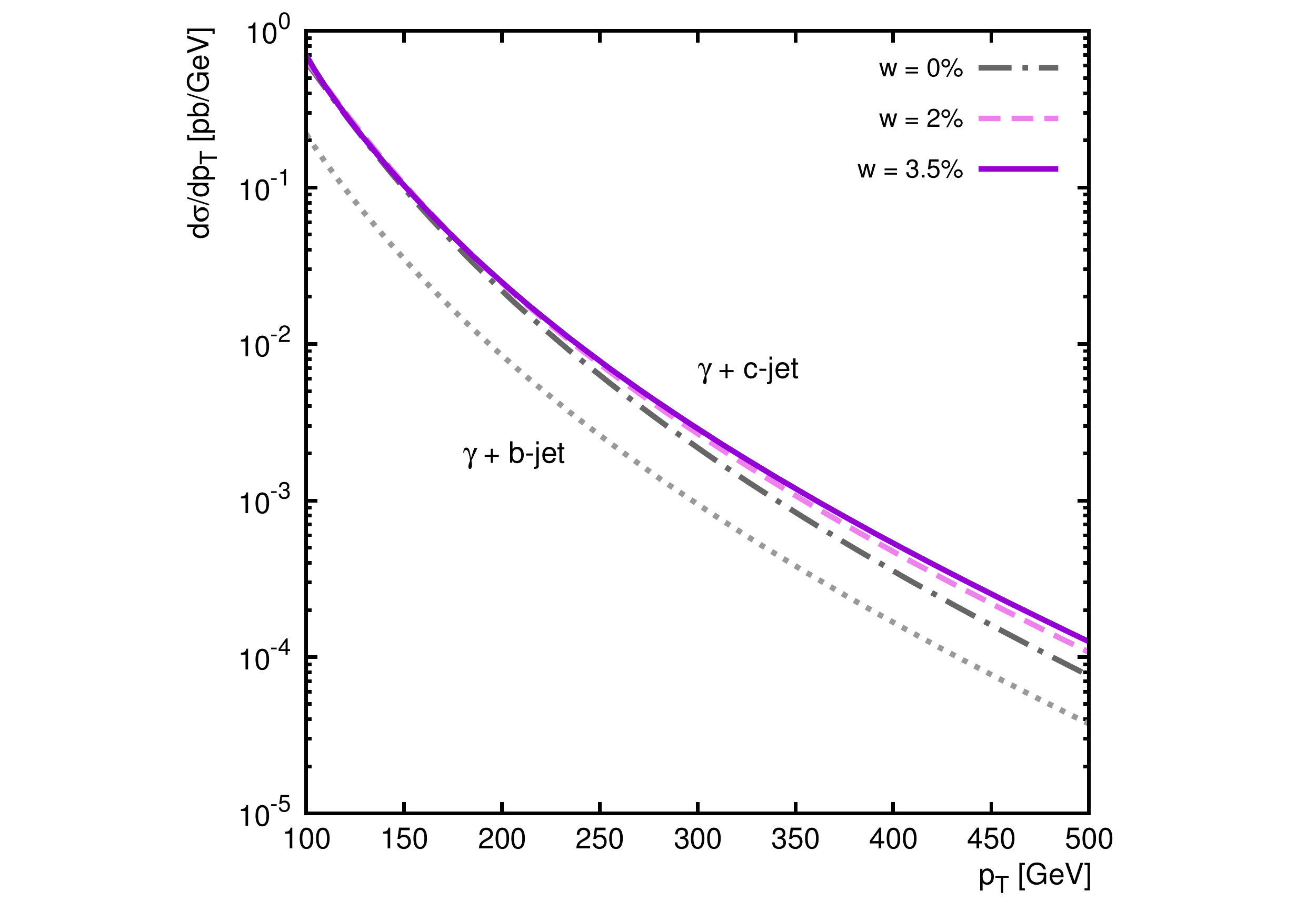}
\caption{The cross-sections of the associated $\gamma + c$ and $\gamma + b$
production in the $pp$ collision calculated as a function of the photon
transverse momentum \pt\ at $\sqrt{s} = 8$~\tev\ (top) and $\sqrt{s} = 13$~\tev\
(bottom) within the \kt-factorization approach. The kinematical conditions are
described in the text.}
\label{fig_ptgam}
\end{figure}

\begin{figure}[h!]
\centering
\includegraphics[width=.65\textwidth,angle=270]{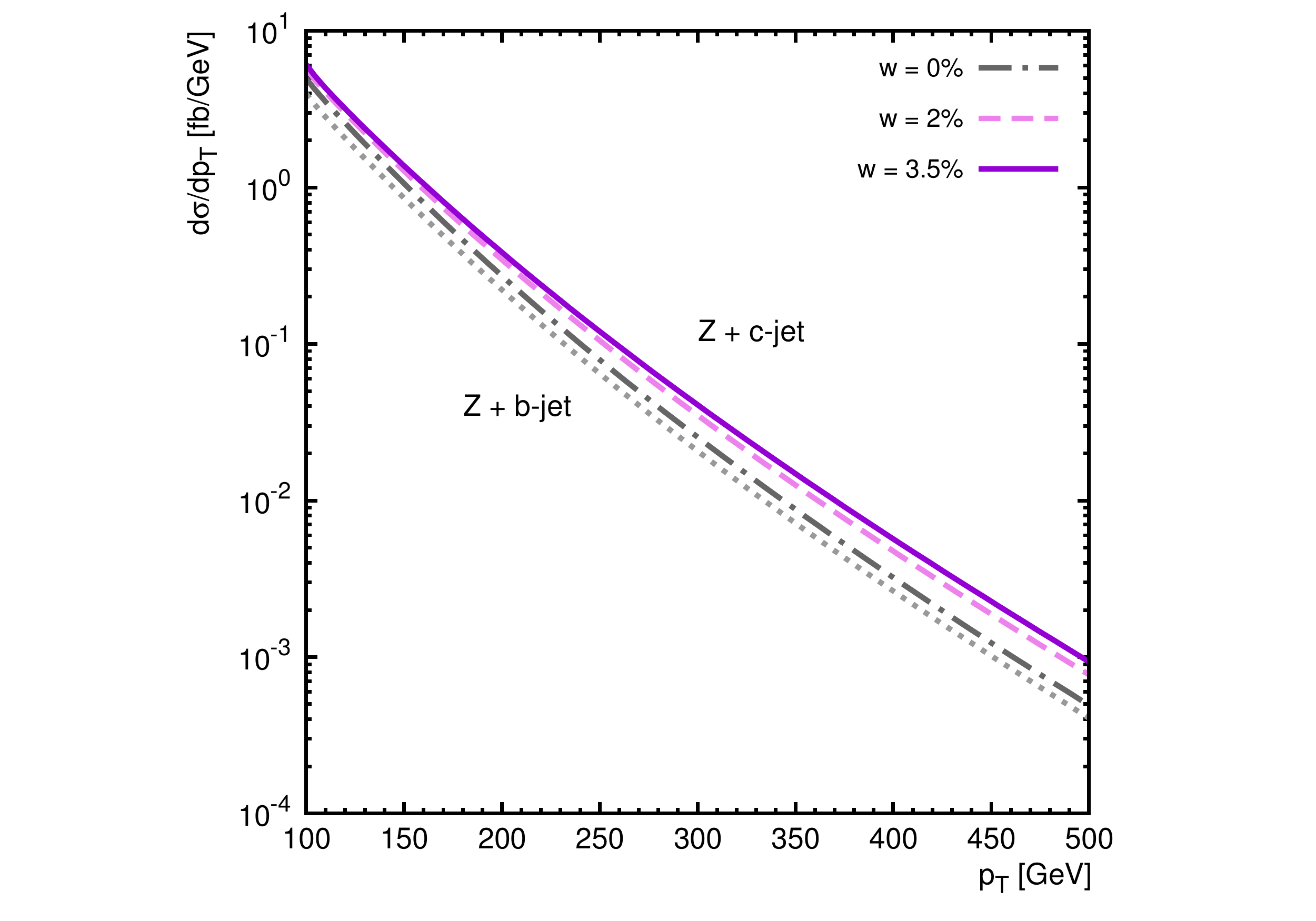}
\includegraphics[width=.65\textwidth,angle=270]{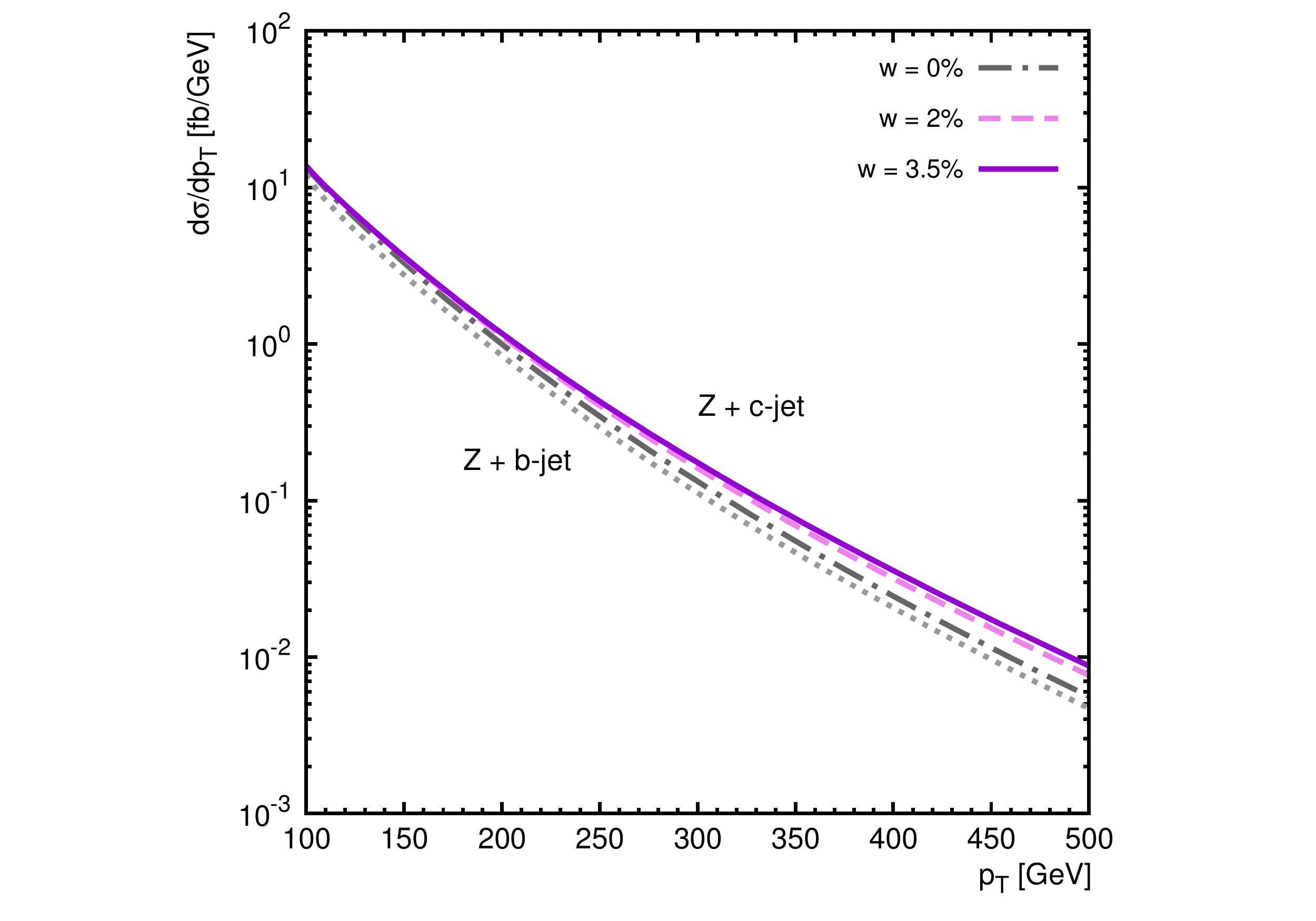}
\caption{The cross-sections of the associated $Z + c$ and $Z + b$ production in
the $pp$ collision calculated as a function of the $Z$ boson transverse
momentum \pt\ at $\sqrt s = 8$~\tev\ (top) and $\sqrt s = 13$~\tev\ (bottom)
within the \kt-factorization approach. The kinematical conditions are described
in the text.}
\label{fig_ptZ}
\end{figure}

\begin{figure}[h!]
\centering
\includegraphics[width=.65\textwidth]{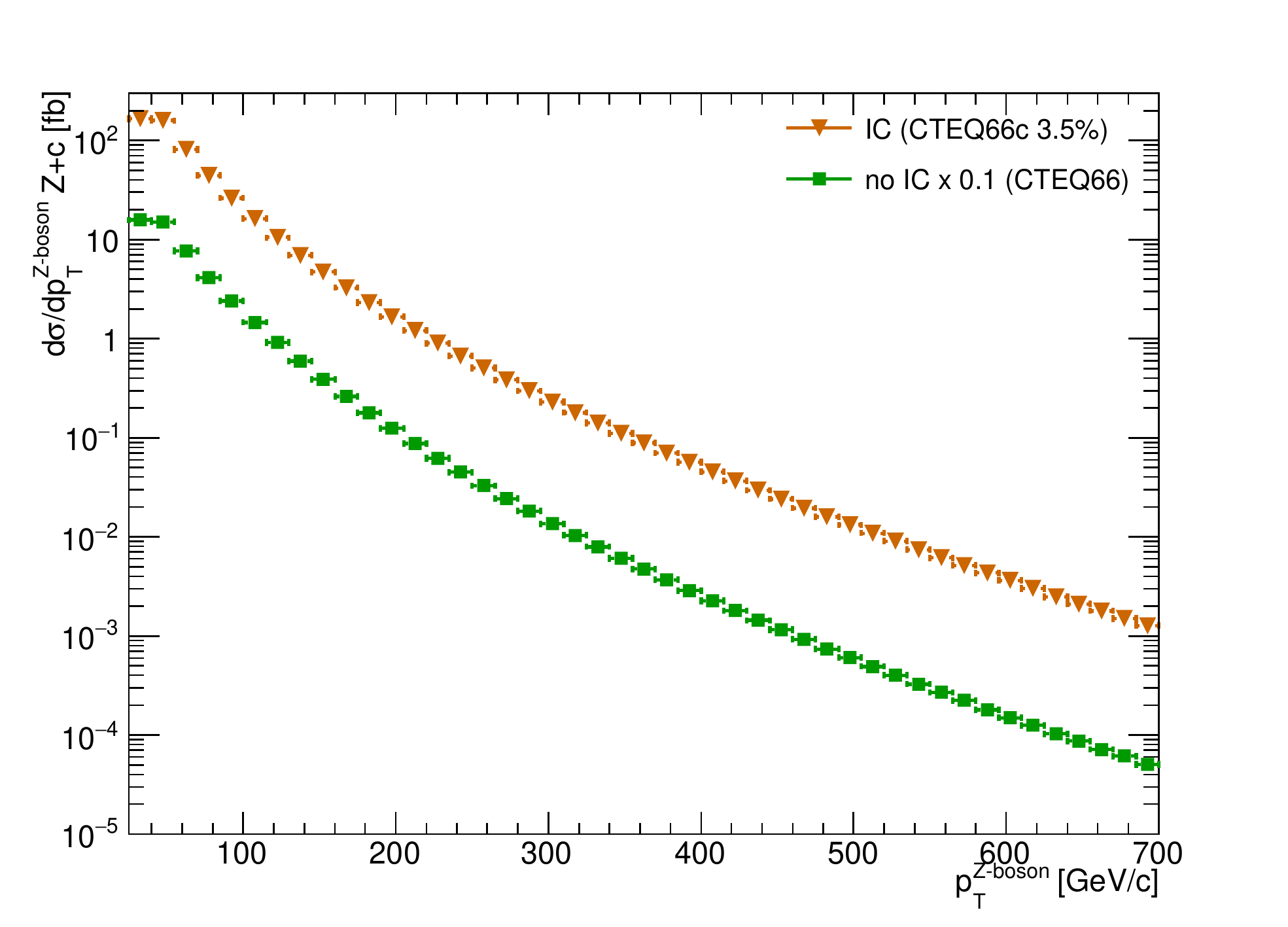}
\includegraphics[width=.65\textwidth]{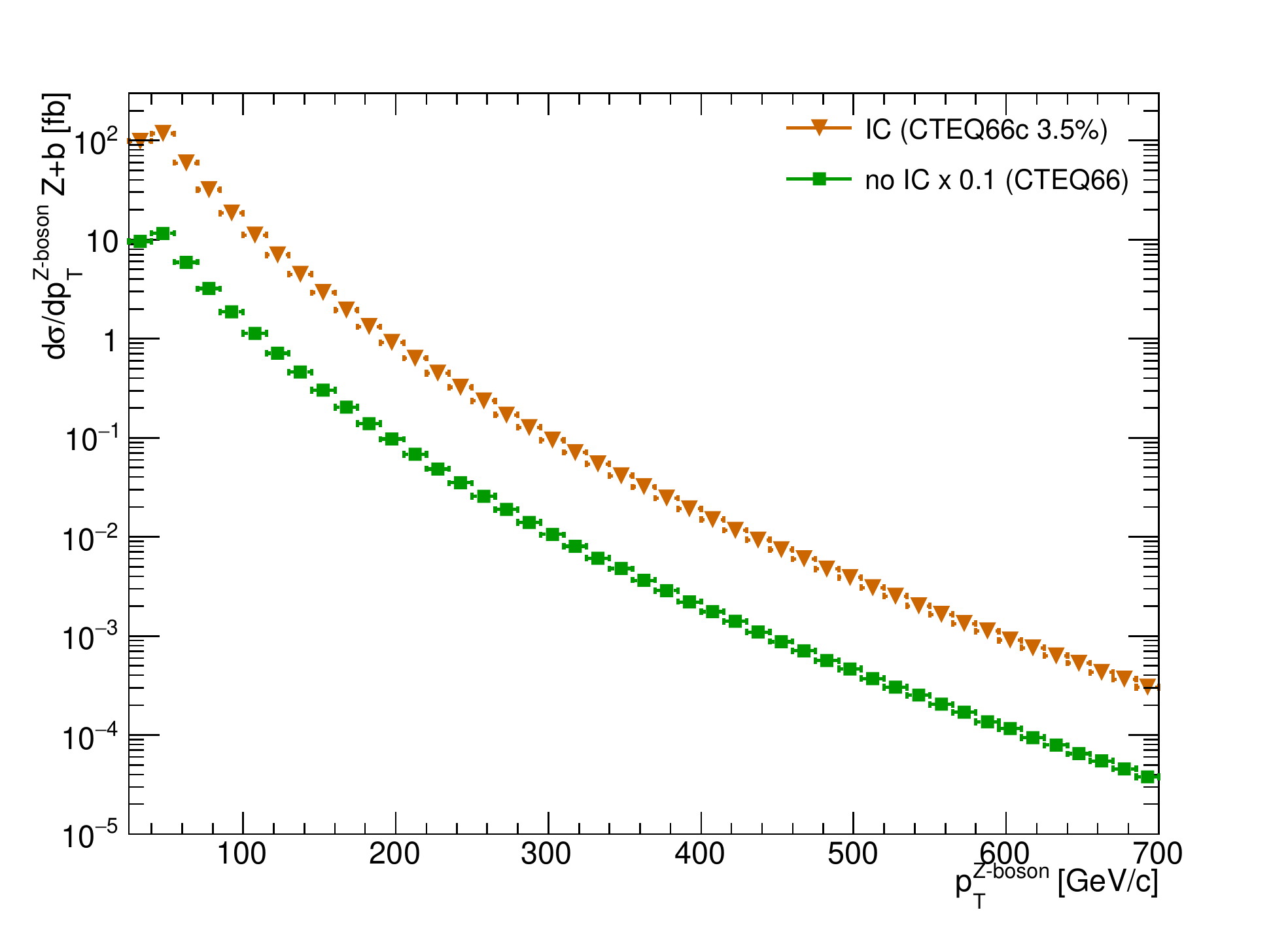}
\caption{The cross-sections of the associated $Z + c$ (top) and $Z + b$
(bottom) production in $pp$ collision calculated as a function of the $Z$ boson
transverse momentum \pt\ at $\sqrt{s} = 13$~\tev\ within the MCFM routine.}
\label{fig_ptZmcfm}
\end{figure}

\begin{figure}[h!]
\centering
\includegraphics[width=.67\textwidth,angle=270]{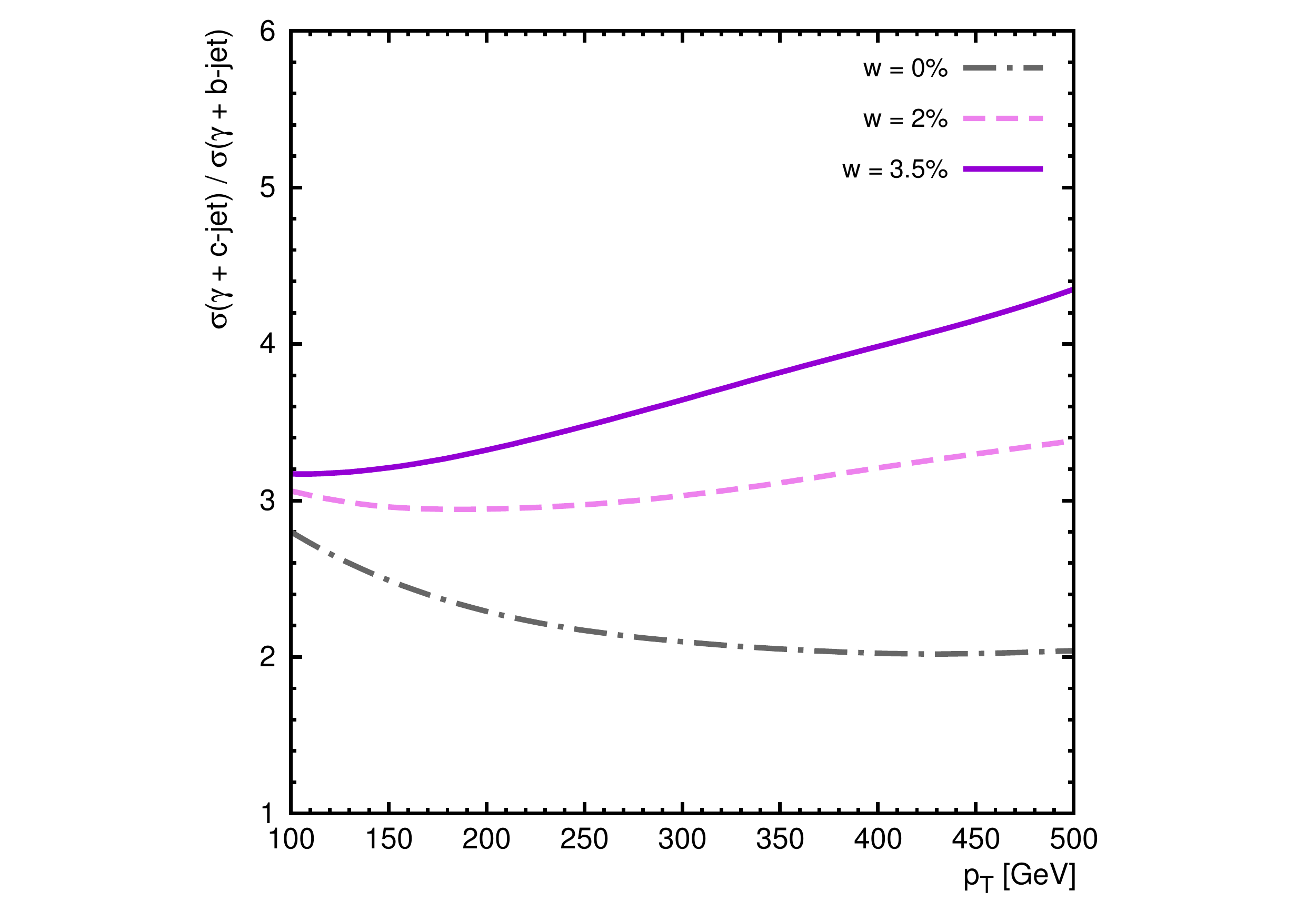}
\includegraphics[width=.67\textwidth,angle=270]{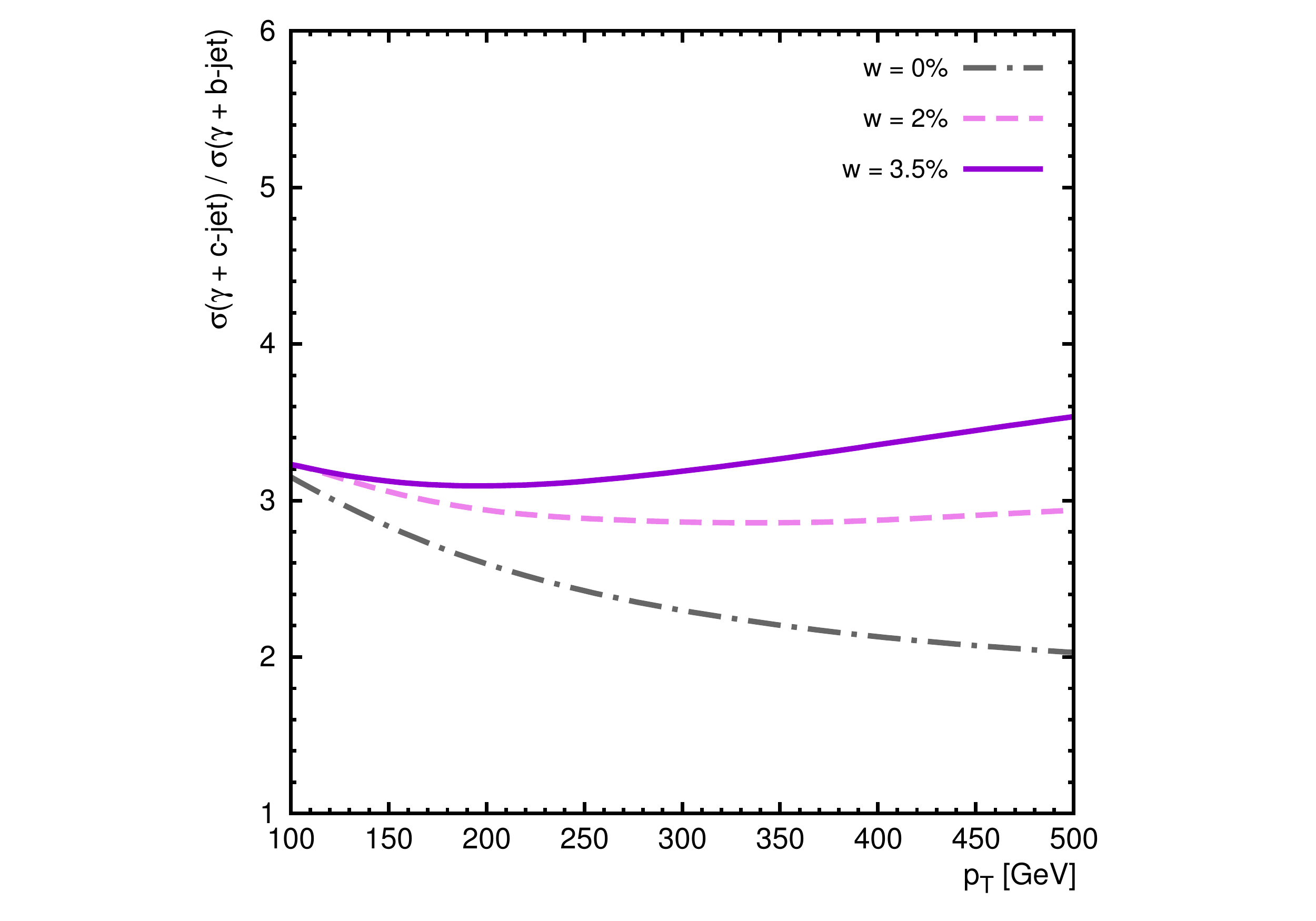}
\caption{The cross-section ratio of the $\gamma + c$ production to the
$\gamma + b$ one in the $pp$ collision calculated as a function of the photon
transverse momentum \pt\ at $\sqrt{s} = 8$~\tev\ (top) and $\sqrt{s} = 13$~\tev\
(bottom) within the \kt-factorization approach.}
\label{fig_ptgamrat}
\end{figure}

\begin{figure}[h!]
\centering
\includegraphics[width=.67\textwidth,angle=270]{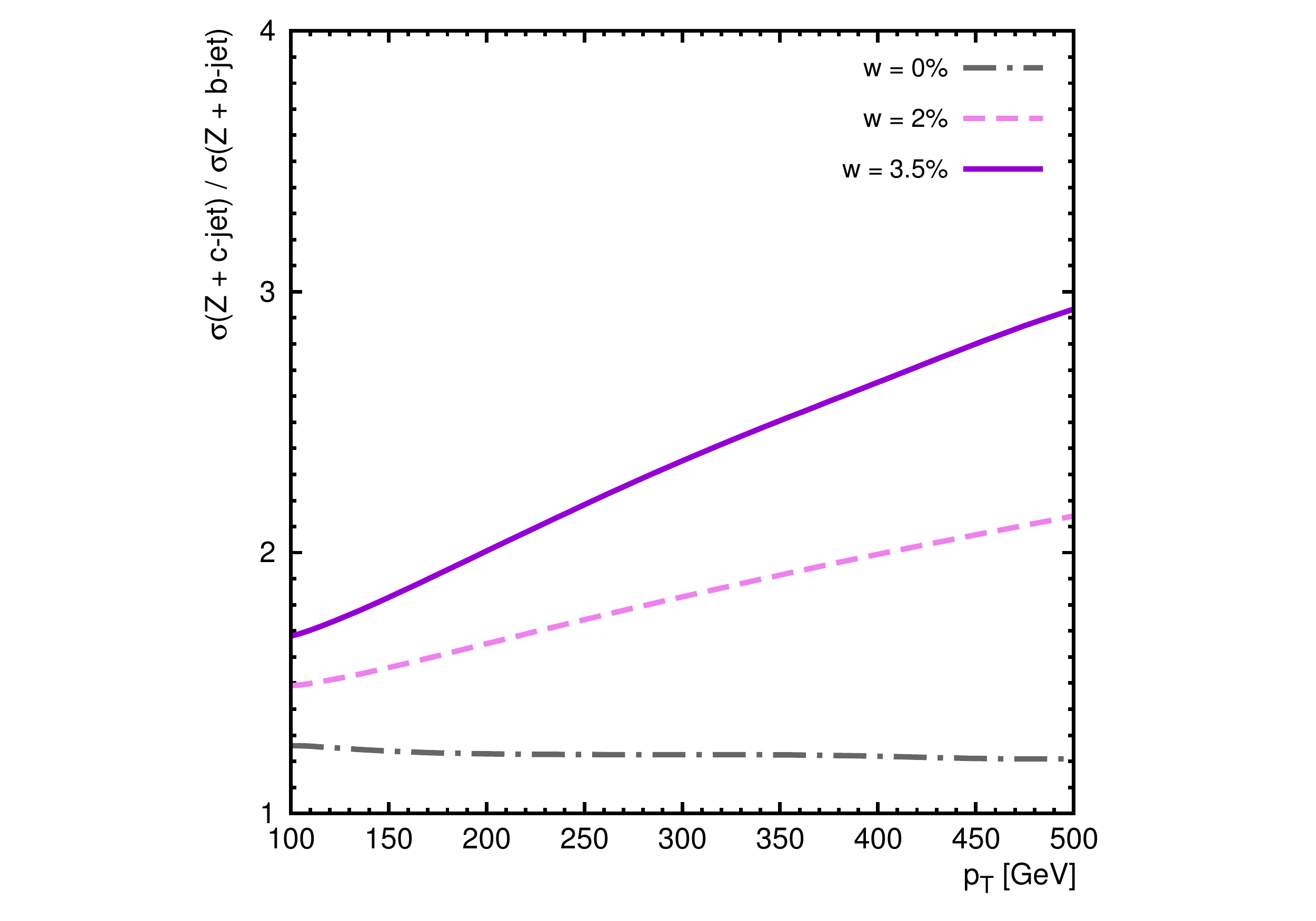}
\includegraphics[width=.67\textwidth,angle=270]{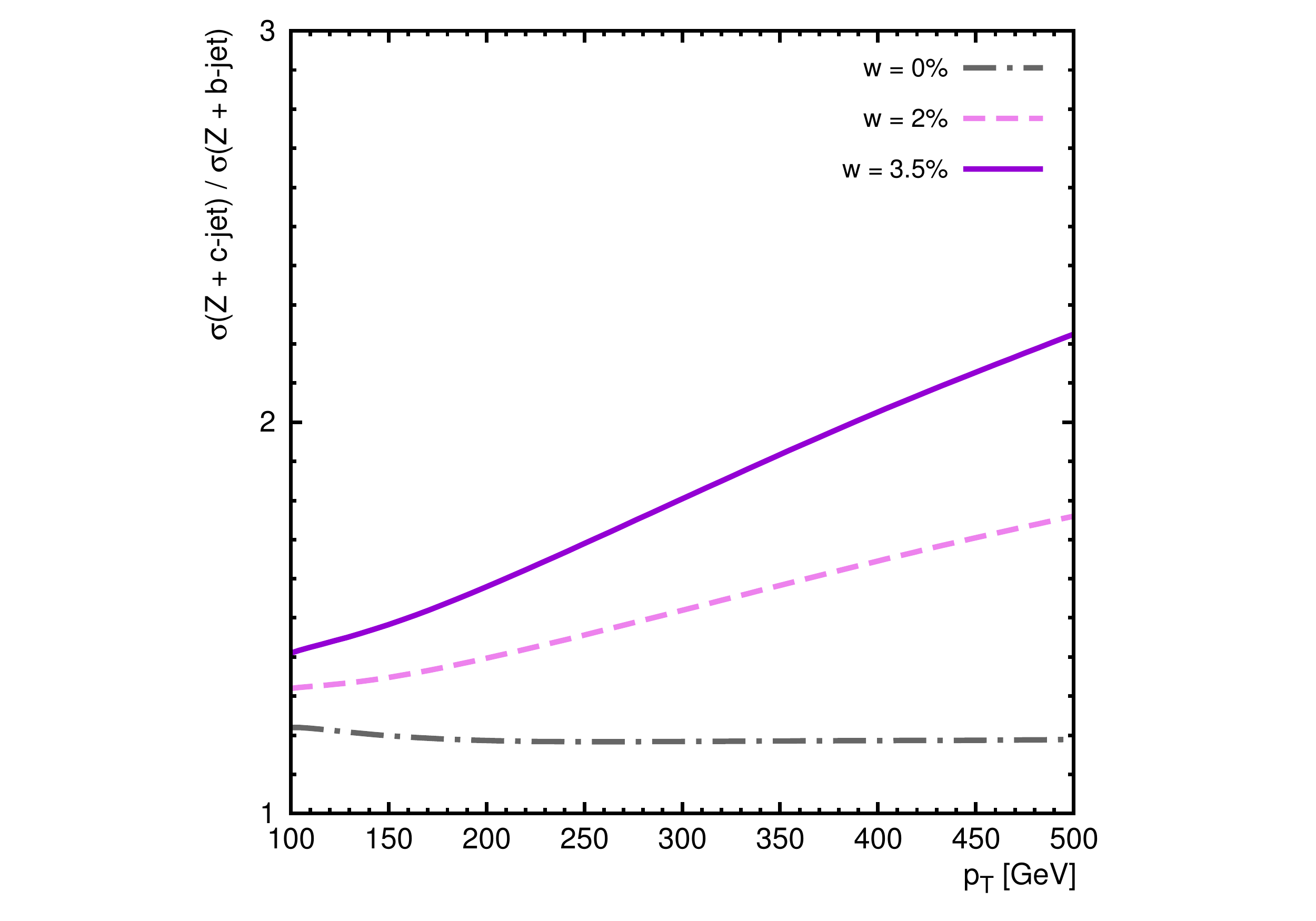}
\caption{The cross-section ratio of the $Z + c$ production to the $Z + b$ one
in the $pp$ collision calculated as a function of the $Z$ boson transverse
momentum \pt\ at $\sqrt{s} = 8$~\tev\ (top) and $\sqrt{s} = 13$~\tev\ (bottom)
within the \kt-factorization approach.}
\label{fig_ptZrat}
\end{figure}

\begin{figure}[h!]
\centering
\includegraphics[width=.68\textwidth,angle=270]{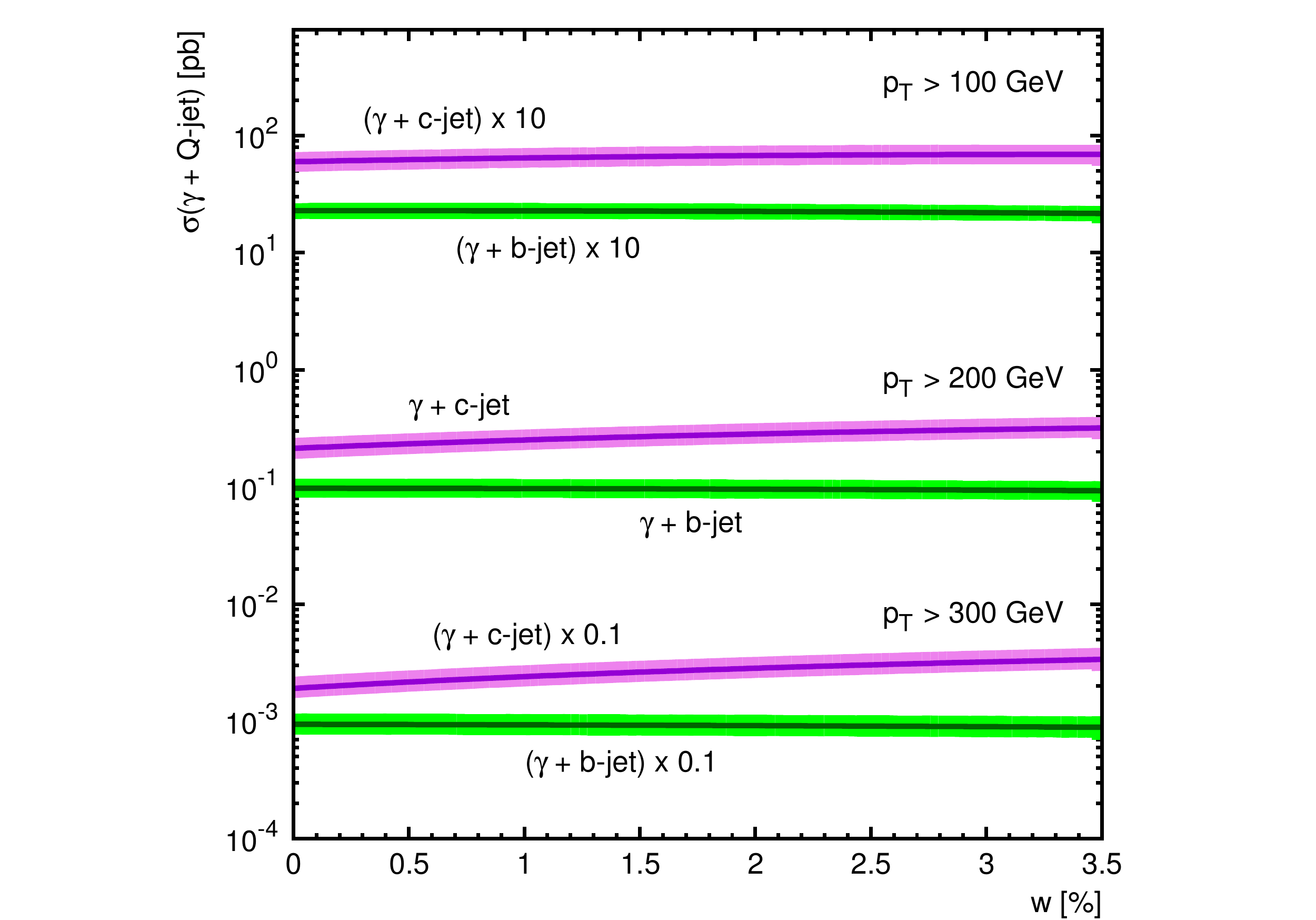}
\includegraphics[width=.68\textwidth,angle=270]{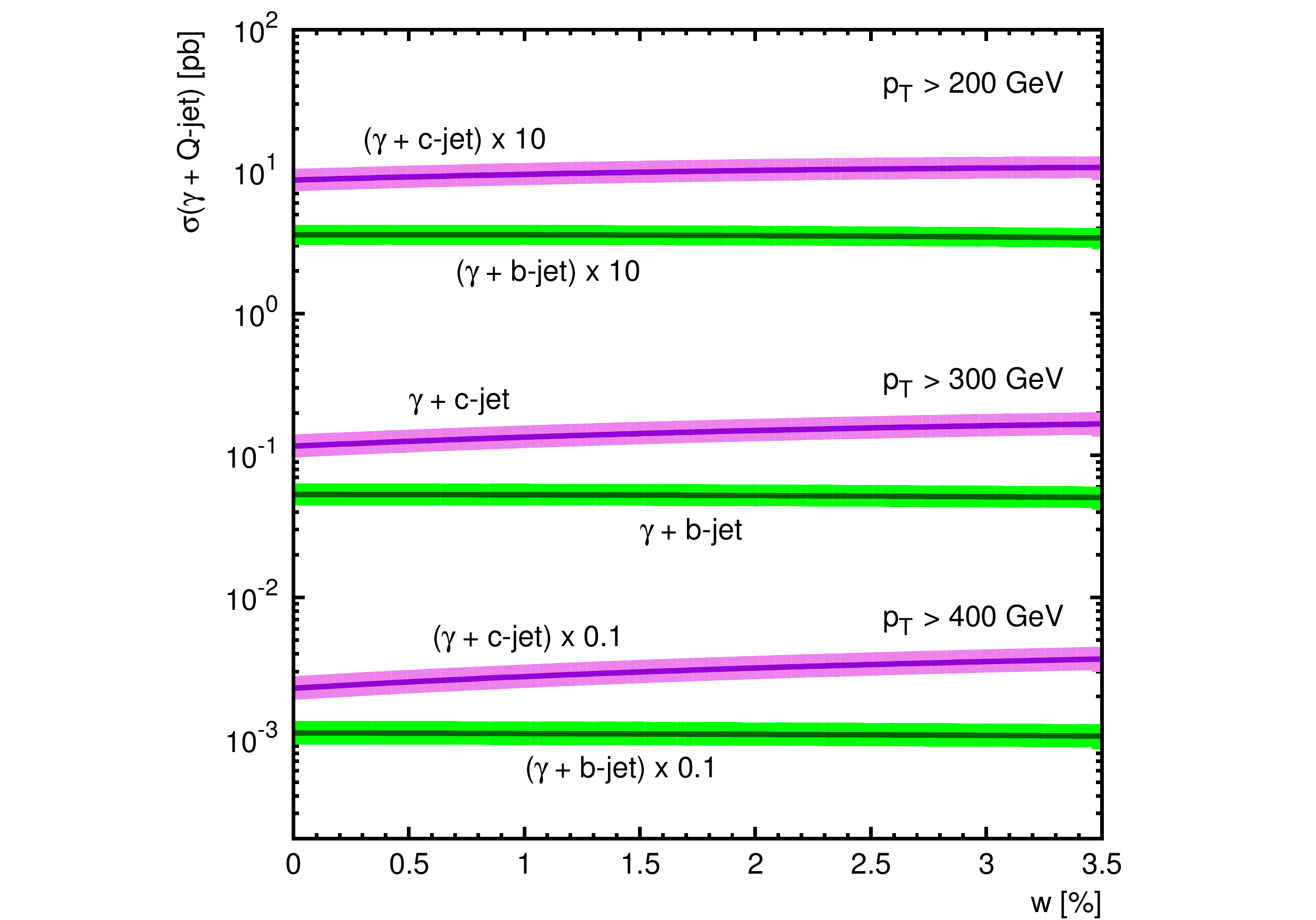}
\caption{The cross-sections of the associated $\gamma + c$ and $\gamma + b$
production in the $pp$ collision as a function of $w$ integrated over the
photon transverse momenta $p_\mathrm{T} > p_\mathrm{T}^{\min}$ for different
$p_\mathrm{T}^{\min}$ at $\sqrt{s} = 8$~\tev\ (\textcolor{purple}{top}) and
$\sqrt{s} = 13$~\tev\ (\textcolor{purple}{bottom}).}
\label{fig_gamma}
\end{figure}

\begin{figure}[h!]
\centering
\includegraphics[width=.68\textwidth,angle=270]{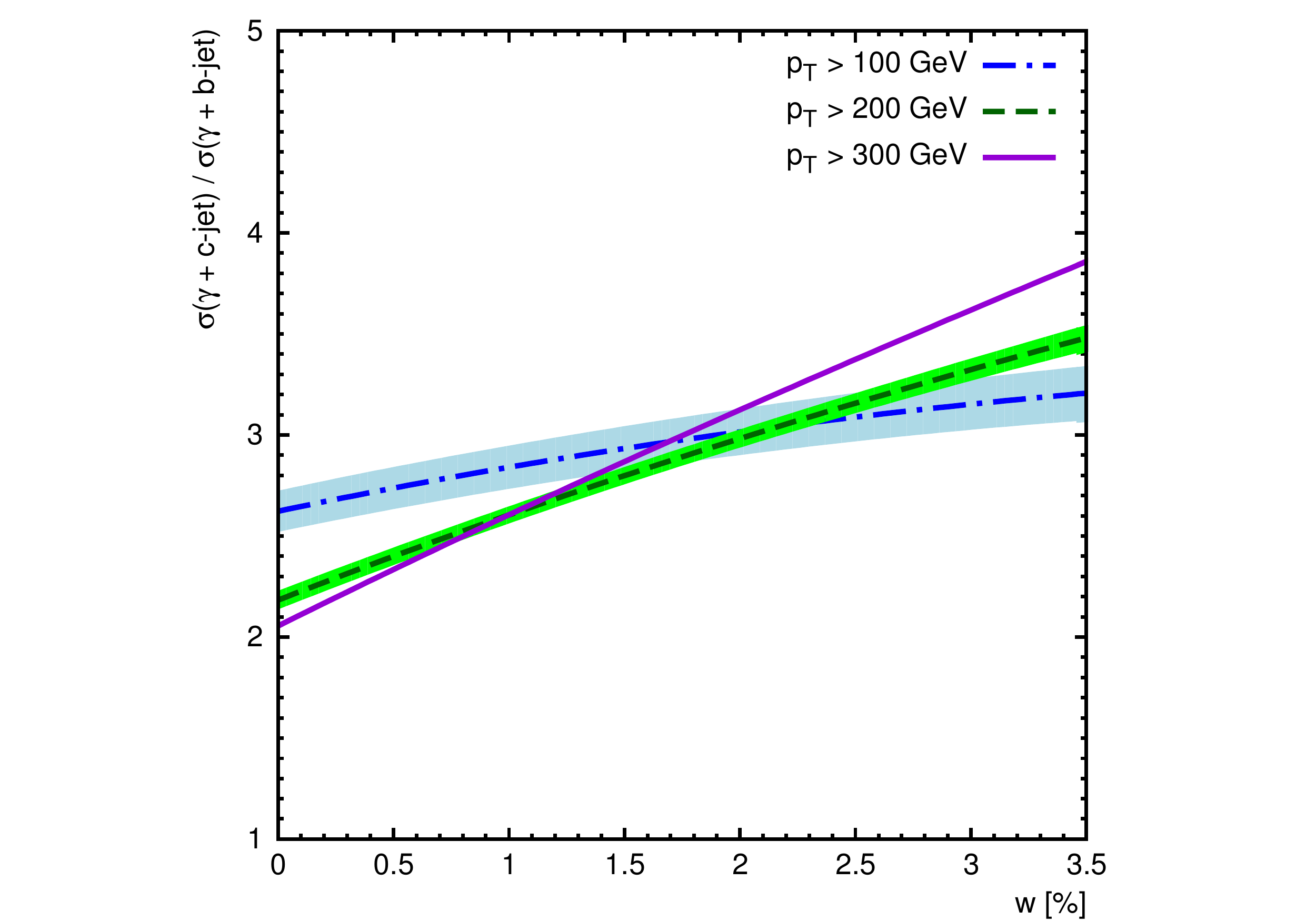}
\includegraphics[width=.68\textwidth,angle=270]{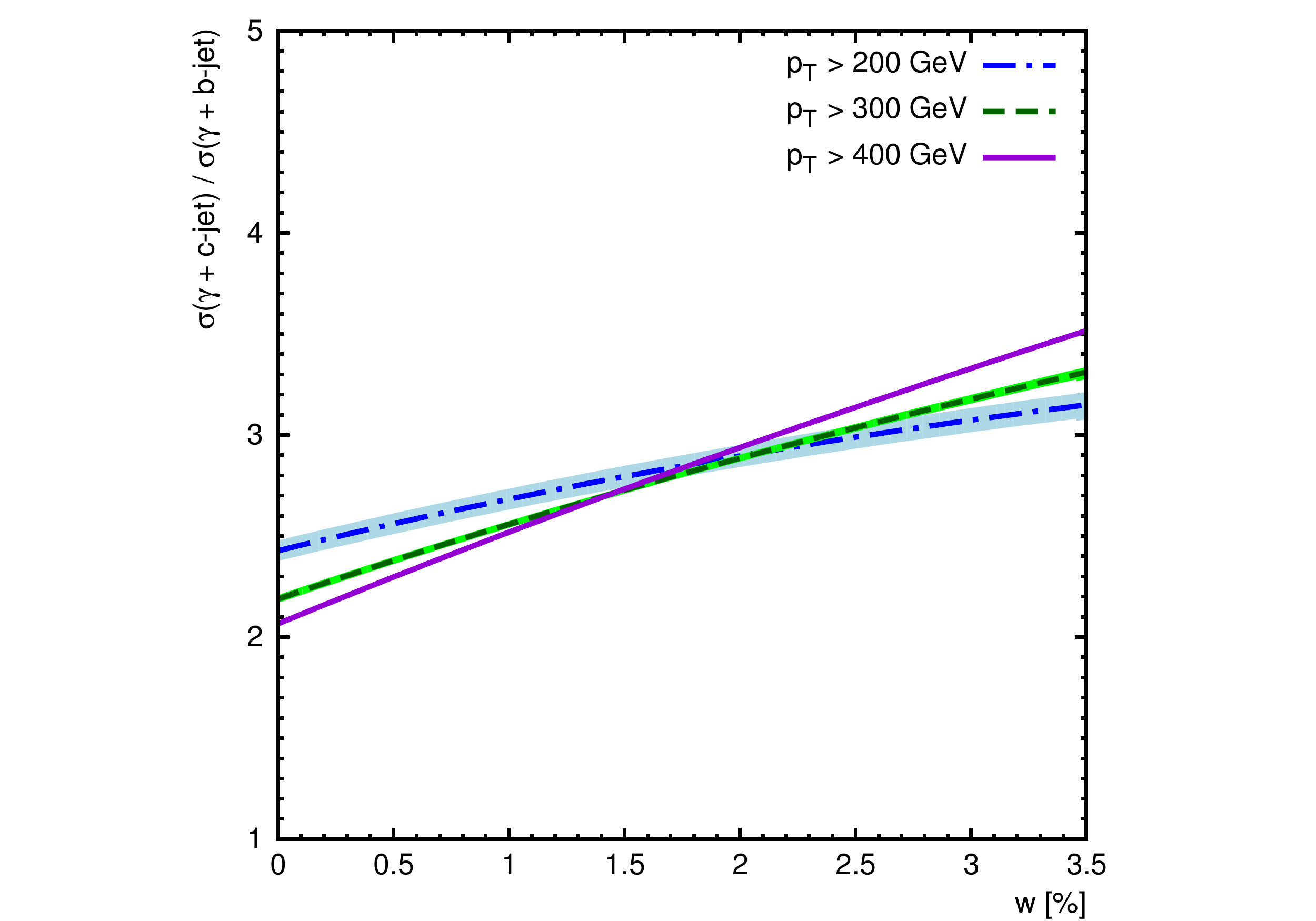}
\caption{The corresponding ratios of these cross-sections. The calculations
were done using the \kt-factorization approach. The bands correspond to the
usual scale variation as it is described in the text.}
\label{fig_gamrat}
\end{figure}

\begin{figure}[h!]
\centering
\includegraphics[width=.68\textwidth,angle=270]{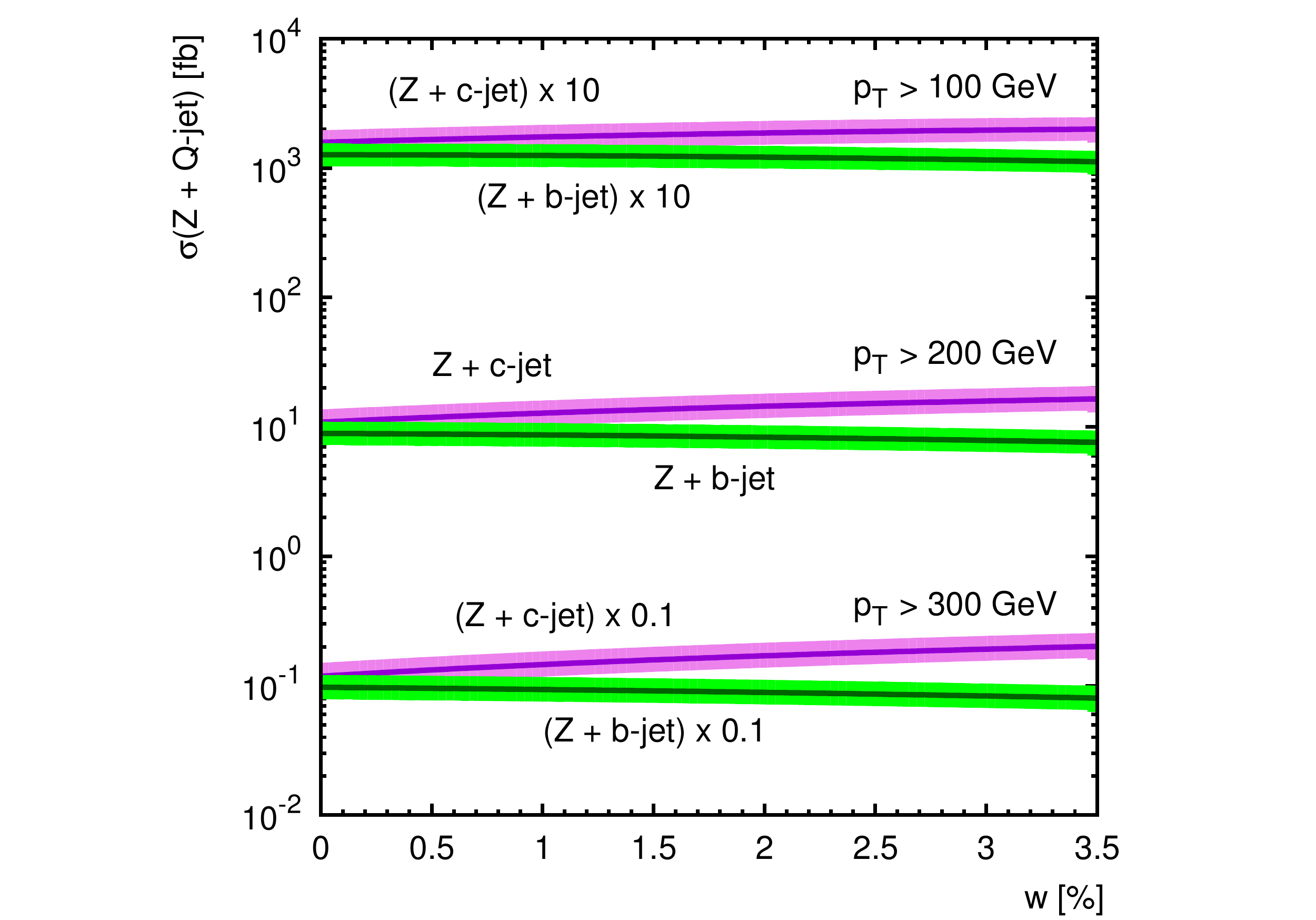}
\includegraphics[width=.68\textwidth,angle=270]{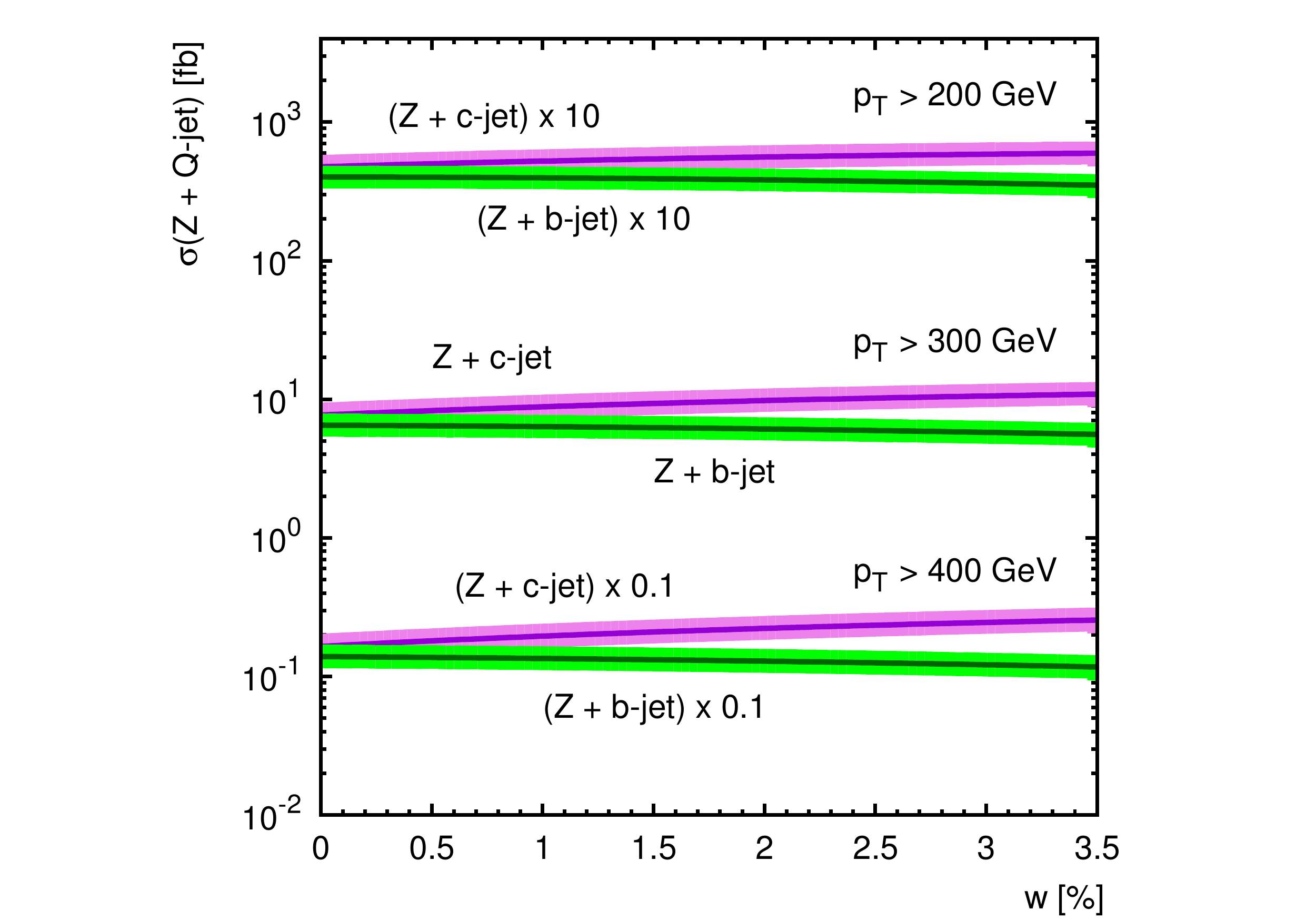}
\caption{The cross-sections of the associated $Z + c$ and $Z + b$ production in
the $pp$ collision as a function of $w$ integrated over the $Z$ boson
transverse momenta $p_\mathrm{T} > p_\mathrm{T}^{\min}$ for different
$p_\mathrm{T}^{\min}$ at $\sqrt{s} = 8$~\tev\ (\textcolor{purple}{top}) and
$\sqrt{s} = 13$~\tev\ (\textcolor{purple}{bottom}). The kinematical
conditions are described in the text.}
\label{fig_Z}
\end{figure}

\begin{figure}[h!]
\centering
\includegraphics[width=.68\textwidth,angle=270]{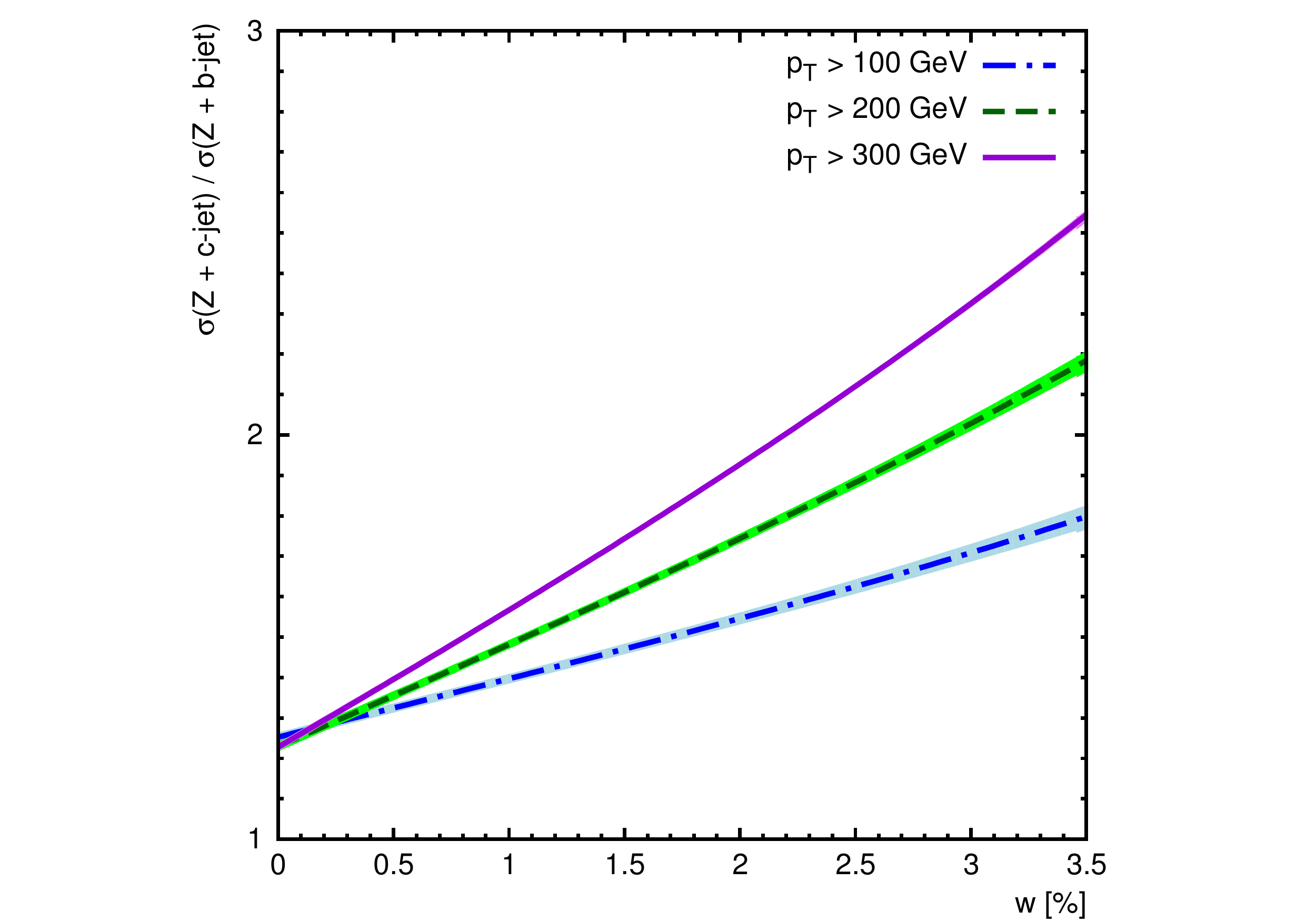}
\includegraphics[width=.68\textwidth,angle=270]{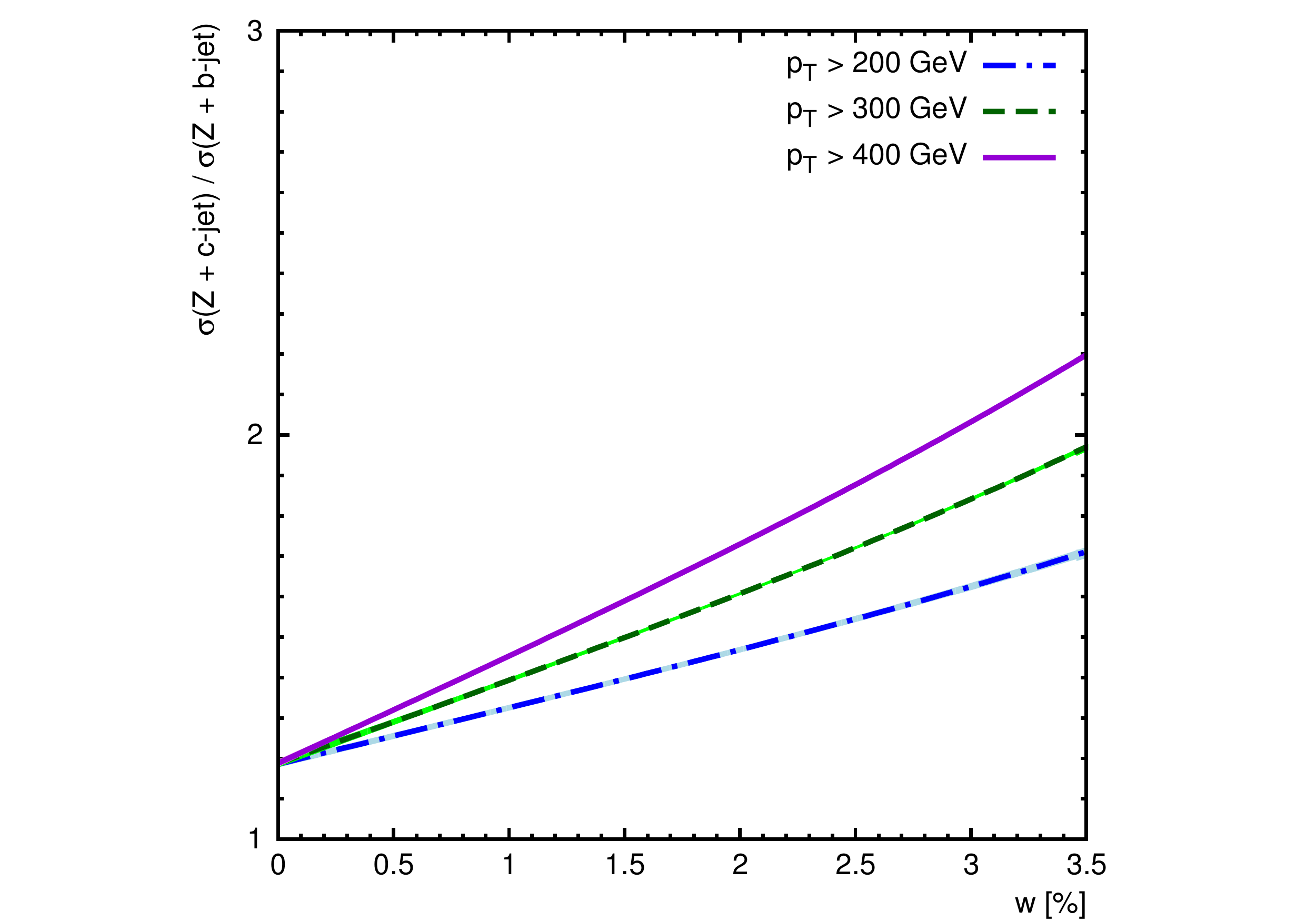}
\caption{The corresponding ratios of these cross-sections. The calculations
were done using the \kt-factorization approach. The bands correspond to the
usual scale variation as it is described in the text.}
\label{fig_Zrat}
\end{figure}

The results of our calculations are shown in
Figs.~\ref{fig_ptgam}~--~\ref{fig_Zrat}.  The transverse momentum distributions
of photons and $Z$ bosons accompanied by the $c$ and $b$ quarks are presented
in Figs.~\ref{fig_ptgam},~\ref{fig_ptZ} and~\ref{fig_ptZmcfm} at the different
IC probability $w$ (namely, $w = 0$~\%, $w = 2$~\% and $w = 3.5$~\%) at
$\sqrt{s} = 8$ and 13~\tev.

One can see in Figs.~\ref{fig_ptZ} and~\ref{fig_ptZmcfm} that the MCFM and
\kt-factorization predictions for $Z + Q$ production are very similar in the
whole \pt\ region, therefore below we will present the observables calculated
within the \kt-factorization approach only. The coincidence of these two
calculations is due to effective allowance for the high-order corrections
within the \kt-factorization formalism (see, for
example,~\cite{Andersson:2002cf,Andersen:2003xj,Andersen:2006pg} for more
information). Both types of calculations predict a significant enhancement of
\pt\ distributions due to the IC terms at $p_\mathrm{T} \geq 100$~\gev, which is
in agreement with the previous studies~\cite{Bednyakov:2013zta,Rostami:2015iva,
Beauchemin:2014rya}.

The \pt\ spectrum ratios $\sigma(\gamma + c)/\sigma(\gamma + b)$ and $\sigma(Z +
c)/\sigma(Z + b)$ versus \pt\ at different $w$ are presented in
Figs.~\ref{fig_ptgamrat} and~\ref{fig_ptZrat}. One can see that in the absence
of the IC contribution the ratio $\sigma(\gamma + c)/\sigma(\gamma + b)$ is
about $3$ at $p_\mathrm{T} \sim 100$~\gev\ and decreases down to $2$ at
$p_\mathrm{T} \sim 500$~\gev. This behavior is the same for both energies
$\sqrt{s} = 8$~\tev\ and $\sqrt{s} = 13$~\tev.

If one takes into account the IC contributions, this ratio becomes
approximately flat at $w = 2$~\% or even increasing up to about $4$ at $w =
3.5$~\%. It is very close to the TEVATRON data~\cite{D0:2012gw}: the constant
ratio $\sigma(\gamma + c) / \sigma (\gamma + b) \sim 3.5$~--~4.5 measured in
the $p\bar{p}$ collisions at $110 < p_\mathrm{T} < 300$~\gev\ and ${\sqrt{s} =
1.96}$~\tev. However, this agreement cannot be treated as the IC indication due
to huge experimental uncertainties (about $50$~\%) and rather different
kinematical conditions. If the IC contribution is included, the ratio $\sigma(Z
+ c)/\sigma(Z + b)$ also increases by a factor about $2$ at $w = 3.5$~\%, when
the $Z$ boson transverse momentum grows from $100$~\gev\ to $500$~\gev\ (see
Fig.~\ref{fig_ptZrat}). In the absence of the IC terms this ratio slowly
decreases.

One can consider other observables which could be useful to detect the IC
signal, the  cross-sections discussed above but integrated over $p_\mathrm{T} >
p_\mathrm{T}^{\min}$, where $p_\mathrm{T}^{\min} \geq 100$~\gev, and their
ratios.  Our predictions for such integrated cross-sections versus the IC
probability $w$ at $p_\mathrm{T}^{\min} = 100$, 200 and 300~\gev\ for $\sqrt{s}
= 8$~\tev\ and $p_\mathrm{T}^{\min} = 200$, $300$ and $400$~\gev\ for $\sqrt s
= 13$~\tev\ are shown in Figs.~\ref{fig_gamma},~\ref{fig_gamrat} and
Figs.~\ref{fig_Z},~\ref{fig_Zrat}.

All the \pt-spectra have a significant scale uncertainty as is shown
in~\cite{Beauchemin:2014rya}. According to~\cite{Beauchemin:2014rya}, the ratio
between the cross-sections for the $Z + Q$ and $W + Q$ production in the $pp$
collision is less sensitive to the scale variation calculated within the
MCFM\@. Nevertheless, the uncertainty in this ratio at large ${p_\mathrm{T} >
250}$~\gev\ is about 40~--~50~\%. In the present paper we check these results
for the ratios $\sigma(\gamma + c)/\sigma(\gamma + b)$ and $\sigma(Z +
c)/\sigma(Z + b)$. In Figs.~\ref{fig_gamma},~\ref{fig_gamrat} and
Figs.~\ref{fig_Z},~\ref{fig_Zrat} we present these ratios versus the IC
probability $w$ calculated at different scales, when the cross-sections of
$\gamma(Z) + Q$ production are integrated within the different intervals of
transverse momentum. One can see a very small QCD scale uncertainty, especially
at $\sqrt{s} = 13$~\tev\ (bottom right), which is less than 1~\%. In contrast,
the scale uncertainty for the integrated $\gamma(Z) + Q$ cross-sections (see
Figs.~\ref{fig_gamma},~\ref{fig_gamrat} and Figs.~\ref{fig_Z},~\ref{fig_Zrat})
is significant and amounts to about $30$~--~40~\%. The sizable difference
between the scale uncertainties for the ratios $\sigma(Z + Q)/\sigma(W + Q)$
and $\sigma(Z + c)/\sigma(Z + b)$ is due to the different matrix elements for
the $Z + Q$ and $W + Q $ production in $pp$ collisions, while the matrix
elements for the $Z + c$ and $Z + b$ production are the same.

It is important that the calculated ratios $\sigma(\gamma + c)/\sigma(\gamma +
b)$ and $\sigma(Z + c)/\sigma(Z + b)$ can be used to determine the IC
probability $w$ from the future LHC data. Moreover, these ratios are
practically independent of the uncertainties of our calculations: actually, the
curves corresponding to the usual scale variations as described above coincide
with each other (see Figs.~\ref{fig_gamrat} and~\ref{fig_Zrat}, bottom).
Therefore, we can recommend these observables as a test for the hypothesis of
the IC component inside the proton.

\subsection{Prompt photon and open charm production}
\label{VII.IV}

\begin{figure}[h!]
\centering
\includegraphics[width=.68\textwidth,angle=270]{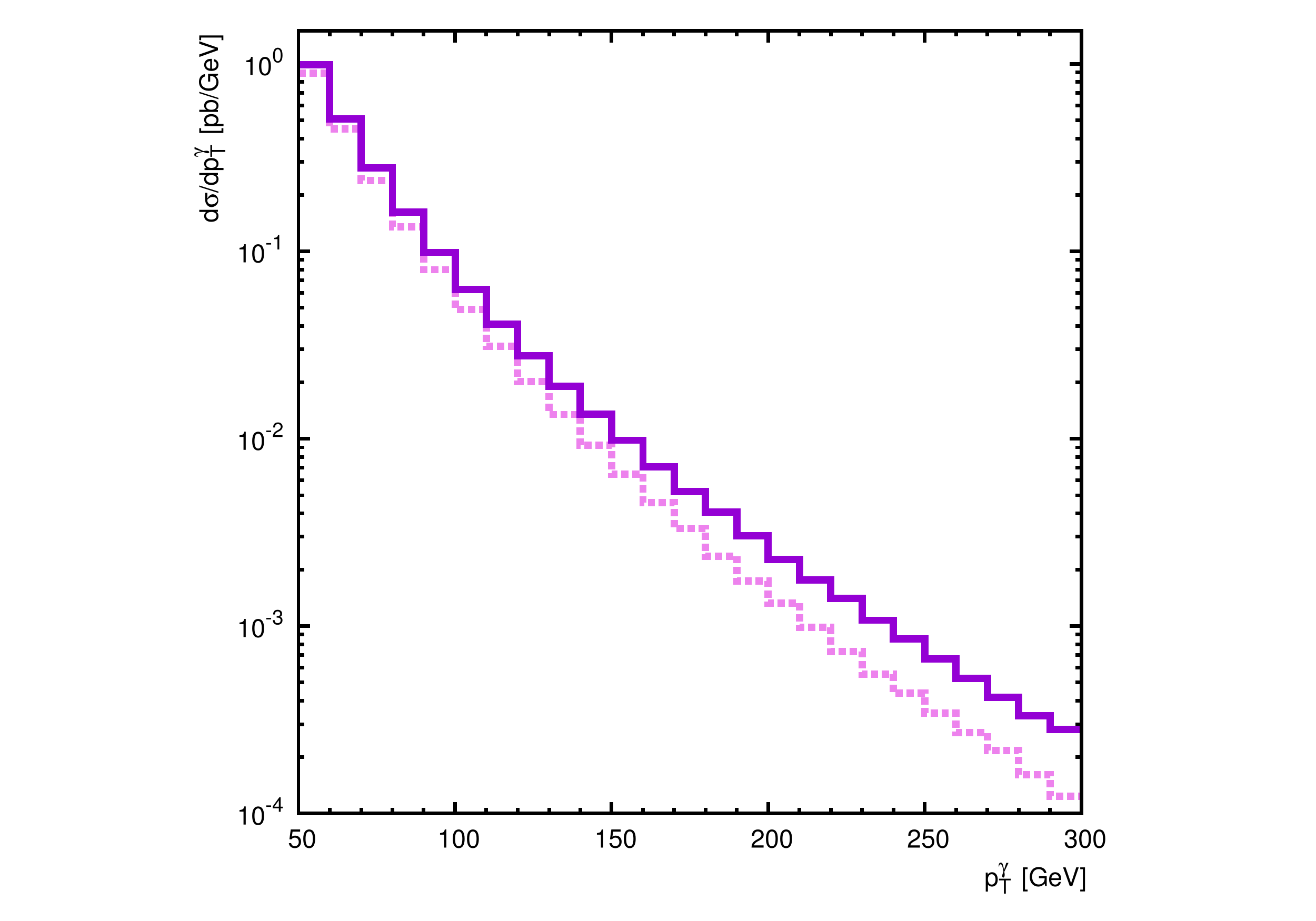}
\includegraphics[width=.68\textwidth,angle=270]{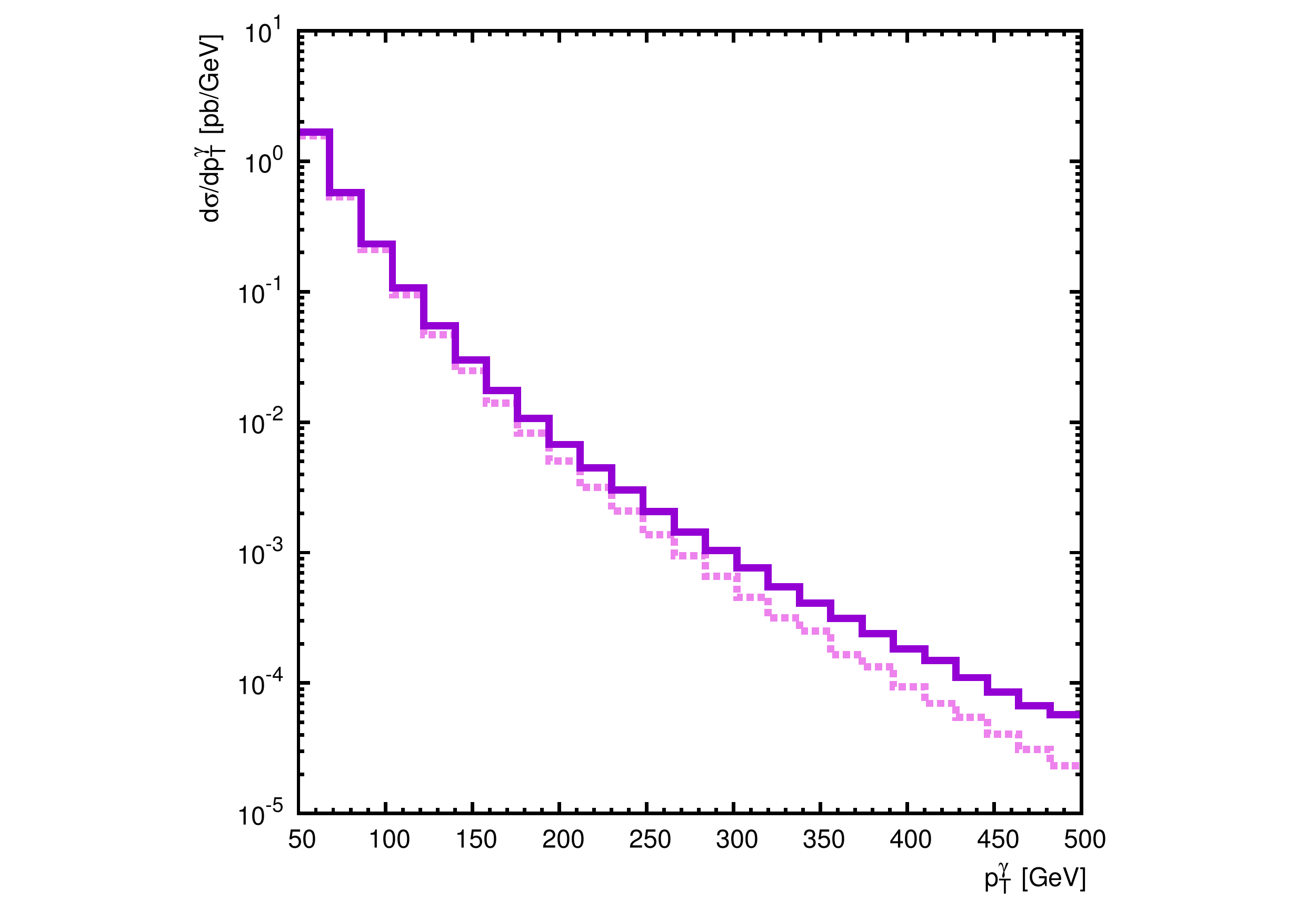}
\caption{The \pt-spectra of the photons accompanied by the $D^*$-mesons in $pp$
collisions with the intrinsic 3.5~\% charm contribution and without it at
$\sqrt{s} = 8$~\tev\ (\textcolor{purple}{top}) and $\sqrt{s} = 13$~\tev\
(\textcolor{purple}{bottom}). Both plots correspond to
$1.5 < |y^\gamma| < 2.4$, $|y^\mathrm{jet}| < 2.4 $.}
\label{fig_Dst_sp}
\end{figure}

\begin{figure}[h!]
\centering
\includegraphics[width=.68\textwidth,angle=270]{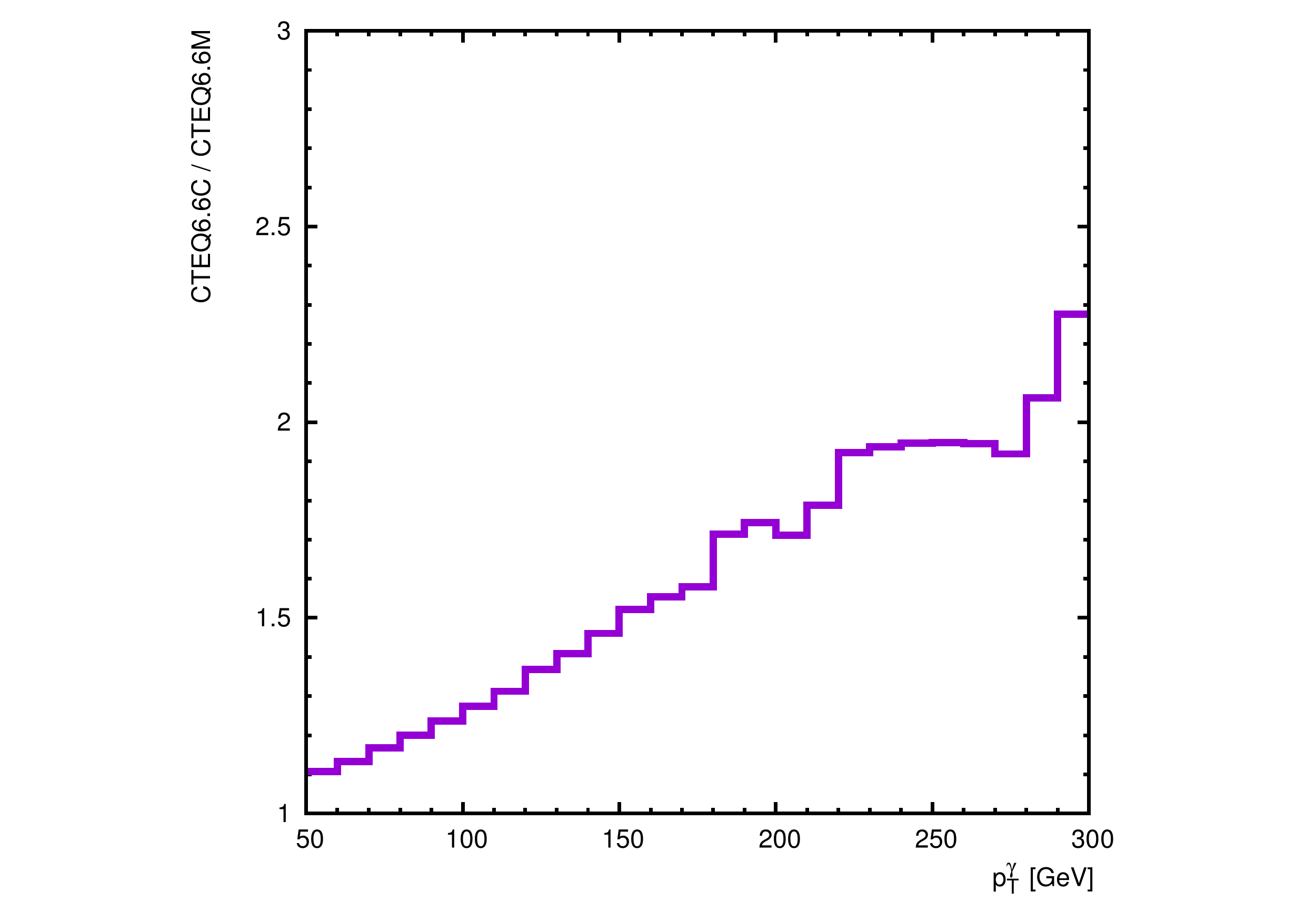}
\includegraphics[width=.68\textwidth,angle=270]{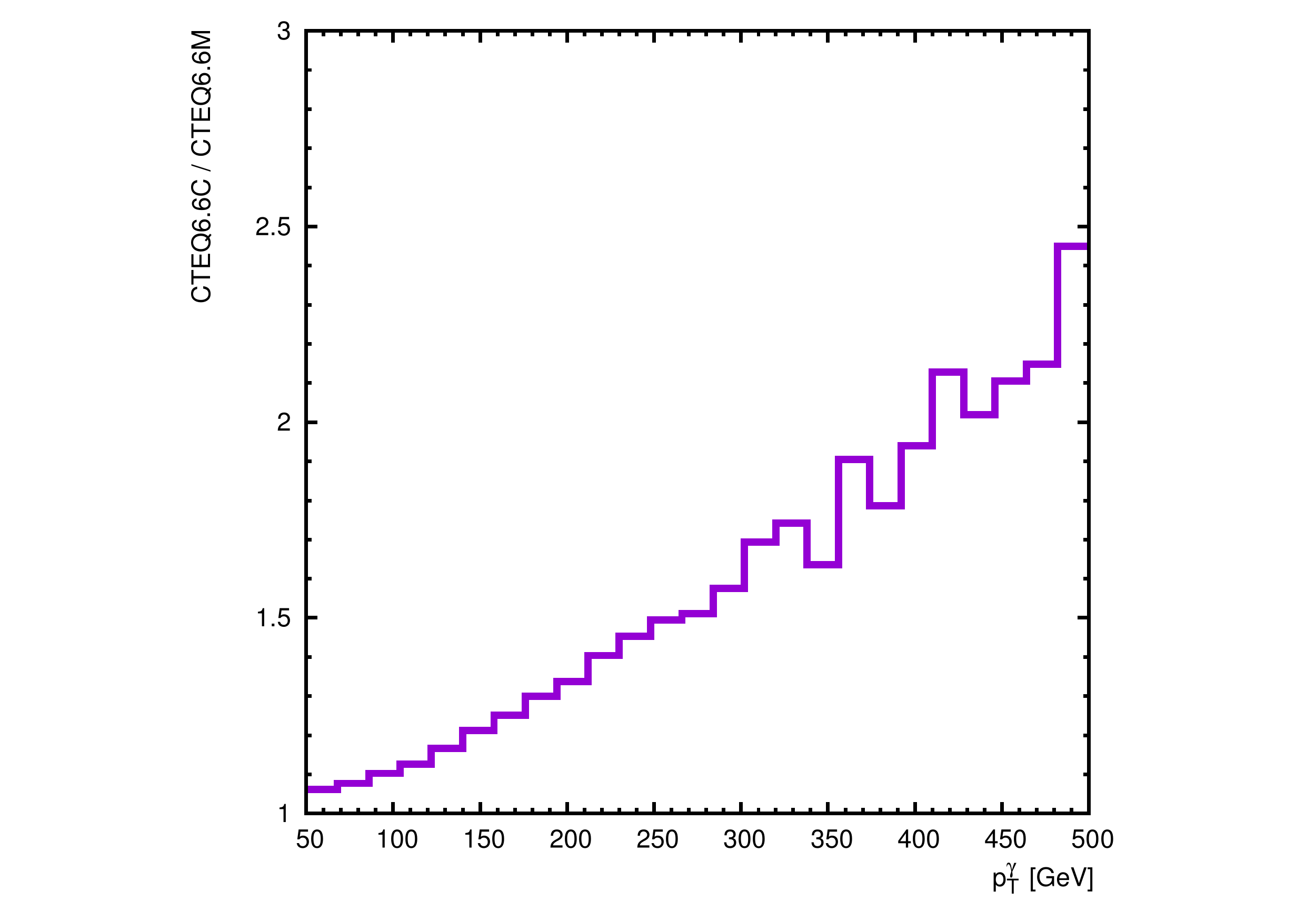}
\caption{The ratios of the \pt-spectra of the photons accompanied by the
$D^*$-mesons in $pp$ collisions with the intrinsic 3.5~\% charm contribution
and without it at $\sqrt{s} = 8$~\tev\ (\textcolor{purple}{top}) and
$\sqrt{s} = 13$~\tev\ (\textcolor{purple}{bottom}). Both plots correspond to
$1.5 < |y^\gamma| < 2.4$, $|y^\mathrm{jet}| < 2.4 $.}
\label{fig_Dst_ratio}
\end{figure}

In Fig.~\ref{fig_Dst_sp} the differential cross-sections of prompt photons
accompanied by the $D^*$-mesons calculated as a function of produced photon
transverse momenta at $\sqrt{s} = 8$~\tev\ (\textcolor{purple}{top}) and
13~\tev\ (\textcolor{purple}{bottom}) are presented with and without the IC
contribution, and the ratios of these spectra are shown in
Fig.~\ref{fig_Dst_ratio}. The calculations were performed using the
\kt-factorization approach and the kinematical requirements applied are the
same as above. We produce $D^*$ mesons from charmed quarks using the Peterson
fragmentation function with a shape parameter $\epsilon_c = 0.06$, and the
branching fraction $f(c \to D^*)$ is equal to $0.255$. In
Fig.~\ref{fig_Dst_ratio} the ratio of the \pt-spectra of the photons
accompanied by the $D^*$ mesons in $pp$ collisions with the intrinsic 3.5~\%
charm contribution and without it at $\sqrt{s} = 8$~\tev\
(\textcolor{purple}{top}) and $\sqrt{s} = 13$~\tev\
(\textcolor{purple}{bottom}). One can see that the IC signal in the $\gamma +
D^*$ cross-section is practically the same as in the case of $\gamma + c$-jet
production.

\subsection{$W(Z)$-boson and $b(c)$-jet production}
\label{VII.V}

\begin{figure}[h]
\centering
\includegraphics[width=1.\textwidth]{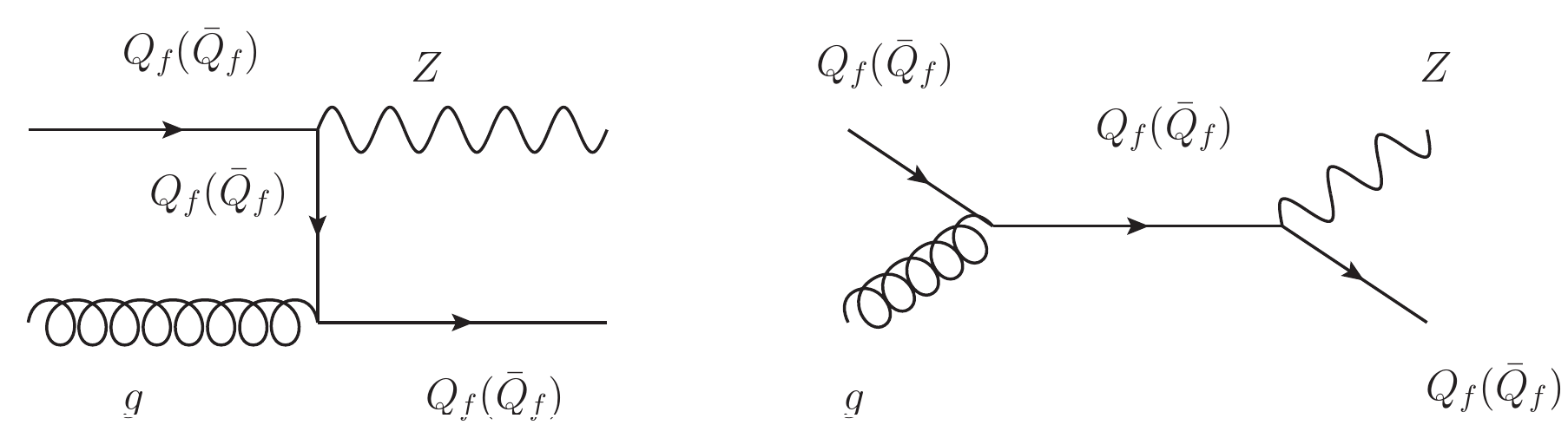}
\caption{LO Feynman diagrams for the process $Q_f(\bar{Q}_f) g \to Z
Q_f(\bar{Q}_f)$, where $Q_f = c, b$.}
\label{Fig_Qg_ZQ}
\end{figure}

Let us analyze another hard processes of the production of vector boson
accompanied by the $b$ and $c$ jets in $pp$ collision, which can give us also
information on the intrinsic charm in proton~\cite{Beauchemin:2014rya}. The LO
QCD diagram for the process ${Q_\mathrm{f}(\bar{Q}_\mathrm{f}) + g \to
Z^0 + Q_\mathrm{f}(\bar{Q}_\mathrm{f})}$ is presented in Fig.~\ref{Fig_Qg_ZQ}.
These hard subprocesses can give the main contribution to the reaction $pp
\to {Z^0(\to l^+ + l^-)} + Q_\mathrm{f}(\bar{Q}_\mathrm{f}) -
jet + X$, which could also give us information on the IC contribution in the
proton.

\begin{figure}[h]
\centering
\includegraphics[width=.6\textwidth]{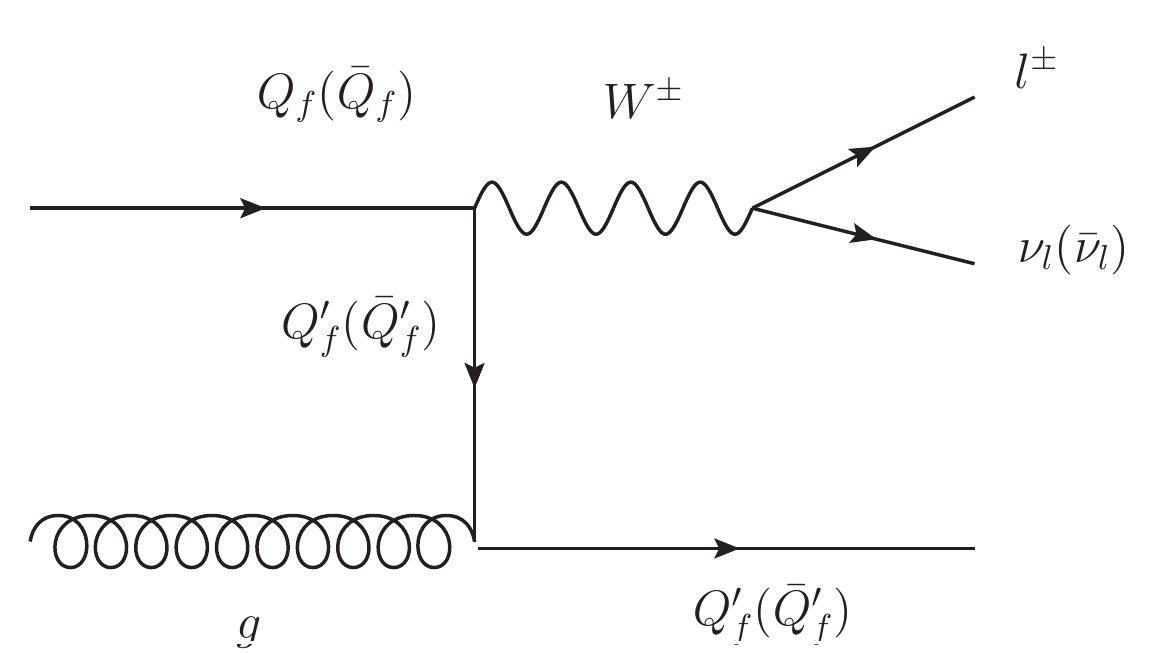}
\caption{Example of an LO Feynman diagram for the process
$Q_\mathrm{f}(\bar{Q}_\mathrm{f}) g \to W^\pm
Q_\mathrm{f}^\prime(\bar{Q}_\mathrm{f}^\prime)$, where $Q_\mathrm{f} = c, b$
and $Q_\mathrm{f}^\prime = b, c$ respectively.}
\label{Fig_Qg_WQ}
\end{figure}

The LO QCD diagram for the process $Q_\mathrm{f}(\bar{Q}_\mathrm{f}) + g
\to W^{\pm} + Q^\prime_\mathrm{f}(\bar{Q}^\prime_\mathrm{f})$ is
presented in Fig.~\ref{Fig_Qg_WQ}, where $Q_\mathrm{f} = c, b$ and
$Q^\prime_\mathrm{f} = b, c$. These hard subprocesses can give the main
contribution to the reaction $pp \to W^\pm(\to l^+ + \nu) +
Q^\prime_\mathrm{f}(\bar{Q}^\prime_\mathrm{f})\textnormal{-jet} + X$,
which could give us information not only on the IC
contribution but also on the IS in the proton.

\begin{figure}[h]
\centering
\includegraphics[width=0.365\textwidth]{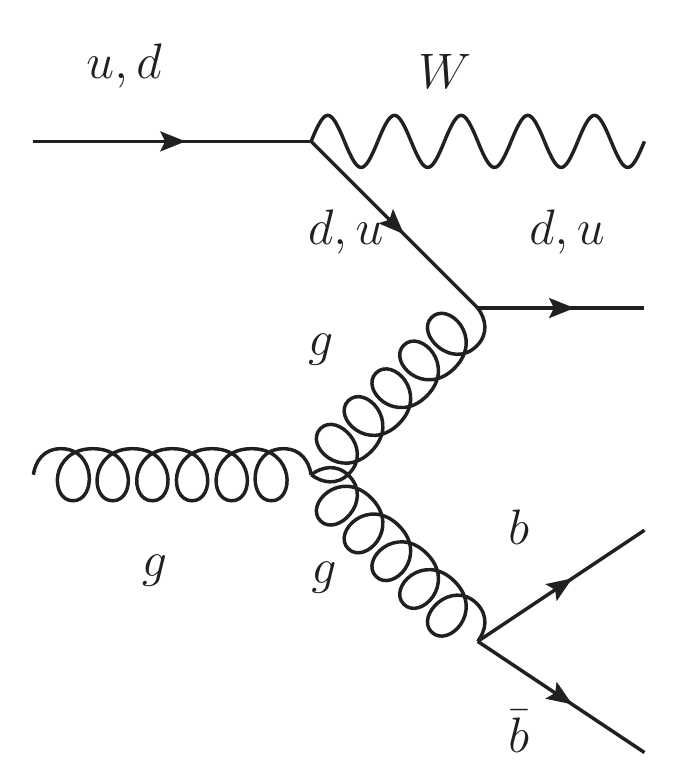}
\includegraphics[width=0.4\textwidth]{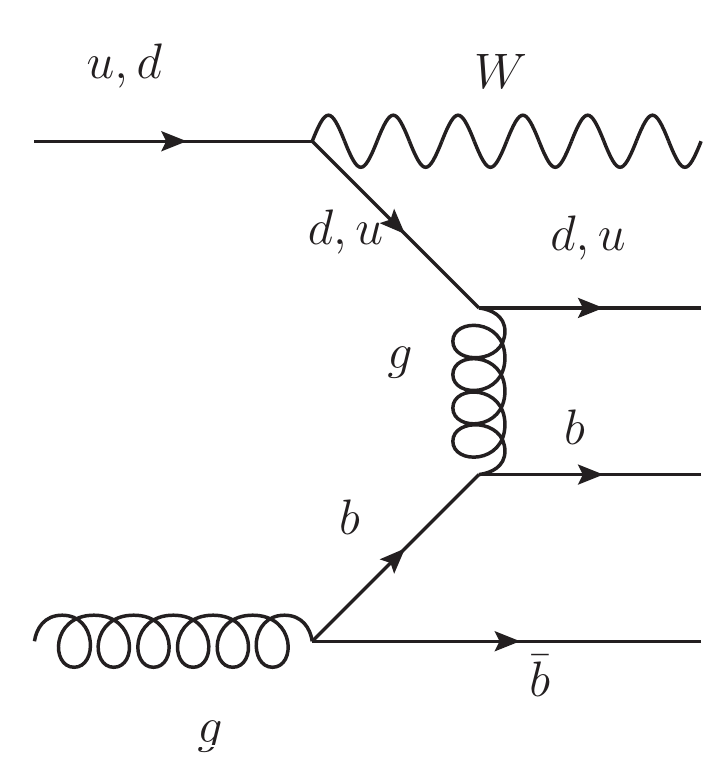}
\caption{Some NLO Feynman diagrams for the process $Q_f(\bar{Q}_f) g
\to W^\pm Q_f^\prime(\bar{Q}_f^\prime)$, where $Q_f = c,b$ and
$Q_f^\prime = b, c$ respectively. Left: gluon-splitting; Right: $t$-channel type
of W-scattering with one gluon exchange in the intermediate state.}
\label{Fig_6Wb}
\end{figure}

At NLO in QCD, $W/Z + Q_f$ diagrams, often more complicated than the ones
presented in Figs.~\ref{Fig_Qg_ZQ} and~\ref{Fig_Qg_WQ}, must also be considered. As
can be  seen in Fig.~\ref{Fig_6Wb}, the heavy flavor jets in the final state of
these diagrams come from a gluon splitting somewhere along the event chain, and
does thus not feature any intrinsic quark contribution. If the cross-sections
of these diagrams is large enough, the conclusions about the sensitivity of a
measurement to intrinsic charm at the LHC will be affected. It is thus
important to consider QCD NLO calculations in the current study.

To this end, we calculated the \pt-spectra of heavy flavor jets ($b$ and $c$)
in association with a vector boson produced at NLO in $pp$ collisions at
$\sqrt{s} = 8$~\tev\ using the parton level Monte Carlo (MC) generator MCFM
version 6.7~\cite{Campbell:2002tg}. The NLO corrections include
the splitting of a gluon into a pair of heavy flavor quarks, and thus provides
a better description of such process than what is yielded by parton showers, at
least for the first splitting. The lack of further parton radiation and of
hadronization in MCFM will affect the shape of the hadronic recoil to vector
bosons and the \pt\ spectra of the leading heavy flavor jet in the various $V +
c$ and $V + b$ ($V = W$ or $Z$) events, but it affects the predictions with and
without an intrinsic charm contributions to the PDF in the exact same way.
Conclusions that will be derived from MCFM about the IC sensitivity studies to
be presented below are thus not affected by the fact that MCFM provides only a
fixed order calculation with no parton shower or further non-perturbative
corrections. For the various processes considered, the vector boson is required
to decay leptonically, in order to allow experimental studied to trigger on
these events, and the pseudo-rapidity of the heavy quark jet is required to
satisfy $|\eta_\mathrm{Q}| < 1.5$, to probe high-$x$ PDFs.

\begin{figure}[h!]
\centering
\includegraphics[width=.68\textwidth]{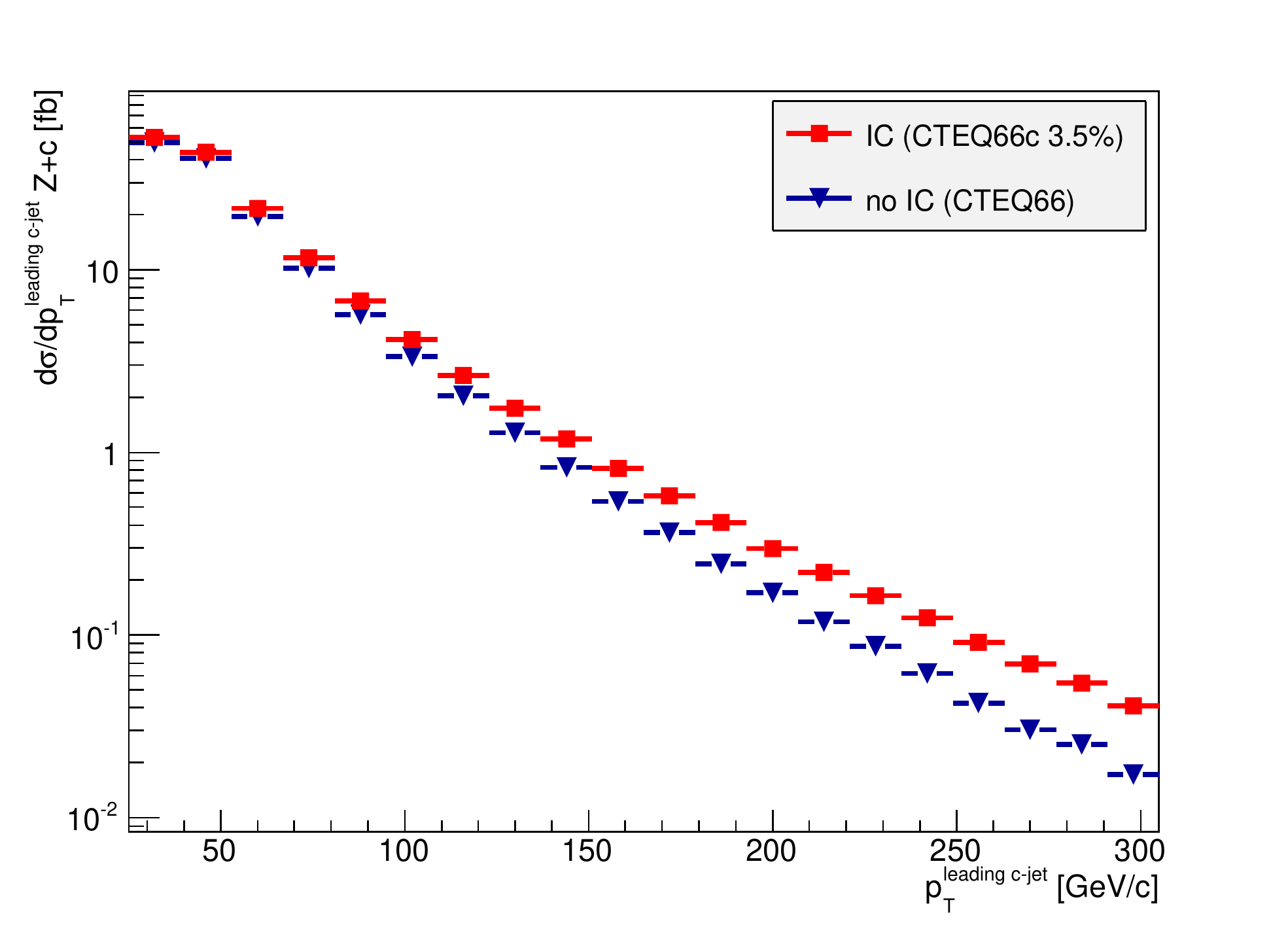}
\includegraphics[width=.68\textwidth]{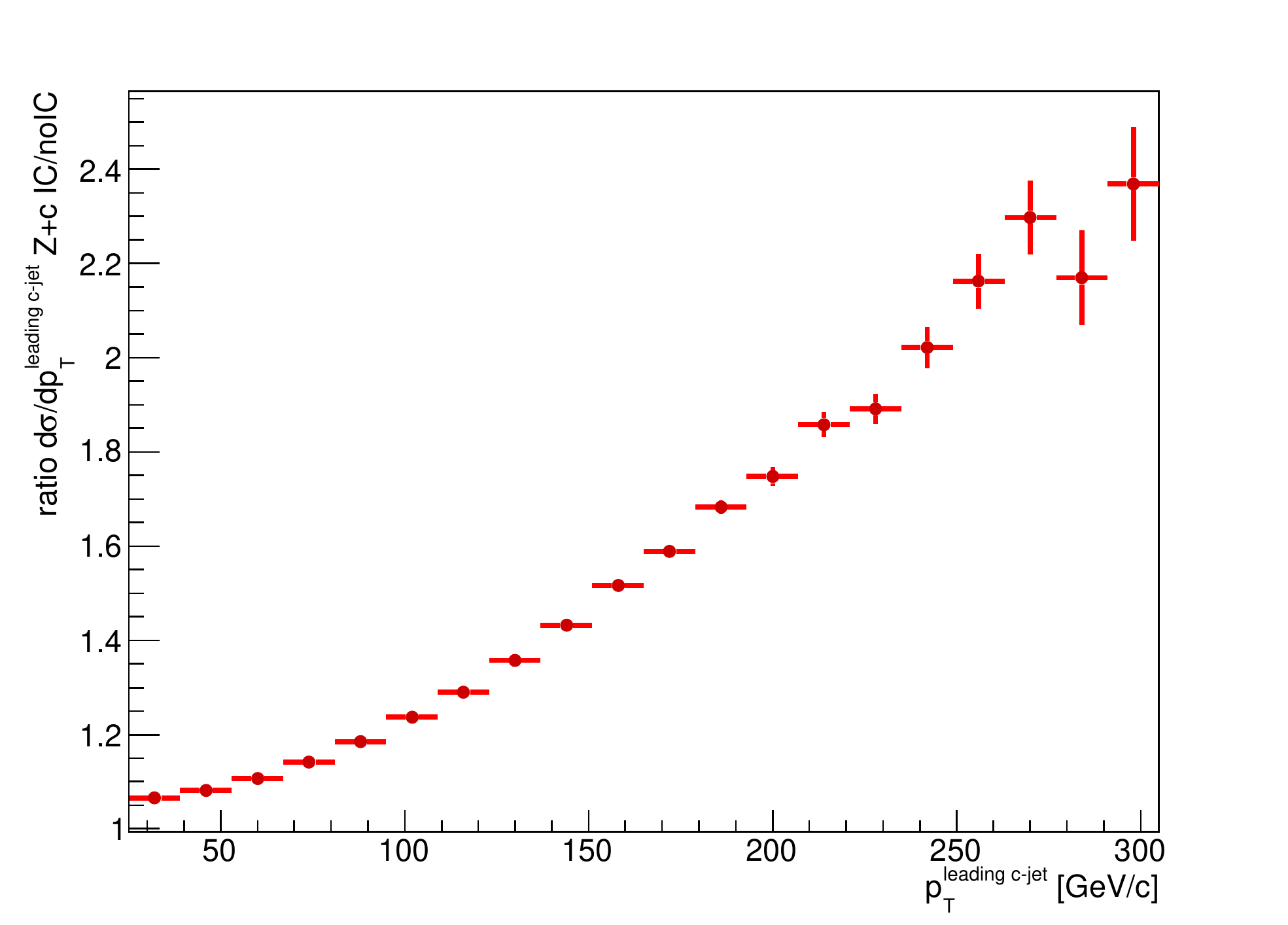}
\caption{Comparison of the \pt-spectra for the NLO $pp \to Z + c$
process 262~\cite{Campbell:2002tg} obtained with PDF including
an intrinsic charm component (CTEQ66c) and PDF having only an extrinsic
component (CTEQ66) (top). Ratio of these two spectra (bottom).}
\label{Fig_4Zc}
\end{figure}

By selecting $Z + c$-jet events, where the $c$-jet is required to be rather
forward ($1.5 < |y_\mathrm{c}| < 2.0$), we can see on the
\textcolor{purple}{top} panel of Fig.~\ref{Fig_4Zc} that the $c$-jet transverse
momentum spectrum of events with a 3.5~\% intrinsic charm contribution to the
PDF (CTEQ66c) features an excess, increasing with the $c$-jet \pt, compared to
the corresponding differential cross-section when only extrinsic heavy flavor
components of the PDF are considered (CTEQ66). These differential cross-section
distributions have been obtained at NLO from the MCFM processes number 262.
From the right panel of the same figure, showing the ratio of the two spectra
obtained with and without an IC contribution, we can see that the excess in the
$c$-jet \pt\ spectrum due to IC is of $\sim 5$~\% for \pt\ of 50~\gev, and
rises to about 220~\% for $p_\mathrm{T} \sim 300$~\gev. This effect can thus be
observed at the LHC if the $c$-jet \pt\ differential cross-section in $Z + c$
events can be measured with sufficient precision.

\begin{figure}[h!]
\centering
\includegraphics[width=.68\textwidth]{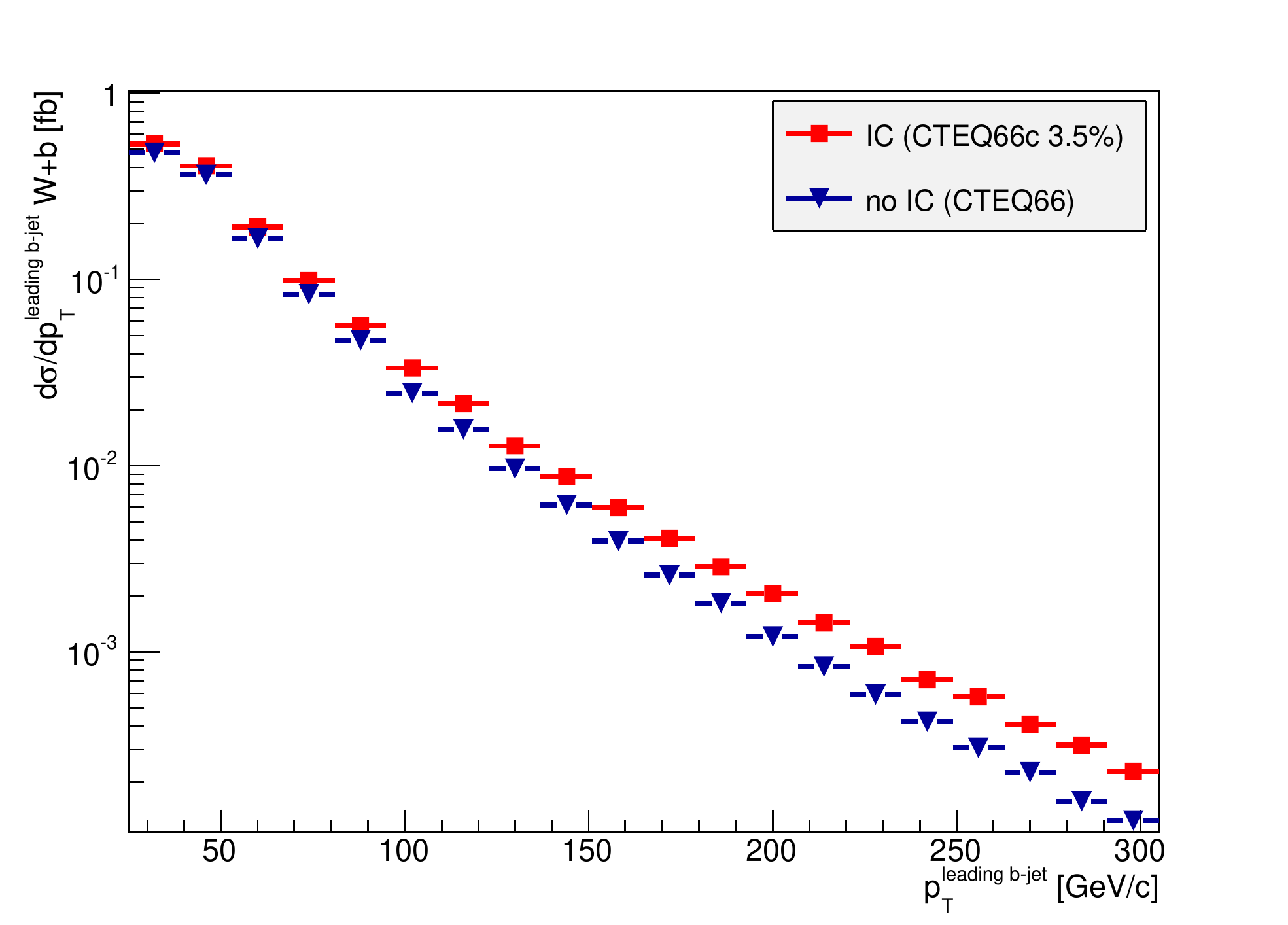}
\includegraphics[width=.68\textwidth]{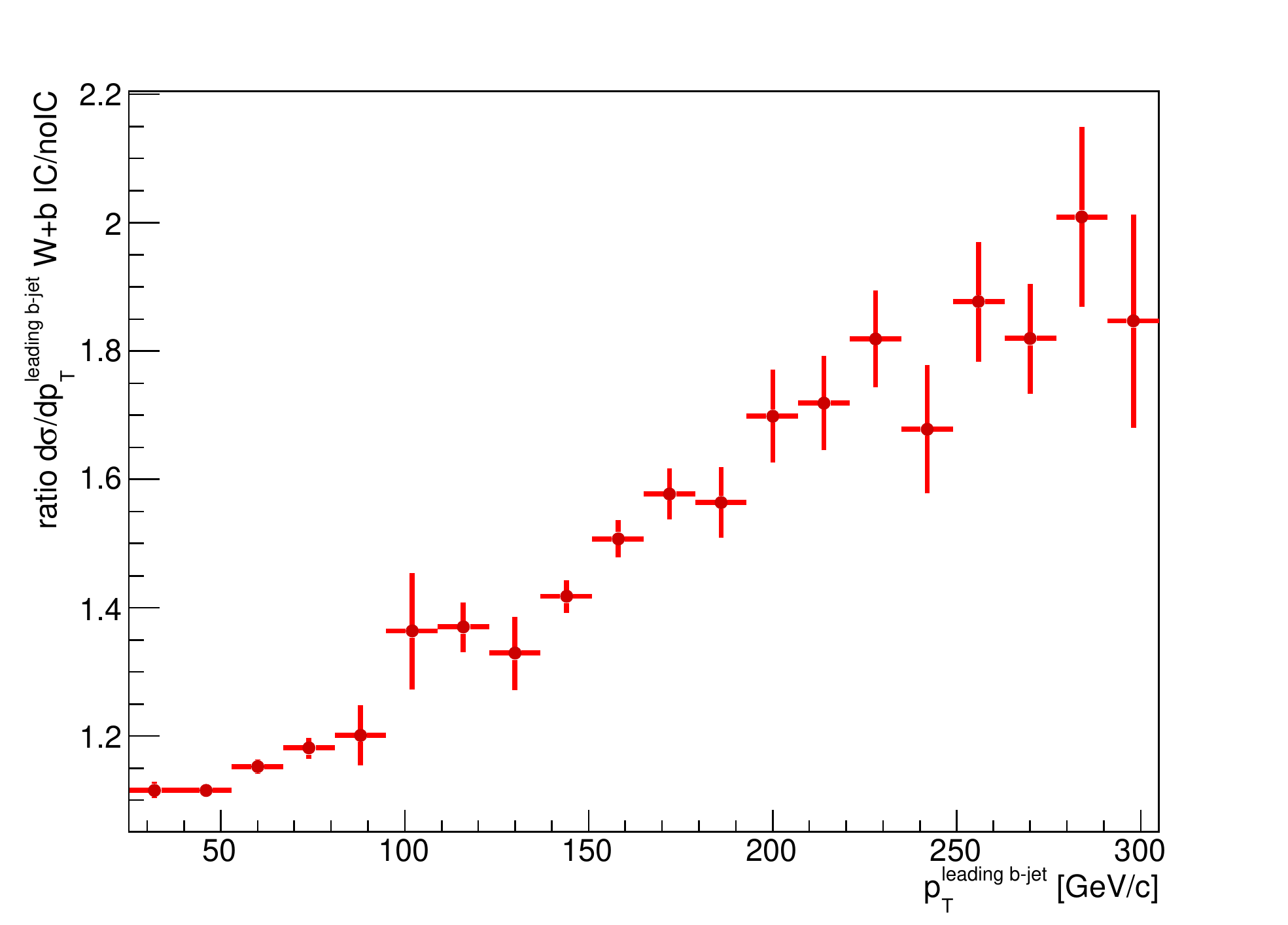}
\caption{Comparison of the \pt-spectra for the NLO $pp \to W + b$,
processes 12 + 17~\cite{Campbell:2002tg} obtained with PDF
including an intrinsic charm component (CTEQ66c) and PDF having only an
extrinsic component (CTEQ66) (top). Ratio of these two spectra (bottom).}
\label{Fig_5Wb}
\end{figure}

In the case of the $W$ production in association with heavy flavor jets, the
intrinsic charm contribution would be observed in a $W + b$-jet final state due
to the change of flavor in the charged current. In MCFM, the NLO $W + b$
Feynman diagrams for which the LO part is depicted in Fig.~\ref{Fig_Qg_WQ},
correspond to the MCFM processes 12 and 17~\cite{Campbell:2002tg}. They provide
the contribution to $W + Q'$ which is sensitive to IC\@. The \pt\ spectrum of
the $b$-jet is presented, for the sum of these processes, in Fig.~\ref{Fig_5Wb}
(\textcolor{purple}{top}), where one calculation (squares) has been obtained at
NLO in QCD with the CTEQ66c PDF that includes an IC contribution (about
3.5~\%), and the other calculation (triangles) uses the CTEQ66 PDF, which does
not include IC\@. On the \textcolor{purple}{bottom} panel of Fig.~\ref{Fig_5Wb},
the ratio of these two spectra (with and without an IC contribution to the PDF
used in the $W + b$ production calculations) is presented. From this figure,
one can see that the inclusion of the IC contribution to the PDF leads to an
increase in the $b$-jet spectrum by a factor of about 1.9 at $p_\mathrm{T} >
250$~\gev. This is comparable to what was observed in the $Z + c$ case of
Fig.~\ref{Fig_4Zc}.

Similarly, the $W + c$ final state would be sensitive to intrinsic strange
while the $Z + b$ final state would be sensitive to intrinsic bottom. These
processes are however suboptimal for finding intrinsic quarks at the LHC\@. As
mentioned above, the contribution of the IB to the PDF is suppressed by a
factor of ${(\frac{m_\mathrm{c}}{m_\mathrm{b}})}^2$ and is thus subdominant
compared to intrinsic charm. The contribution of the intrinsic strangeness (IS)
can be of the same order of magnitude as the IC according
to~\cite{Peng:2012rn,Lykasov:2013rva}.  The $Q^2$ evolution for this component
has however not been calculated up to now, and thus contains many unknowns. This
is why this paper concentrate on the intrinsic charm component of the proton.

The above results of Figs.~\ref{Fig_4Zc} and~\ref{Fig_5Wb} seem a priori very
encouraging regarding the capacity of the LHC to provide an observation of an
intrinsic charm contribution to the PDFs in $W/Z + Q_\mathrm{f}$ events, but
the real situation is unfortunately more complex than this. The $W$ boson plus
one or more $b$-quark jets production, calculated at NLO in the 4-flavor scheme
(4FNS), for which two of the diagrams  are represented in Fig.~\ref{Fig_6Wb}
must also be included. These correspond to the MCFM processes number 401/406
and 402/407~\cite{Campbell:2002tg}. Their total cross-section is about 50 times
larger that the $W + b$ processes sensitive to IC\@. As a result, the total $W
+ b$ production is not sensitive to an intrinsic charm component of the PDF, as
it was shown in~\cite{Beauchemin:2014rya}, where the sum of all processes
contributing to $W + b$ has been taken into account. Fortunately the $Z + c$
processes do not suffer from a similar large dilution of the intrinsic quark
component because the $Q_\mathrm{f} + g \to Z + Q_\mathrm{f}$
processes, which are sensitive to IC, are not Cabibbo-suppressed.

Another difficulty lies in the experimental identification of heavy flavor jets
in the way how to select them, for example, $Z + c$-jet events in a very large
$Z$-jets sample. Algorithms typically disentangling heavy flavor jets from
light-quark jets exploit the longer lifetime of heavy-quark hadrons that decay
away from the primary vertex of the main process, but close enough to allow for
a reconstruction of the tracks of the decay products of the heavy-flavor hadron
in the inner part of the detector.  Such algorithms are typically not capable
of explicitly distinguishing $c$-jets from $b$-jets; only the efficiency for
identifying the heavy flavor nature of the jet would differ between $c$-jets
and $b$-jets. For example, one of the ATLAS heavy-flavor tagging algorithm
(MV1) yields an efficiency of 85~\% for $b$-jet identification and 50~\% for
$c$-jet (for a working point where the light flavor rejection is 10), see
References in~\cite{Beauchemin:2014rya}. As a result of such heavy flavor jet
tagging algorithm, the selected $Z + Q$ final state will be a mixture of $Z +
c$ and $Z + b$.

A priori, one would expect that $Z + b$ events are sensitive to intrinsic
bottom and therefore act only as a small background to intrinsic charm studies,
when the two processes cannot be experimentally distinguished. The situation is
however more complicated than this. Because of sum rules, an intrinsic charm
component would affect the total $b$-quark contribution to the proton, and the
$Z + b$-jet final state therefore becomes sensitive to intrinsic charm as well.
As was shown in~\cite{Beauchemin:2014rya}, this contribution is in the opposite
direction of the intrinsic charm effect on $Z + c$ processes presented in
Fig.~\ref{Fig_4Zc}. In addition, the heavy flavor tagging efficiency is lower
for $c$-jets than it is for $b$-jets, therefore increasing the weight of the
negative $Z + b$ contribution to the total $Z$ plus heavy flavor tagged jets
signal. The question is thus: are $Z + Q$-jet events still sensitive to
intrinsic charm?

To avoid these difficulties in~\cite{Beauchemin:2014rya} the following idea to
search the IC signal in the $Z/W$ boson production accompanied by heavy flavor
jets was suggested. The new idea is to use the ratio of the leading heavy
flavor spectra in inclusive heavy flavor $Z + Q$ to $W + Q$ events to verify
the predictions about an IC contribution to the proton. The background related
to the $Z/W$ production with association of light jets is included, which does
not change the main conclusion on the IC signal in that ratio. We would like to
stress also that studying the ratios discussed above we avoid many
uncertainties related to a background due to the light jet production,
rescaling in QCD and many others because they are almost cancelled.  Such
measurements can already be made with ATLAS and CMS available data.

\begin{figure}[h!]
\centering
\includegraphics[width=.68\textwidth]{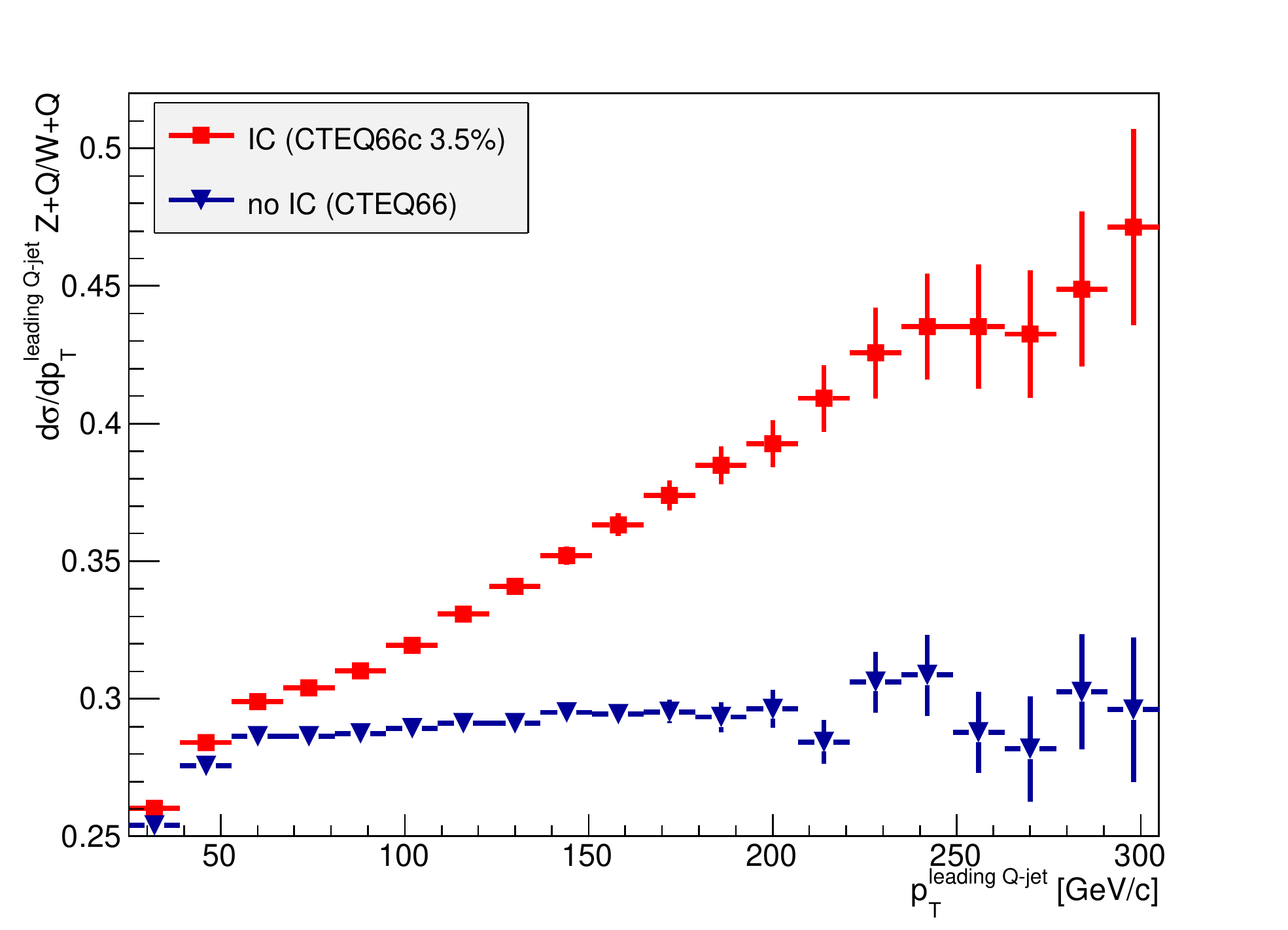}
\includegraphics[width=.68\textwidth]{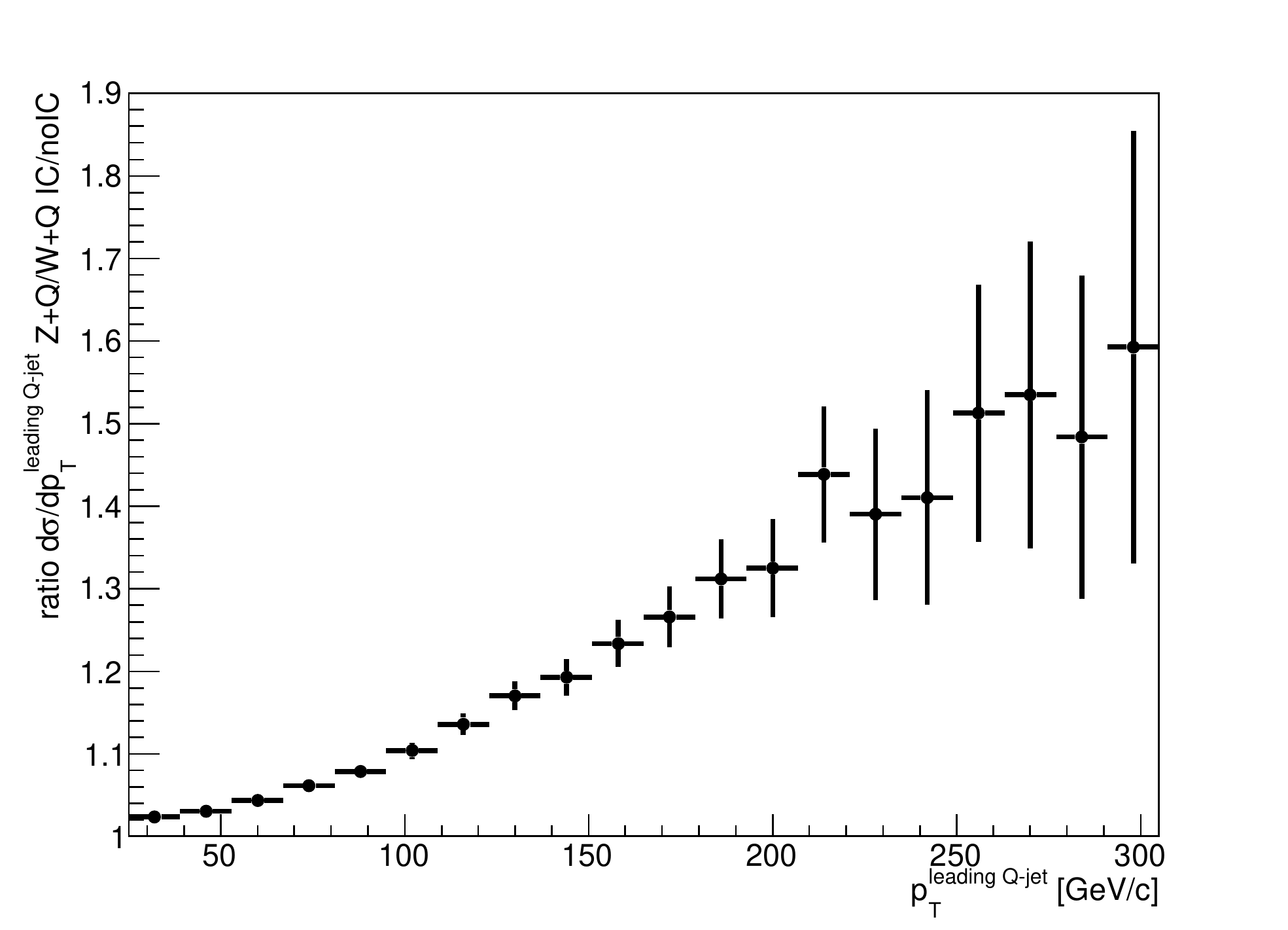}
\caption{Comparison of the ratio of the \pt-spectra for the $Z + Q$ to $W + Q$
NLO processes obtained with PDF including an intrinsic charm component
(CTEQ66c) and PDF having only an extrinsic component (CTEQ66). Heavy flavor jet
tagging efficiencies have been applied to the $c$-jets and the $b$-jets (top).
Ratio of these two ratios of spectra (bottom).}
\label{Fig_11Rall}
\end{figure}

To verify this, the ratio of the \pt\ spectra of the leading heavy flavor jet
($b$, $c$) produced in $Zb + Zc$ and $Wb + Wc + Wbj$ processes has been
to the proton. The of the calculation is presented in Fig.~\ref{Fig_11Rall}. As
can be seen in this figure, the sensitivity to IC signal observed in $Z + Q$ is
maintained in the ratio, which can amount to about 160~\% of the extrinsic only
contribution at \pt\ of about 270~--~300~\gev. This ratio measurement would, at
least partially, cancel a number of large experimental systematic
uncertainties, especially since in our proposal, $V + c$-jets and $V + b$-jets
are both considered as signal and not treated as a background with respect to
the other. This would allow for a clear signal at the LHC, if the IC
contribution is sufficiently high (here we considered a 3.5~\% contribution).
In the case where no excess is observed, limits on the IC contribution to the
proton can be obtained from such measurement. Note that ratio predictions
obtained with MCFM~\cite{Beauchemin:2014rya} would agree with predictions that
include a parton shower and a modeling of the hadronization, because such
effects cancel in the ratio for jets above $\sim 100$~\gev.

As discussed above, a high \pt\ and relatively high rapidity heavy-flavor jet
enhances the probability to have a heavy flavor quark in the initial state with
a high-$x$ fraction, ensuring that the effect of intrinsic quarks on the
cross-section is more prominent. This is the reason why we proposed to measure
the ratio of $Z + Q$ to $W + Q$ differential cross-sections as a function of
the transverse momentum of the leading heavy-flavor jet measured within a
specific rapidity interval. As indicated by Eq.~\ref{def:xFptteta}, a large-$x$
heavy flavor quark will in general be achieved by a high-value of the Feynman
variable $x_\mathrm{F}^\mathrm{V}$ of the final state vector boson $V$
recoiling to the hadronic system. While such variable cannot be reconstructed
at the detector level in $W + Q$ events because of the presence of an
undetectable neutrino  in the final state, it is possible to construct a
quantity highly correlated to such Feynman variable by using the leading
heavy-flavor jet in the final state, rather than the vector boson. We therefore
propose to investigate the sensitivity to IC of the ratio of the $Z + Q$ to $W
+ Q$ differential cross-sections as a function of the pseudo-Feynman variable
of the leading heavy-flavor jet defined as:
\begin{equation}
x_\mathrm{F}^\mathrm{Q} =
\frac{2 p_\mathrm{T}^{\mathrm{Lead~}Q\mathrm{\textnormal{-}jet}}}{\sqrt{s}}
\sinh(\eta_Q),
\label{def:xFQ}
\end{equation}
where $p_\mathrm{T}^{\mathrm{Lead~}Q\mathrm{\textnormal{-}jet}}$ is the
transverse momentum of the leading heavy flavor jet in the final state and
$\eta_Q$ is the pseudo-rapidity of this jet.

\begin{figure}[h!]
\centering
\includegraphics[width=.68\textwidth]{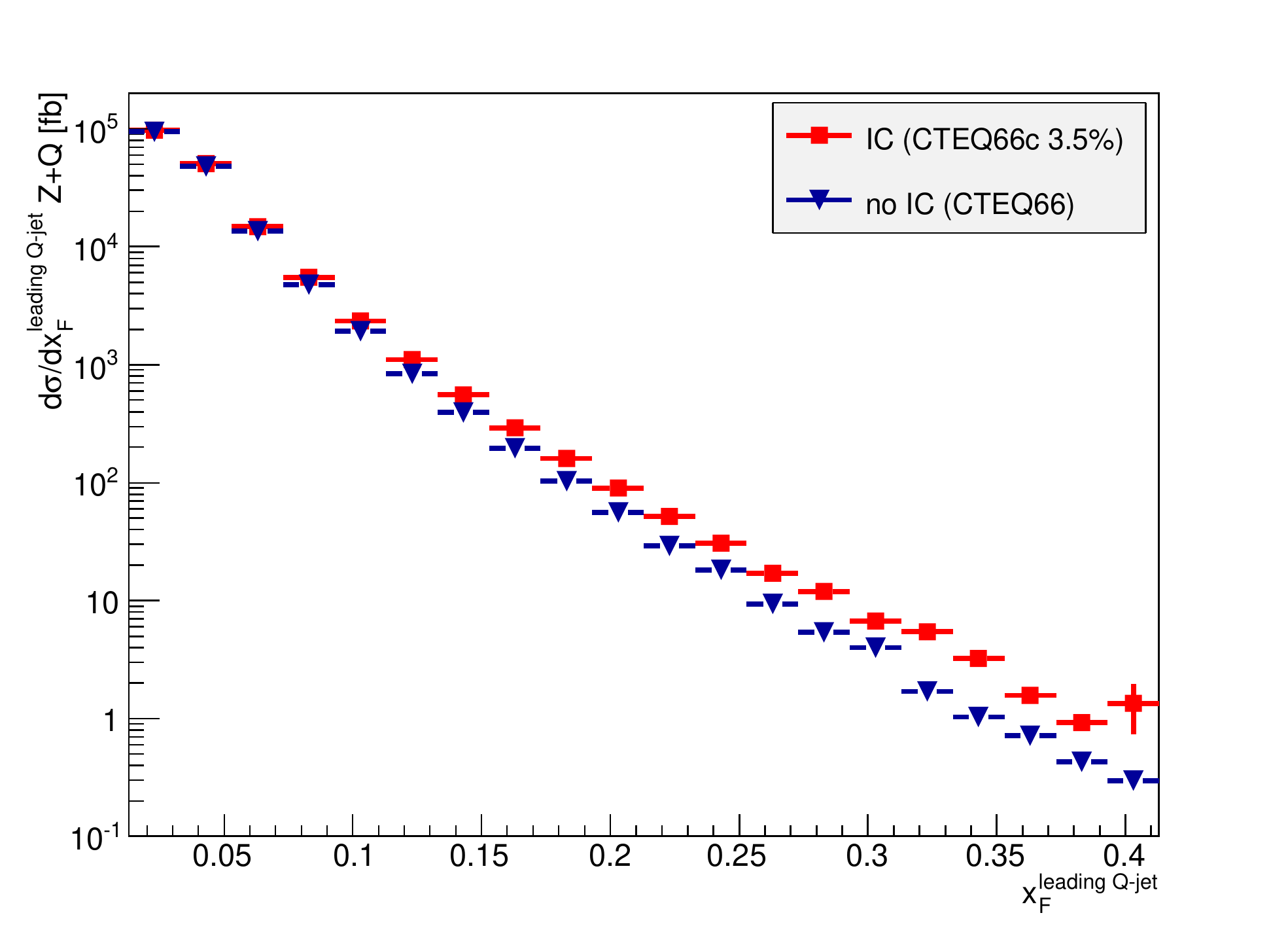}
\includegraphics[width=.68\textwidth]{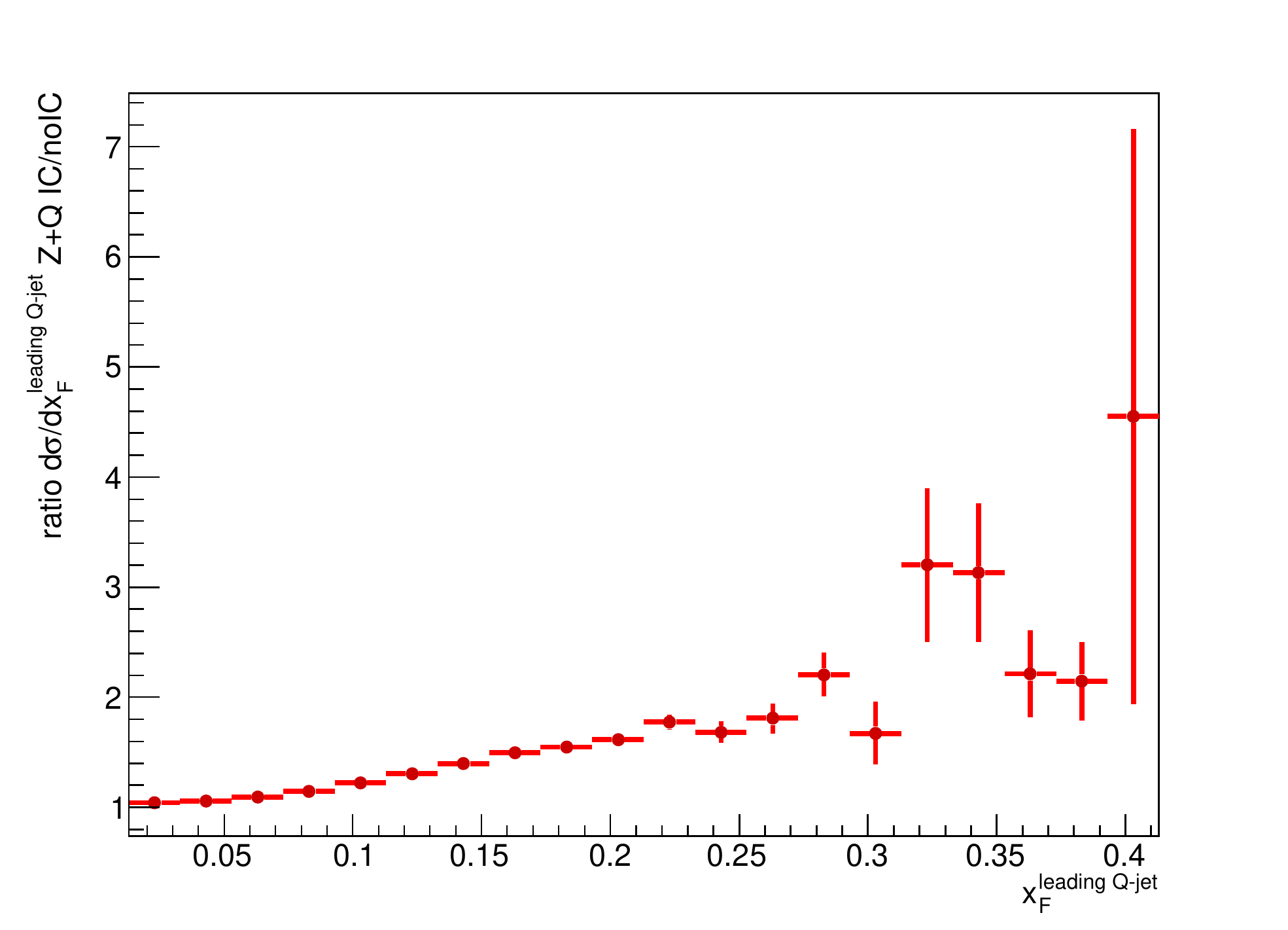}
\caption{Comparison of the \xfq-spectra for the total NLO $pp \to Z +
b(\bar{b})$ process plus $pp \to Z + c(\bar{c})$ (processes 261,
262~\cite{Campbell:2002tg}) obtained with PDF including an intrinsic charm
component (CTEQ66c) and PDF having only an extrinsic component (CTEQ66) (top).
Ratio of these two spectra (bottom).}
\label{Fig_ZxFspectra}
\end{figure}

First, the sensitivity of $Z+Q$ events to intrinsic charm is presented in the
\textcolor{purple}{top} panel of Fig.~\ref{Fig_ZxFspectra} as a function of
this pseudo-Feynman variable of the leading heavy-flavor jet. The \xfq-spectrum
has been obtained from the total NLO ${pp \to Z + b(\bar{b})}$ plus ${pp \to Z
+ c(\bar{c})}$ contributions calculated with MCFM (processes 261,
262~\cite{Campbell:2002tg}) for collisions at $\sqrt{s} = 8$~\tev. The
distribution displayed with red square has been obtained using the CTEQ66c PDF
distribution displayed with red square has been obtained using the CTEQ66c PDF
that includes an intrinsic charm component, while the distribution displayed as
blue inverted triangles have been obtained with the CTEQ66 PDF that only
contains an extrinsic charm component. \textcolor{purple}{In the bottom panel
the ratio of the two spectra is shown.}

\begin{figure}[h!]
\centering
\includegraphics[width=.68\textwidth]{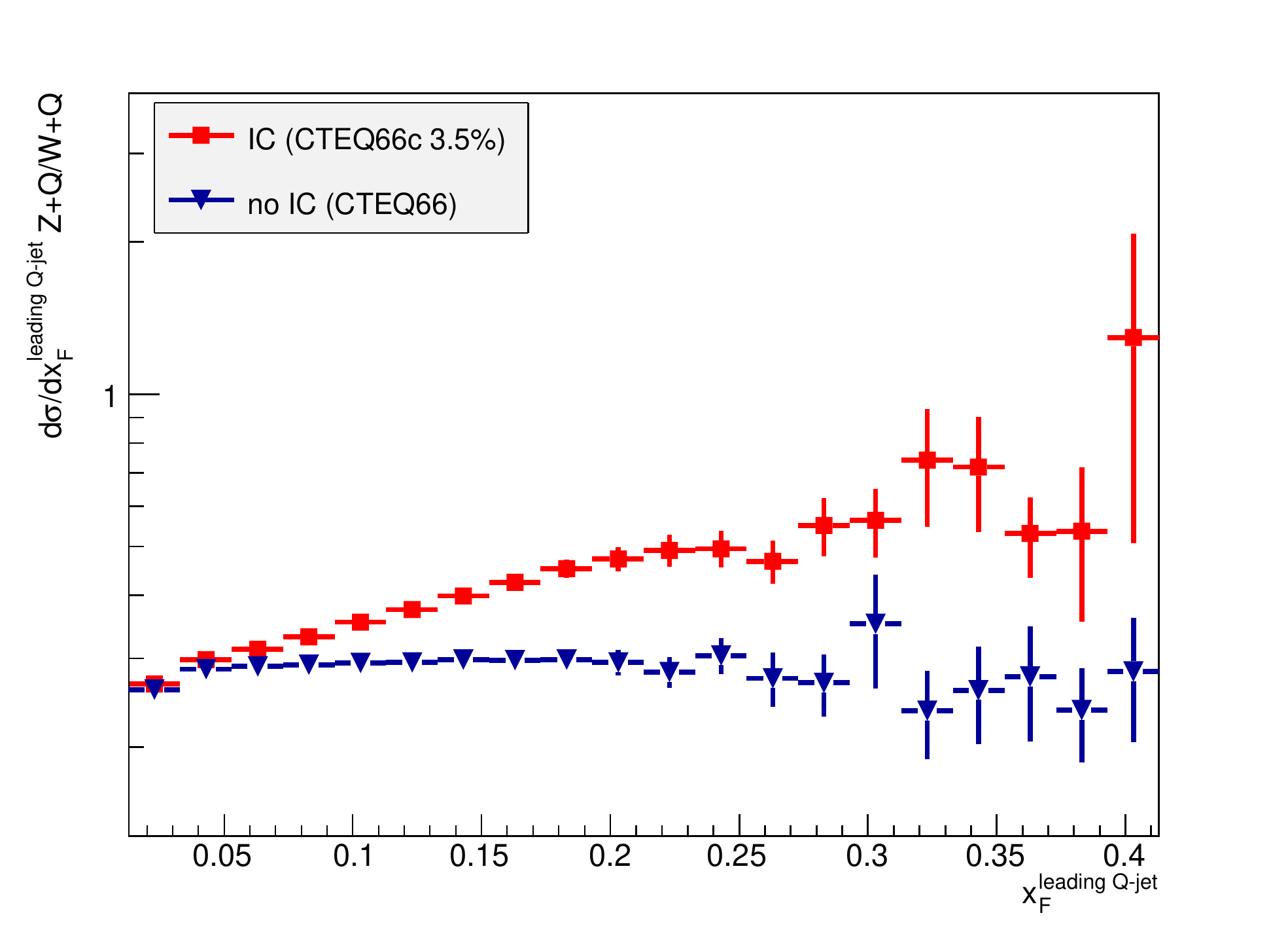}
\includegraphics[width=.68\textwidth]{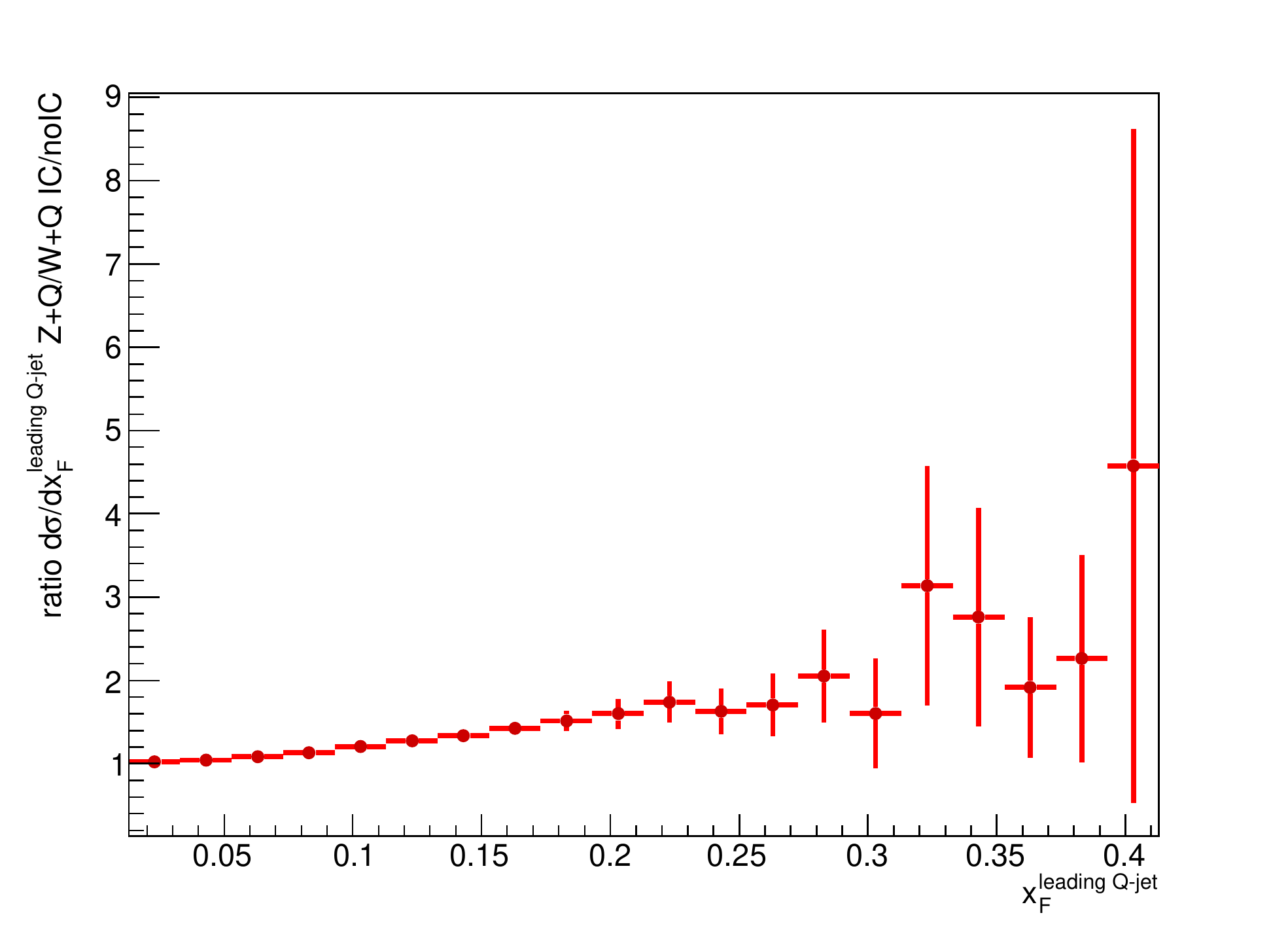}
\caption{Comparison of the ratio of the \xfq-spectra for the
$Z + Q$ to $W + Q$ NLO processes obtained with PDF including an intrinsic charm
component (CTEQ66c) and PDF having only an extrinsic component (CTEQ66) (top).
Ratio of these two ratios of spectra (bottom).}
\label{Fig_xFRatio}
\end{figure}

Fig.~\ref{Fig_xFRatio} presents the sensitivity to an intrinsic charm component
to the PDF for the ratio of $Z + Q$ to $W + Q$ events as a function of the
pseudo-Feynman variable \xfq. In this figure, heavy flavor tagging has been
applied to both $Z + Q$ and $W + Q$ processes. An IC contribution of 3.5~\%
yields a change by a factor of 2 to 4 in the $Z + Q$ to $W + Q$ cross-section
ratio at $x_\mathrm{F}^\mathrm{Q} \simeq 0.3$~--~0.4 compared to the
calculation where the PDF do not include any IC component. The number of events
in that kinematic region runs from about a few hundred up to a few thousand
events, for both $Z + Q$ and $W + Q$ processes. This results in a reduced
statistical uncertainty on the $Z + Q$ to $W + Q$ ratio compared to the
proposed ratio measured as a function of the transverse momentum of the leading
heavy flavor jet in the phase space region discussed above. Because the
pseudo-Feynman variable and the $Q$-jet transverse momentum observables
correspond to significantly different distributions in shape, both sensitive to
an intrinsic charm contribution to the proton, but with different sensitivity
to the various systematic uncertainties, we thus have here two complementary
ratio observables to be measured at the LHC in order to observed an IC
contribution to the proton, or determine an upper limit on it.

As discussed above, the leading heavy flavor jet transverse momentum and
rapidity distributions are similar for $b$-jet and $c$-jet in $Z + Q$ and $W +
Q$ events. As a consequence, the experimental uncertainties on $Q$-jet energy
measurements and heavy flavor tagging efficiencies will get significantly
reduced in the ratio measurements proposed above. The Feynman diagrams
contributing to $Z + Q$ and $W + Q$ processes are however quite different. It
is therefore important to verify that a similar cancellation of the theory
uncertainty also occurs in this ratio, therefore not impeding the conclusion
about IC that can be obtained with such ratio. The dominant theoretical
systematic uncertainty on a NLO cross-section calculation obtained at fixed
order in perturbative QCD comes, by far, from the uncertainty introduced by the
choice of renormalization ($\mu_\mathrm{R}$) and factorization
($\mu_\mathrm{F}$) scales in the calculation. In the current calculations
performed with MCFM~\cite{Campbell:2002tg}, the central predictions were
obtained with a dynamic scale $\mu_\mathrm{R} = \mu_\mathrm{F} = H_\mathrm{T}$,
where $H_\mathrm{T}$ is the scalar sum of the transverse momentum of all the
particles ($p_{\mathrm{T}i}$) in the final state ($H_T = \sum_i^n
p_{\mathrm{T}i}$). In order to assess the sensitivity of the calculations to
this choice of scale, cross-sections have been calculated with two other
choices of scale, $H_\mathrm{T} \cdot 2$ and $H_\mathrm{T}/2$, and results
compared to the nominal predictions.

It was shown in~\cite{Beauchemin:2014rya} that the choice of scale made in both
predictions is the same, therefore leaving a ratio of predictions with IC to
predictions without IC independent of the choice of scale.

\section{Conclusion}
\label{sec:conclusion}

The quark and gluon distribution functions of the proton encode the
color-confining dynamics of QCD\@. These probability distributions are directly
related to the frame-independent light-front wavefunctions, the eigensolutions
of the QCD Hamiltonian. Because of QCD factorization, the PDFs allow the
computation of the cross sections for virtually all high energy collision
processes studied at the LHC\@. However, the heavy quark contribution to the
proton PDFs is still a major uncertainty. As we have discussed, QCD predicts
two sources of heavy quarks to the constituent structure of the light
hadrons~--~the standard small-x extrinsic contribution at from gluon splitting
$g \to Q \bar Q$~--~plus the intrinsic contribution at large $x$ which arises
from diagrams where the $Q \bar{Q}$ pair is connected by two or more gluons to
the valence quarks. The maximum configuration of such intrinsic contributions
occur at minimal off-shellness; i.e., when all of the quarks in the
$|uudQ\bar{Q}\rangle$ Fock state are at rest in the proton's rest frame,
corresponding to equal rapidity $y_i$ in the moving proton, and thus where the
momentum fraction carried by each quark is proportional to its transverse mass:
$ x_i \propto \sqrt{m_{\mathrm{T} i}^2 + {\vec k}_{\mathrm{T} i}^2 }$.

The hypothesis of \emph{intrinsic} quark components in the proton at high $x$
was originally suggested in~\cite{Brodsky:1980pb} was motivated possible
explanation of the large cross-section for the forward open charm production in
$pp$ collision at ISR energies~\cite{Drijard:1978gv,Giboni:1979rm,
Lockman:1979aj,Drijard:1979vd}. However, the accuracy of the experimental data
on the open charm production at large $x$ does not provide precise constraints
on the intrinsic heavy quark probability. We have discussed a number of
experiments such as as deep inelastic scattering measurements of $c(c, Q)$ at
the EMC and HERA\@; soft processes of the open charm or strangeness production
and even the Higgs boson production in $pp$ collisions at LHC\@. We have also
shown that the inclusive production of open charm or strangeness at high
transverse momentum can serve as a tool for the search for the heavy quark
$Q\bar{Q}$ Fock states in the nucleon. The IC or IS signal can be visible in
the \pt-spectrum of $D$-mesons or $K$-mesons produced, for example, in $pp$
collisions at their large rapidities $y$ and transverse momenta \pt. It can be
verified at CERN experiments such as the LHC and NA61.

The semi-inclusive production of prompt photons or gauge vector bosons in
association with heavy flavor jets $c$ or $b$ provides an ideal method for
verifying the IC contribution in the proton PDF\@. The increase of \pt\
spectrum of these hadrons or jets produced at large \pt\ and the forward
rapidity region of ATLAS or CMS ($1.5 < |y| < 2.4$) due to the IC enhancement
in the PDF is predicted. We also have found observables very sensitive to the
non-zero intrinsic charm contribution to the proton density.  They are the
$\gamma/Z + b$ production.  These ratios should be decreasing in the absence of
the IC contribution to the PDF and they should be flat or increasing if the IC
is included, when \pt\ grows. This prediction, can also verify the IC
hypothesis at LHC\@.

We argued that the ratio of the cross-sections $\gamma/Z + c$ and $\gamma/Z +
b$ integrated over ${p_\mathrm{T} > p_\mathrm{T}^{\min}}$ with
${p_\mathrm{T}^{\min} \geq 100}$~\gev\ can be used to determine the IC
probability $w$ from the future LHC data because we calculated it as a function
of $w$. The advantage of the proposed ratios is that the theoretical
uncertainties are very small, while the uncertainties for the \pt-spectra of
photons or $Z$ bosons produced in association with the $c$ or $b$ jets are
large. Therefore, the search for the IC signal by analyzing the ratio
$\sigma(\gamma/Z + c) / \sigma(\gamma/Z + b)$ can be more promising.

We have also shown that because of dominance of gluon-splitting processes, the
production of $W$-bosons accompanied by heavy flavor jets is not directly
sensitive to intrinsic heavy quarks. However, we can take advantage of this
fact to propose a promising new method which reduces the expected systematic
uncertainties by comparing the differential cross-section for $W$ plus heavy
flavor jets to that in $Z + Q$ events. The ratio of the leading heavy flavor
spectra in inclusive heavy flavor $Z + Q$ to $W + Q$ events can thus be used to
determine the intrinsic heavy quark contribution to the proton PDF\@. Such
measurements can already be made with available data from ATLAS and CMS\@.

We have also discussed the fact that the existence of intrinsic heavy charm and
bottom quarks in the proton wavefunction implies that the Higgs boson will be
produced at high momentum fractions $x_\mathrm{F} > 0.8$ in $pp \to H X$
collisions at the LHC, via the same mechanism that produces quarkonium states
such as $p p \to J/\psi X$ at high \xf. In addition, we have noted that fixed
target experiments at the LHC such as the AFTER facility and the SMOG nuclear
target at LHCb can materialize the intrinsic heavy quark Fock states in a novel
way. Since the momentum distributions of the intrinsic heavy quarks in the
$|uud Q \bar{Q}\rangle$ Fock state are maximal when all of the constituents
have the same rapidity, the collision of the LHC proton beam with nucleons in
the nuclear target will lead to the production of heavy hadrons such as the
$\Lambda_\mathrm{b}$ and exotic heavy quark states, such as tetraquarks and
pentaquarks,  at \emph{small rapidities} relative to the rapidity of the
target. Each of these novel processes will illuminate one of the most
interesting features of QCD bound state dynamics~--~intrinsic heavy quarks.

%
%
%
%
%
%
%

\section{Acknowledgements}

We thank A.~Glazov for extremely helpful discussions and recommendations for
the predictions on the search for the possible intrinsic heavy flavor
components in $pp$ collisions at high energies. We are also grateful to
P-H.~Beauchmin, T.~Dado, M.~Demichev, N.~Husseynov, A.V.~Lipatov, T.~Stavreva,
Yu.~Stepanneko and M.~Stockton for very productive collaboration, H.~Jung,
R.~Keys, B.Z.~Kopeliovich, A.~Likhoded, S.~Prince, S.~Rostami, L.~Rottoli,
O.V.~Teryaev and M.~Williams for very helpful discussions. This research was
supported in part by the Department of Energy contract
DE{}--{}AC02{}--{}76SF00515 SLAC-PUB-16844.

\section{References}
\bibliography{bibfile}

\begin{thebibliography}{104}
\expandafter\ifx\csname natexlab\endcsname\relax\def\natexlab#1{#1}\fi
\providecommand{\url}[1]{\texttt{#1}}
\providecommand{\href}[2]{#2}
\providecommand{\path}[1]{#1}
\providecommand{\DOIprefix}{doi:}
\providecommand{\ArXivprefix}{arXiv:}
\providecommand{\URLprefix}{URL: }
\providecommand{\Pubmedprefix}{pmid:}
\providecommand{\doi}[1]{\href{http://dx.doi.org/#1}{\path{#1}}}
\providecommand{\Pubmed}[1]{\href{pmid:#1}{\path{#1}}}
\providecommand{\bibinfo}[2]{#2}
\ifx\xfnm\relax \def\xfnm[#1]{\unskip,\space#1}\fi
\bibitem[{Gribov and Lipatov(1972)}]{Gribov:1972ri}
\bibinfo{author}{V.~N. Gribov}, \bibinfo{author}{L.~N. Lipatov},
  \bibinfo{journal}{Sov. J. Nucl. Phys.} \bibinfo{volume}{15}
  (\bibinfo{year}{1972}) \bibinfo{pages}{438--450}. \bibinfo{note}{[Yad.
  Fiz.15,781(1972)]}.
\bibitem[{Altarelli and Parisi(1977)}]{Altarelli:1977zs}
\bibinfo{author}{G.~Altarelli}, \bibinfo{author}{G.~Parisi},
  \bibinfo{journal}{Nucl. Phys.} \bibinfo{volume}{B126} (\bibinfo{year}{1977})
  \bibinfo{pages}{298}. \DOIprefix\doi{10.1016/0550-3213(77)90384-4}.
\bibitem[{Dokshitzer(1977)}]{Dokshitzer:1977sg}
\bibinfo{author}{Y.~L. Dokshitzer}, \bibinfo{journal}{Sov. Phys. JETP}
  \bibinfo{volume}{46} (\bibinfo{year}{1977}) \bibinfo{pages}{641--653}.
  \bibinfo{note}{[Zh. Eksp. Teor. Fiz.73,1216(1977)]}.
\bibitem[{Brodsky et~al.(1984)Brodsky, Collins, Ellis, Gunion, and
  Mueller}]{Brodsky:1984nx}
\bibinfo{author}{S.~J. Brodsky}, \bibinfo{author}{J.~C. Collins},
  \bibinfo{author}{S.~D. Ellis}, \bibinfo{author}{J.~F. Gunion},
  \bibinfo{author}{A.~H. Mueller}, in: \bibinfo{booktitle}{{ELECTROWEAK
  SYMMETRY BREAKING. PROCEEDINGS, WORKSHOP, BERKELEY, USA, JUNE 3-22, 1984}}.
  \URLprefix
  \url{http://www-public.slac.stanford.edu/sciDoc/docMeta.aspx?slacPubNumber=SLAC-PUB-15471}.
\bibitem[{Franz et~al.(2000)Franz, Polyakov, and Goeke}]{Franz:2000ee}
\bibinfo{author}{M.~Franz}, \bibinfo{author}{M.~V. Polyakov},
  \bibinfo{author}{K.~Goeke}, \bibinfo{journal}{Phys. Rev.}
  \bibinfo{volume}{D62} (\bibinfo{year}{2000}) \bibinfo{pages}{074024}.
  \DOIprefix\doi{10.1103/PhysRevD.62.074024}.
  \href{http://arxiv.org/abs/hep-ph/0002240}{\tt arXiv:hep-ph/0002240}.
\bibitem[{Brodsky et~al.(1980)Brodsky, Hoyer, Peterson, and
  Sakai}]{Brodsky:1980pb}
\bibinfo{author}{S.~J. Brodsky}, \bibinfo{author}{P.~Hoyer},
  \bibinfo{author}{C.~Peterson}, \bibinfo{author}{N.~Sakai},
  \bibinfo{journal}{Phys. Lett.} \bibinfo{volume}{B93} (\bibinfo{year}{1980})
  \bibinfo{pages}{451--455}. \DOIprefix\doi{10.1016/0370-2693(80)90364-0}.
\bibitem[{Drijard et~al.(1979)}]{Drijard:1978gv}
\bibinfo{author}{D.~Drijard}, et~al. (\bibinfo{collaboration}{CERN-College de
  France-Heidelberg-Karlsruhe}), \bibinfo{journal}{Phys. Lett.}
  \bibinfo{volume}{B81} (\bibinfo{year}{1979}) \bibinfo{pages}{250--254}.
  \DOIprefix\doi{10.1016/0370-2693(79)90535-5}.
\bibitem[{Giboni et~al.(1979)}]{Giboni:1979rm}
\bibinfo{author}{K.~L. Giboni}, et~al., \bibinfo{journal}{Phys. Lett.}
  \bibinfo{volume}{B85} (\bibinfo{year}{1979}) \bibinfo{pages}{437--442}.
  \DOIprefix\doi{10.1016/0370-2693(79)91291-7}.
\bibitem[{Lockman et~al.(1979)Lockman, Meyer, Rander, Schlein, Webb, Erhan, and
  Zsembery}]{Lockman:1979aj}
\bibinfo{author}{W.~S. Lockman}, \bibinfo{author}{T.~Meyer},
  \bibinfo{author}{J.~Rander}, \bibinfo{author}{P.~Schlein},
  \bibinfo{author}{R.~Webb}, \bibinfo{author}{S.~Erhan},
  \bibinfo{author}{J.~Zsembery}, \bibinfo{journal}{Phys. Lett.}
  \bibinfo{volume}{B85} (\bibinfo{year}{1979}) \bibinfo{pages}{443--446}.
  \DOIprefix\doi{10.1016/0370-2693(79)91292-9}.
\bibitem[{Drijard et~al.(1979)}]{Drijard:1979vd}
\bibinfo{author}{D.~Drijard}, et~al. (\bibinfo{collaboration}{ACCDHW}),
  \bibinfo{journal}{Phys. Lett.} \bibinfo{volume}{B85} (\bibinfo{year}{1979})
  \bibinfo{pages}{452--457}. \DOIprefix\doi{10.1016/0370-2693(79)91294-2}.
\bibitem[{Laha and Brodsky(2016)}]{Laha:2016dri}
\bibinfo{author}{R.~Laha}, \bibinfo{author}{S.~J. Brodsky}
  (\bibinfo{year}{2016}). \href{http://arxiv.org/abs/1607.08240}{\tt
  arXiv:1607.08240}.
\bibitem[{Aktas et~al.(2006)}]{Aktas:2005iw}
\bibinfo{author}{A.~Aktas}, et~al. (\bibinfo{collaboration}{H1}),
  \bibinfo{journal}{Eur. Phys. J.} \bibinfo{volume}{C45} (\bibinfo{year}{2006})
  \bibinfo{pages}{23--33}. \DOIprefix\doi{10.1140/epjc/s2005-02415-6}.
  \href{http://arxiv.org/abs/hep-ex/0507081}{\tt arXiv:hep-ex/0507081}.
\bibitem[{Brodsky et~al.(1981)Brodsky, Peterson, and Sakai}]{Brodsky:1981se}
\bibinfo{author}{S.~J. Brodsky}, \bibinfo{author}{C.~Peterson},
  \bibinfo{author}{N.~Sakai}, \bibinfo{journal}{Phys. Rev.}
  \bibinfo{volume}{D23} (\bibinfo{year}{1981}) \bibinfo{pages}{2745}.
  \DOIprefix\doi{10.1103/PhysRevD.23.2745}.
\bibitem[{Brodsky(2005)}]{Brodsky:2004er}
\bibinfo{author}{S.~J. Brodsky}, \bibinfo{journal}{Few Body Syst.}
  \bibinfo{volume}{36} (\bibinfo{year}{2005}) \bibinfo{pages}{35--52}.
  \DOIprefix\doi{10.1007/s00601-004-0077-8}.
  \href{http://arxiv.org/abs/hep-ph/0411056}{\tt arXiv:hep-ph/0411056}.
\bibitem[{Brodsky et~al.(2015)Brodsky, Kusina, Lyonnet, Schienbein,
  Spiesberger, and Vogt}]{Brodsky:2015fna}
\bibinfo{author}{S.~J. Brodsky}, \bibinfo{author}{A.~Kusina},
  \bibinfo{author}{F.~Lyonnet}, \bibinfo{author}{I.~Schienbein},
  \bibinfo{author}{H.~Spiesberger}, \bibinfo{author}{R.~Vogt},
  \bibinfo{journal}{Adv. High Energy Phys.} \bibinfo{volume}{2015}
  (\bibinfo{year}{2015}) \bibinfo{pages}{231547}.
  \DOIprefix\doi{10.1155/2015/231547}.
  \href{http://arxiv.org/abs/1504.06287}{\tt arXiv:1504.06287}.
\bibitem[{Pumplin(2006)}]{Pumplin:2005yf}
\bibinfo{author}{J.~Pumplin}, \bibinfo{journal}{Phys. Rev.}
  \bibinfo{volume}{D73} (\bibinfo{year}{2006}) \bibinfo{pages}{114015}.
  \DOIprefix\doi{10.1103/PhysRevD.73.114015}.
  \href{http://arxiv.org/abs/hep-ph/0508184}{\tt arXiv:hep-ph/0508184}.
\bibitem[{Aubert et~al.(1983)}]{Aubert:1982tt}
\bibinfo{author}{J.~J. Aubert}, et~al. (\bibinfo{collaboration}{European
  Muon}), \bibinfo{journal}{Nucl. Phys.} \bibinfo{volume}{B213}
  (\bibinfo{year}{1983}) \bibinfo{pages}{31--64}.
  \DOIprefix\doi{10.1016/0550-3213(83)90174-8}.
\bibitem[{Ball et~al.(2016)Ball, Bertone, Bonvini, Carrazza, Forte, Guffanti,
  Hartland, Rojo, and Rottoli}]{Ball:2016neh}
\bibinfo{author}{R.~D. Ball}, \bibinfo{author}{V.~Bertone},
  \bibinfo{author}{M.~Bonvini}, \bibinfo{author}{S.~Carrazza},
  \bibinfo{author}{S.~Forte}, \bibinfo{author}{A.~Guffanti},
  \bibinfo{author}{N.~P. Hartland}, \bibinfo{author}{J.~Rojo},
  \bibinfo{author}{L.~Rottoli} (\bibinfo{collaboration}{NNPDF})
  (\bibinfo{year}{2016}). \href{http://arxiv.org/abs/1605.06515}{\tt
  arXiv:1605.06515}.
\bibitem[{Ball et~al.(2015)}]{Ball:2014uwa}
\bibinfo{author}{R.~D. Ball}, et~al. (\bibinfo{collaboration}{NNPDF}),
  \bibinfo{journal}{JHEP} \bibinfo{volume}{04} (\bibinfo{year}{2015})
  \bibinfo{pages}{040}. \DOIprefix\doi{10.1007/JHEP04(2015)040}.
  \href{http://arxiv.org/abs/1410.8849}{\tt arXiv:1410.8849}.
\bibitem[{Dulat et~al.(2016)Dulat, Hou, Gao, Guzzi, Huston, Nadolsky, Pumplin,
  Schmidt, Stump, and Yuan}]{Dulat:2015mca}
\bibinfo{author}{S.~Dulat}, \bibinfo{author}{T.-J. Hou},
  \bibinfo{author}{J.~Gao}, \bibinfo{author}{M.~Guzzi},
  \bibinfo{author}{J.~Huston}, \bibinfo{author}{P.~Nadolsky},
  \bibinfo{author}{J.~Pumplin}, \bibinfo{author}{C.~Schmidt},
  \bibinfo{author}{D.~Stump}, \bibinfo{author}{C.~P. Yuan},
  \bibinfo{journal}{Phys. Rev.} \bibinfo{volume}{D93} (\bibinfo{year}{2016})
  \bibinfo{pages}{033006}. \DOIprefix\doi{10.1103/PhysRevD.93.033006}.
  \href{http://arxiv.org/abs/1506.07443}{\tt arXiv:1506.07443}.
\bibitem[{Adamovich et~al.(1993)}]{Adamovich:1993kc}
\bibinfo{author}{M.~Adamovich}, et~al. (\bibinfo{collaboration}{WA82}),
  \bibinfo{journal}{Phys. Lett.} \bibinfo{volume}{B305} (\bibinfo{year}{1993})
  \bibinfo{pages}{402--406}. \DOIprefix\doi{10.1016/0370-2693(93)91074-W}.
\bibitem[{Alves et~al.(1994)}]{Alves:1993mp}
\bibinfo{author}{G.~A. Alves}, et~al. (\bibinfo{collaboration}{E769}),
  \bibinfo{journal}{Phys. Rev. Lett.} \bibinfo{volume}{72}
  (\bibinfo{year}{1994}) \bibinfo{pages}{812--815}.
  \DOIprefix\doi{10.1103/PhysRevLett.72.812}, \bibinfo{note}{[Erratum: Phys.
  Rev. Lett.72,1946(1994)]}.
\bibitem[{Aitala et~al.(1996)}]{Aitala:1996hf}
\bibinfo{author}{E.~M. Aitala}, et~al. (\bibinfo{collaboration}{E791}),
  \bibinfo{journal}{Phys. Lett.} \bibinfo{volume}{B371} (\bibinfo{year}{1996})
  \bibinfo{pages}{157--162}. \DOIprefix\doi{10.1016/0370-2693(96)00093-7}.
  \href{http://arxiv.org/abs/hep-ex/9601001}{\tt arXiv:hep-ex/9601001}.
\bibitem[{Kaidalov(1982)}]{Kaidalov:1982xg}
\bibinfo{author}{A.~B. Kaidalov}, \bibinfo{journal}{Phys. Lett.}
  \bibinfo{volume}{B116} (\bibinfo{year}{1982}) \bibinfo{pages}{459--463}.
  \DOIprefix\doi{10.1016/0370-2693(82)90168-X}.
\bibitem[{Capella et~al.(1994)Capella, Sukhatme, Tan, and Tran
  Thanh~Van}]{Capella:1992yb}
\bibinfo{author}{A.~Capella}, \bibinfo{author}{U.~Sukhatme},
  \bibinfo{author}{C.-I. Tan}, \bibinfo{author}{J.~Tran Thanh~Van},
  \bibinfo{journal}{Phys. Rept.} \bibinfo{volume}{236} (\bibinfo{year}{1994})
  \bibinfo{pages}{225--329}. \DOIprefix\doi{10.1016/0370-1573(94)90064-7}.
\bibitem[{Kaidalov and Piskunova(1986)}]{Kaidalov:1985jg}
\bibinfo{author}{A.~B. Kaidalov}, \bibinfo{author}{O.~I. Piskunova},
  \bibinfo{journal}{Z. Phys.} \bibinfo{volume}{C30} (\bibinfo{year}{1986})
  \bibinfo{pages}{145}. \DOIprefix\doi{10.1007/BF01560688}.
\bibitem[{Shabelski(1995)}]{Shabelski:1995ei}
\bibinfo{author}{{\relax Yu}.~M. Shabelski}, \bibinfo{journal}{Surveys High
  Energ. Phys.} \bibinfo{volume}{9} (\bibinfo{year}{1995})
  \bibinfo{pages}{1--88}. \DOIprefix\doi{10.1080/01422419508225681}.
\bibitem[{Lykasov et~al.(1999)Lykasov, Arakelian, and
  Sergeenko}]{Lykasov:1999fj}
\bibinfo{author}{G.~I. Lykasov}, \bibinfo{author}{G.~H. Arakelian},
  \bibinfo{author}{M.~N. Sergeenko}, \bibinfo{journal}{Phys. Part. Nucl.}
  \bibinfo{volume}{30} (\bibinfo{year}{1999}) \bibinfo{pages}{343--368}.
  \DOIprefix\doi{10.1134/1.953111}, \bibinfo{note}{[Fiz. Elem. Chast. Atom.
  Yadra30,817(1999)]}.
\bibitem[{Barlag et~al.(1990)}]{Barlag:1990hg}
\bibinfo{author}{S.~Barlag}, et~al. (\bibinfo{collaboration}{ACCMOR}),
  \bibinfo{journal}{Phys. Lett.} \bibinfo{volume}{B247} (\bibinfo{year}{1990})
  \bibinfo{pages}{113--120}. \DOIprefix\doi{10.1016/0370-2693(90)91058-J}.
\bibitem[{Bari et~al.(1991)}]{Bari:1991in}
\bibinfo{author}{G.~Bari}, et~al., \bibinfo{journal}{Nuovo Cim.}
  \bibinfo{volume}{A104} (\bibinfo{year}{1991}) \bibinfo{pages}{571--599}.
  \DOIprefix\doi{10.1007/BF02813593}.
\bibitem[{Abazov et~al.(2009)}]{Abazov:2009de}
\bibinfo{author}{V.~M. Abazov}, et~al. (\bibinfo{collaboration}{D0}),
  \bibinfo{journal}{Phys. Rev. Lett.} \bibinfo{volume}{102}
  (\bibinfo{year}{2009}) \bibinfo{pages}{192002}.
  \DOIprefix\doi{10.1103/PhysRevLett.102.192002}.
  \href{http://arxiv.org/abs/0901.0739}{\tt arXiv:0901.0739}.
\bibitem[{Abazov et~al.(2012)}]{Abazov:2012ea}
\bibinfo{author}{V.~M. Abazov}, et~al. (\bibinfo{collaboration}{D0}),
  \bibinfo{journal}{Phys. Lett.} \bibinfo{volume}{B714} (\bibinfo{year}{2012})
  \bibinfo{pages}{32--39}. \DOIprefix\doi{10.1016/j.physletb.2012.06.056}.
  \href{http://arxiv.org/abs/1203.5865}{\tt arXiv:1203.5865}.
\bibitem[{Abazov et~al.(2013)}]{D0:2012gw}
\bibinfo{author}{V.~M. Abazov}, et~al. (\bibinfo{collaboration}{D0}),
  \bibinfo{journal}{Phys. Lett.} \bibinfo{volume}{B719} (\bibinfo{year}{2013})
  \bibinfo{pages}{354--361}. \DOIprefix\doi{10.1016/j.physletb.2013.01.033}.
  \href{http://arxiv.org/abs/1210.5033}{\tt arXiv:1210.5033}.
\bibitem[{Aaltonen et~al.(2010)}]{Aaltonen:2009wc}
\bibinfo{author}{T.~Aaltonen}, et~al. (\bibinfo{collaboration}{CDF}),
  \bibinfo{journal}{Phys. Rev.} \bibinfo{volume}{D81} (\bibinfo{year}{2010})
  \bibinfo{pages}{052006}. \DOIprefix\doi{10.1103/PhysRevD.81.052006}.
  \href{http://arxiv.org/abs/0912.3453}{\tt arXiv:0912.3453}.
\bibitem[{Aaltonen et~al.(2013)}]{Aaltonen:2013ama}
\bibinfo{author}{T.~Aaltonen}, et~al. (\bibinfo{collaboration}{CDF}),
  \bibinfo{journal}{Phys. Rev. Lett.} \bibinfo{volume}{111}
  (\bibinfo{year}{2013}) \bibinfo{pages}{042003}.
  \DOIprefix\doi{10.1103/PhysRevLett.111.042003}.
  \href{http://arxiv.org/abs/1303.6136}{\tt arXiv:1303.6136}.
\bibitem[{Polyakov et~al.(1999)Polyakov, Schafer, and
  Teryaev}]{Polyakov:1998rb}
\bibinfo{author}{M.~V. Polyakov}, \bibinfo{author}{A.~Schafer},
  \bibinfo{author}{O.~V. Teryaev}, \bibinfo{journal}{Phys. Rev.}
  \bibinfo{volume}{D60} (\bibinfo{year}{1999}) \bibinfo{pages}{051502}.
  \DOIprefix\doi{10.1103/PhysRevD.60.051502}.
  \href{http://arxiv.org/abs/hep-ph/9812393}{\tt arXiv:hep-ph/9812393}.
\bibitem[{Stavreva and Owens(2009)}]{Stavreva:2009vi}
\bibinfo{author}{T.~P. Stavreva}, \bibinfo{author}{J.~F. Owens},
  \bibinfo{journal}{Phys. Rev.} \bibinfo{volume}{D79} (\bibinfo{year}{2009})
  \bibinfo{pages}{054017}. \DOIprefix\doi{10.1103/PhysRevD.79.054017}.
  \href{http://arxiv.org/abs/0901.3791}{\tt arXiv:0901.3791}.
\bibitem[{Bednyakov et~al.(2014)Bednyakov, Demichev, Lykasov, Stavreva, and
  Stockton}]{Bednyakov:2013zta}
\bibinfo{author}{V.~A. Bednyakov}, \bibinfo{author}{M.~A. Demichev},
  \bibinfo{author}{G.~I. Lykasov}, \bibinfo{author}{T.~Stavreva},
  \bibinfo{author}{M.~Stockton}, \bibinfo{journal}{Phys. Lett.}
  \bibinfo{volume}{B728} (\bibinfo{year}{2014}) \bibinfo{pages}{602--606}.
  \DOIprefix\doi{10.1016/j.physletb.2013.12.031}.
  \href{http://arxiv.org/abs/1305.3548}{\tt arXiv:1305.3548}.
\bibitem[{Beauchemin et~al.(2015)Beauchemin, Bednyakov, Lykasov, and
  Stepanenko}]{Beauchemin:2014rya}
\bibinfo{author}{P.-H. Beauchemin}, \bibinfo{author}{V.~A. Bednyakov},
  \bibinfo{author}{G.~I. Lykasov}, \bibinfo{author}{{\relax Yu}.~{\relax Yu}.
  Stepanenko}, \bibinfo{journal}{Phys. Rev.} \bibinfo{volume}{D92}
  (\bibinfo{year}{2015}) \bibinfo{pages}{034014}.
  \DOIprefix\doi{10.1103/PhysRevD.92.034014}.
  \href{http://arxiv.org/abs/1410.2616}{\tt arXiv:1410.2616}.
\bibitem[{Lipatov et~al.(2016)Lipatov, Lykasov, Stepanenko, and
  Bednyakov}]{Lipatov:2016feu}
\bibinfo{author}{A.~V. Lipatov}, \bibinfo{author}{G.~I. Lykasov},
  \bibinfo{author}{{\relax Yu}.~{\relax Yu}. Stepanenko},
  \bibinfo{author}{V.~A. Bednyakov}, \bibinfo{journal}{Phys. Rev.}
  \bibinfo{volume}{D94} (\bibinfo{year}{2016}) \bibinfo{pages}{053011}.
  \DOIprefix\doi{10.1103/PhysRevD.94.053011}.
  \href{http://arxiv.org/abs/1606.04882}{\tt arXiv:1606.04882}.
\bibitem[{Brodsky et~al.(2006)Brodsky, Kopeliovich, Schmidt, and
  Soffer}]{Brodsky:2006wb}
\bibinfo{author}{S.~J. Brodsky}, \bibinfo{author}{B.~Kopeliovich},
  \bibinfo{author}{I.~Schmidt}, \bibinfo{author}{J.~Soffer},
  \bibinfo{journal}{Phys. Rev.} \bibinfo{volume}{D73} (\bibinfo{year}{2006})
  \bibinfo{pages}{113005}. \DOIprefix\doi{10.1103/PhysRevD.73.113005}.
  \href{http://arxiv.org/abs/hep-ph/0603238}{\tt arXiv:hep-ph/0603238}.
\bibitem[{Brodsky et~al.(2009)Brodsky, Goldhaber, Kopeliovich, and
  Schmidt}]{Brodsky:2007yz}
\bibinfo{author}{S.~J. Brodsky}, \bibinfo{author}{A.~S. Goldhaber},
  \bibinfo{author}{B.~Z. Kopeliovich}, \bibinfo{author}{I.~Schmidt},
  \bibinfo{journal}{Nucl. Phys.} \bibinfo{volume}{B807} (\bibinfo{year}{2009})
  \bibinfo{pages}{334--347}. \DOIprefix\doi{10.1016/j.nuclphysb.2008.09.014}.
  \href{http://arxiv.org/abs/0707.4658}{\tt arXiv:0707.4658}.
\bibitem[{Brodsky et~al.(2013)Brodsky, Fleuret, Hadjidakis, and
  Lansberg}]{Brodsky:2012vg}
\bibinfo{author}{S.~J. Brodsky}, \bibinfo{author}{F.~Fleuret},
  \bibinfo{author}{C.~Hadjidakis}, \bibinfo{author}{J.~P. Lansberg},
  \bibinfo{journal}{Phys. Rept.} \bibinfo{volume}{522} (\bibinfo{year}{2013})
  \bibinfo{pages}{239--255}. \DOIprefix\doi{10.1016/j.physrep.2012.10.001}.
  \href{http://arxiv.org/abs/1202.6585}{\tt arXiv:1202.6585}.
\bibitem[{Negele et~al.(2004)}]{Negele:2004iu}
\bibinfo{author}{J.~W. Negele}, et~al., \bibinfo{journal}{Nucl. Phys. Proc.
  Suppl.} \bibinfo{volume}{128} (\bibinfo{year}{2004})
  \bibinfo{pages}{170--178}. \DOIprefix\doi{10.1016/S0920-5632(03)02474-5}.
  \href{http://arxiv.org/abs/hep-lat/0404005}{\tt arXiv:hep-lat/0404005},
  \bibinfo{note}{[,170(2004)]}.
\bibitem[{Schroers(2005)}]{Schroers:2005rm}
\bibinfo{author}{W.~Schroers}, \bibinfo{journal}{Nucl. Phys.}
  \bibinfo{volume}{A755} (\bibinfo{year}{2005}) \bibinfo{pages}{333--336}.
  \DOIprefix\doi{10.1016/j.nuclphysa.2005.03.034}.
  \href{http://arxiv.org/abs/hep-ph/0501156}{\tt arXiv:hep-ph/0501156}.
\bibitem[{Pumplin et~al.(2002)Pumplin, Stump, Huston, Lai, Nadolsky, and
  Tung}]{Pumplin:2002vw}
\bibinfo{author}{J.~Pumplin}, \bibinfo{author}{D.~R. Stump},
  \bibinfo{author}{J.~Huston}, \bibinfo{author}{H.~L. Lai},
  \bibinfo{author}{P.~M. Nadolsky}, \bibinfo{author}{W.~K. Tung},
  \bibinfo{journal}{JHEP} \bibinfo{volume}{07} (\bibinfo{year}{2002})
  \bibinfo{pages}{012}. \DOIprefix\doi{10.1088/1126-6708/2002/07/012}.
  \href{http://arxiv.org/abs/hep-ph/0201195}{\tt arXiv:hep-ph/0201195}.
\bibitem[{Stump et~al.(2003)Stump, Huston, Pumplin, Tung, Lai, Kuhlmann, and
  Owens}]{Stump:2003yu}
\bibinfo{author}{D.~Stump}, \bibinfo{author}{J.~Huston},
  \bibinfo{author}{J.~Pumplin}, \bibinfo{author}{W.-K. Tung},
  \bibinfo{author}{H.~L. Lai}, \bibinfo{author}{S.~Kuhlmann},
  \bibinfo{author}{J.~F. Owens}, \bibinfo{journal}{JHEP} \bibinfo{volume}{10}
  (\bibinfo{year}{2003}) \bibinfo{pages}{046}.
  \DOIprefix\doi{10.1088/1126-6708/2003/10/046}.
  \href{http://arxiv.org/abs/hep-ph/0303013}{\tt arXiv:hep-ph/0303013}.
\bibitem[{Thorne et~al.(2004)Thorne, Martin, Stirling, and
  Roberts}]{Thorne:2004ci}
\bibinfo{author}{R.~S. Thorne}, \bibinfo{author}{A.~D. Martin},
  \bibinfo{author}{W.~J. Stirling}, \bibinfo{author}{R.~G. Roberts}, in:
  \bibinfo{booktitle}{{Deep inelastic scattering. Proceedings, 12th
  International Workshop, DIS 2004, Strbske Pleso, Slovakia, April 14-18, 2004.
  Vol. 1 + 2}}, pp. \bibinfo{pages}{427--432}. \URLprefix
  \url{http://www.saske.sk/dis04/proceedings/A/thorne.ps.gz}.
  \href{http://arxiv.org/abs/hep-ph/0407311}{\tt arXiv:hep-ph/0407311}.
\bibitem[{Brodsky and Ma(1996)}]{Brodsky:1996hc}
\bibinfo{author}{S.~J. Brodsky}, \bibinfo{author}{B.-Q. Ma},
  \bibinfo{journal}{Phys. Lett.} \bibinfo{volume}{B381} (\bibinfo{year}{1996})
  \bibinfo{pages}{317--324}. \DOIprefix\doi{10.1016/0370-2693(96)00597-7}.
  \href{http://arxiv.org/abs/hep-ph/9604393}{\tt arXiv:hep-ph/9604393}.
\bibitem[{Pumplin et~al.(2007)Pumplin, Lai, and Tung}]{Pumplin:2007wg}
\bibinfo{author}{J.~Pumplin}, \bibinfo{author}{H.~L. Lai},
  \bibinfo{author}{W.~K. Tung}, \bibinfo{journal}{Phys. Rev.}
  \bibinfo{volume}{D75} (\bibinfo{year}{2007}) \bibinfo{pages}{054029}.
  \DOIprefix\doi{10.1103/PhysRevD.75.054029}.
  \href{http://arxiv.org/abs/hep-ph/0701220}{\tt arXiv:hep-ph/0701220}.
\bibitem[{Donoghue and Golowich(1977)}]{Donoghue:1977qp}
\bibinfo{author}{J.~F. Donoghue}, \bibinfo{author}{E.~Golowich},
  \bibinfo{journal}{Phys. Rev.} \bibinfo{volume}{D15} (\bibinfo{year}{1977})
  \bibinfo{pages}{3421}. \DOIprefix\doi{10.1103/PhysRevD.15.3421}.
\bibitem[{Paiva et~al.(1998)Paiva, Nielsen, Navarra, Duraes, and
  Barz}]{Paiva:1996dd}
\bibinfo{author}{S.~Paiva}, \bibinfo{author}{M.~Nielsen},
  \bibinfo{author}{F.~S. Navarra}, \bibinfo{author}{F.~O. Duraes},
  \bibinfo{author}{L.~L. Barz}, \bibinfo{journal}{Mod. Phys. Lett.}
  \bibinfo{volume}{A13} (\bibinfo{year}{1998}) \bibinfo{pages}{2715--2724}.
  \DOIprefix\doi{10.1142/S0217732398002886}.
  \href{http://arxiv.org/abs/hep-ph/9610310}{\tt arXiv:hep-ph/9610310}.
\bibitem[{Melnitchouk and Thomas(1997)}]{Melnitchouk:1997ig}
\bibinfo{author}{W.~Melnitchouk}, \bibinfo{author}{A.~W. Thomas},
  \bibinfo{journal}{Phys. Lett.} \bibinfo{volume}{B414} (\bibinfo{year}{1997})
  \bibinfo{pages}{134--139}. \DOIprefix\doi{10.1016/S0370-2693(97)01150-7}.
  \href{http://arxiv.org/abs/hep-ph/9707387}{\tt arXiv:hep-ph/9707387}.
\bibitem[{Steffens et~al.(1999)Steffens, Melnitchouk, and
  Thomas}]{Steffens:1999hx}
\bibinfo{author}{F.~M. Steffens}, \bibinfo{author}{W.~Melnitchouk},
  \bibinfo{author}{A.~W. Thomas}, \bibinfo{journal}{Eur. Phys. J.}
  \bibinfo{volume}{C11} (\bibinfo{year}{1999}) \bibinfo{pages}{673--683}.
  \DOIprefix\doi{10.1007/s100529900189, 10.1007/s100520050663}.
  \href{http://arxiv.org/abs/hep-ph/9903441}{\tt arXiv:hep-ph/9903441}.
\bibitem[{Hobbs et~al.(2014)Hobbs, Londergan, and Melnitchouk}]{Hobbs:2013bia}
\bibinfo{author}{T.~J. Hobbs}, \bibinfo{author}{J.~T. Londergan},
  \bibinfo{author}{W.~Melnitchouk}, \bibinfo{journal}{Phys. Rev.}
  \bibinfo{volume}{D89} (\bibinfo{year}{2014}) \bibinfo{pages}{074008}.
  \DOIprefix\doi{10.1103/PhysRevD.89.074008}.
  \href{http://arxiv.org/abs/1311.1578}{\tt arXiv:1311.1578}.
\bibitem[{Nadolsky et~al.(2008)Nadolsky, Lai, Cao, Huston, Pumplin, Stump,
  Tung, and Yuan}]{Nadolsky:2008zw}
\bibinfo{author}{P.~M. Nadolsky}, \bibinfo{author}{H.-L. Lai},
  \bibinfo{author}{Q.-H. Cao}, \bibinfo{author}{J.~Huston},
  \bibinfo{author}{J.~Pumplin}, \bibinfo{author}{D.~Stump},
  \bibinfo{author}{W.-K. Tung}, \bibinfo{author}{C.~P. Yuan},
  \bibinfo{journal}{Phys. Rev.} \bibinfo{volume}{D78} (\bibinfo{year}{2008})
  \bibinfo{pages}{013004}. \DOIprefix\doi{10.1103/PhysRevD.78.013004}.
  \href{http://arxiv.org/abs/0802.0007}{\tt arXiv:0802.0007}.
\bibitem[{Brodsky and Gardner(2016)}]{Brodsky:2015uwa}
\bibinfo{author}{S.~J. Brodsky}, \bibinfo{author}{S.~Gardner},
  \bibinfo{journal}{Phys. Rev. Lett.} \bibinfo{volume}{116}
  (\bibinfo{year}{2016}) \bibinfo{pages}{019101}.
  \DOIprefix\doi{10.1103/PhysRevLett.116.019101}.
  \href{http://arxiv.org/abs/1504.00969}{\tt arXiv:1504.00969}.
\bibitem[{Ammar et~al.(1987)}]{Ammar:1986nk}
\bibinfo{author}{R.~Ammar}, et~al. (\bibinfo{collaboration}{LEBC-MPS}),
  \bibinfo{journal}{Phys. Lett.} \bibinfo{volume}{B183} (\bibinfo{year}{1987})
  \bibinfo{pages}{110}. \DOIprefix\doi{10.1016/0370-2693(87)91427-4},
  \bibinfo{note}{[Erratum: Phys. Lett.B192,478(1987)]}.
\bibitem[{Russ et~al.(1998)}]{Russ:1998rr}
\bibinfo{author}{J.~Russ}, et~al. (\bibinfo{collaboration}{SELEX}), in:
  \bibinfo{booktitle}{{High-energy physics. Proceedings, 29th International
  Conference, ICHEP'98, Vancouver, Canada, July 23-29, 1998. Vol. 1, 2}}, pp.
  \bibinfo{pages}{1259--1262}. \URLprefix
  \url{http://alice.cern.ch/format/showfull?sysnb=0300703}.
  \href{http://arxiv.org/abs/hep-ex/9812031}{\tt arXiv:hep-ex/9812031}.
\bibitem[{Brodsky(2015)}]{Brodsky:2014hia}
\bibinfo{author}{S.~J. Brodsky}, \bibinfo{journal}{Nucl. Part. Phys. Proc.}
  \bibinfo{volume}{258-259} (\bibinfo{year}{2015}) \bibinfo{pages}{23--30}.
  \DOIprefix\doi{10.1016/j.nuclphysbps.2015.01.007}.
  \href{http://arxiv.org/abs/1410.0404}{\tt arXiv:1410.0404}.
\bibitem[{Blümlein(2016)}]{Blumlein:2015qcn}
\bibinfo{author}{J.~Blümlein}, \bibinfo{journal}{Phys. Lett.}
  \bibinfo{volume}{B753} (\bibinfo{year}{2016}) \bibinfo{pages}{619--621}.
  \DOIprefix\doi{10.1016/j.physletb.2015.12.068}.
  \href{http://arxiv.org/abs/1511.00229}{\tt arXiv:1511.00229}.
\bibitem[{Catani et~al.(2004)Catani, de~Florian, Rodrigo, and
  Vogelsang}]{Catani:2004nc}
\bibinfo{author}{S.~Catani}, \bibinfo{author}{D.~de~Florian},
  \bibinfo{author}{G.~Rodrigo}, \bibinfo{author}{W.~Vogelsang},
  \bibinfo{journal}{Phys. Rev. Lett.} \bibinfo{volume}{93}
  (\bibinfo{year}{2004}) \bibinfo{pages}{152003}.
  \DOIprefix\doi{10.1103/PhysRevLett.93.152003}.
  \href{http://arxiv.org/abs/hep-ph/0404240}{\tt arXiv:hep-ph/0404240}.
\bibitem[{Garcia et~al.(2002)}]{Garcia:2001xj}
\bibinfo{author}{F.~G. Garcia}, et~al. (\bibinfo{collaboration}{SELEX}),
  \bibinfo{journal}{Phys. Lett.} \bibinfo{volume}{B528} (\bibinfo{year}{2002})
  \bibinfo{pages}{49--57}. \DOIprefix\doi{10.1016/S0370-2693(01)01484-8}.
  \href{http://arxiv.org/abs/hep-ex/0109017}{\tt arXiv:hep-ex/0109017}.
\bibitem[{Rostami et~al.(2016)Rostami, Khorramian, Aleedaneshvar, and
  Goharipour}]{Rostami:2015iva}
\bibinfo{author}{S.~Rostami}, \bibinfo{author}{A.~Khorramian},
  \bibinfo{author}{A.~Aleedaneshvar}, \bibinfo{author}{M.~Goharipour},
  \bibinfo{journal}{J. Phys.} \bibinfo{volume}{G43} (\bibinfo{year}{2016})
  \bibinfo{pages}{055001}. \DOIprefix\doi{10.1088/0954-3899/43/5/055001}.
  \href{http://arxiv.org/abs/1510.08421}{\tt arXiv:1510.08421}.
\bibitem[{Catani et~al.(1991)Catani, Ciafaloni, and Hautmann}]{Catani:1990eg}
\bibinfo{author}{S.~Catani}, \bibinfo{author}{M.~Ciafaloni},
  \bibinfo{author}{F.~Hautmann}, \bibinfo{journal}{Nucl. Phys.}
  \bibinfo{volume}{B366} (\bibinfo{year}{1991}) \bibinfo{pages}{135--188}.
  \DOIprefix\doi{10.1016/0550-3213(91)90055-3}.
\bibitem[{Lykasov et~al.(2012)Lykasov, Bednyakov, Pikelner, and
  Zimine}]{Lykasov:2012hf}
\bibinfo{author}{G.~I. Lykasov}, \bibinfo{author}{V.~A. Bednyakov},
  \bibinfo{author}{A.~F. Pikelner}, \bibinfo{author}{N.~I. Zimine},
  \bibinfo{journal}{Europhys. Lett.} \bibinfo{volume}{99}
  (\bibinfo{year}{2012}) \bibinfo{pages}{21002}.
  \DOIprefix\doi{10.1209/0295-5075/99/21002}.
  \href{http://arxiv.org/abs/1205.1131}{\tt arXiv:1205.1131}.
\bibitem[{Goncalves and Navarra(2010)}]{Goncalves:2008sw}
\bibinfo{author}{V.~P. Goncalves}, \bibinfo{author}{F.~S. Navarra},
  \bibinfo{journal}{Nucl. Phys.} \bibinfo{volume}{A842} (\bibinfo{year}{2010})
  \bibinfo{pages}{59--71}. \DOIprefix\doi{10.1016/j.nuclphysa.2010.04.004}.
  \href{http://arxiv.org/abs/0805.0810}{\tt arXiv:0805.0810}.
\bibitem[{Peng and Chang(2012)}]{Peng:2012rn}
\bibinfo{author}{J.-C. Peng}, \bibinfo{author}{W.-C. Chang},
  \bibinfo{journal}{PoS} \bibinfo{volume}{QNP2012} (\bibinfo{year}{2012})
  \bibinfo{pages}{012}. \href{http://arxiv.org/abs/1207.2193}{\tt
  arXiv:1207.2193}.
\bibitem[{Airapetian et~al.(2008)}]{Airapetian:2008qf}
\bibinfo{author}{A.~Airapetian}, et~al. (\bibinfo{collaboration}{HERMES}),
  \bibinfo{journal}{Phys. Lett.} \bibinfo{volume}{B666} (\bibinfo{year}{2008})
  \bibinfo{pages}{446--450}. \DOIprefix\doi{10.1016/j.physletb.2008.07.090}.
  \href{http://arxiv.org/abs/0803.2993}{\tt arXiv:0803.2993}.
\bibitem[{Litvine and Likhoded(1999)}]{Litvine:1999sv}
\bibinfo{author}{V.~A. Litvine}, \bibinfo{author}{A.~K. Likhoded},
  \bibinfo{journal}{Phys. Atom. Nucl.} \bibinfo{volume}{62}
  (\bibinfo{year}{1999}) \bibinfo{pages}{679--692}. \bibinfo{note}{[Yad.
  Fiz.62,728(1999)]}.
\bibitem[{Lipatov et~al.(2012)Lipatov, Malyshev, and Zotov}]{Lipatov:2012rg}
\bibinfo{author}{A.~V. Lipatov}, \bibinfo{author}{M.~A. Malyshev},
  \bibinfo{author}{N.~P. Zotov}, \bibinfo{journal}{JHEP} \bibinfo{volume}{05}
  (\bibinfo{year}{2012}) \bibinfo{pages}{104}.
  \DOIprefix\doi{10.1007/JHEP05(2012)104}.
  \href{http://arxiv.org/abs/1204.3828}{\tt arXiv:1204.3828}.
\bibitem[{Lipatov and Zotov(2007)}]{Lipatov:2005wk}
\bibinfo{author}{A.~V. Lipatov}, \bibinfo{author}{N.~P. Zotov},
  \bibinfo{journal}{J. Phys.} \bibinfo{volume}{G34} (\bibinfo{year}{2007})
  \bibinfo{pages}{219}. \DOIprefix\doi{10.1088/0954-3899/34/2/005}.
  \href{http://arxiv.org/abs/hep-ph/0507243}{\tt arXiv:hep-ph/0507243}.
\bibitem[{Vogt(2000)}]{Vogt:2000sk}
\bibinfo{author}{R.~Vogt}, \bibinfo{journal}{Prog. Part. Nucl. Phys.}
  \bibinfo{volume}{45} (\bibinfo{year}{2000}) \bibinfo{pages}{S105--S169}.
  \DOIprefix\doi{10.1016/S0146-6410(00)90012-7}.
  \href{http://arxiv.org/abs/hep-ph/0011298}{\tt arXiv:hep-ph/0011298}.
\bibitem[{Navarra et~al.(1996)Navarra, Nielsen, Nunes, and
  Teixeira}]{Navarra:1995rq}
\bibinfo{author}{F.~S. Navarra}, \bibinfo{author}{M.~Nielsen},
  \bibinfo{author}{C.~A.~A. Nunes}, \bibinfo{author}{M.~Teixeira},
  \bibinfo{journal}{Phys. Rev.} \bibinfo{volume}{D54} (\bibinfo{year}{1996})
  \bibinfo{pages}{842--846}. \DOIprefix\doi{10.1103/PhysRevD.54.842}.
  \href{http://arxiv.org/abs/hep-ph/9504388}{\tt arXiv:hep-ph/9504388}.
\bibitem[{Collins et~al.(1989)Collins, Soper, and Sterman}]{Collins:1989gx}
\bibinfo{author}{J.~C. Collins}, \bibinfo{author}{D.~E. Soper},
  \bibinfo{author}{G.~F. Sterman}, \bibinfo{journal}{Adv. Ser. Direct. High
  Energy Phys.} \bibinfo{volume}{5} (\bibinfo{year}{1989})
  \bibinfo{pages}{1--91}. \DOIprefix\doi{10.1142/9789814503266_0001}.
  \href{http://arxiv.org/abs/hep-ph/0409313}{\tt arXiv:hep-ph/0409313}.
\bibitem[{Feynman et~al.(1978)Feynman, Field, and Fox}]{Feynman:1978dt}
\bibinfo{author}{R.~P. Feynman}, \bibinfo{author}{R.~D. Field},
  \bibinfo{author}{G.~C. Fox}, \bibinfo{journal}{Phys. Rev.}
  \bibinfo{volume}{D18} (\bibinfo{year}{1978}) \bibinfo{pages}{3320}.
  \DOIprefix\doi{10.1103/PhysRevD.18.3320}.
\bibitem[{Bednyakov et~al.(2012)Bednyakov, Grinyuk, Lykasov, and
  Poghosyan}]{Bednyakov:2011hj}
\bibinfo{author}{V.~A. Bednyakov}, \bibinfo{author}{A.~A. Grinyuk},
  \bibinfo{author}{G.~I. Lykasov}, \bibinfo{author}{M.~Poghosyan},
  \bibinfo{journal}{Int. J. Mod. Phys.} \bibinfo{volume}{A27}
  (\bibinfo{year}{2012}) \bibinfo{pages}{1250042}.
  \DOIprefix\doi{10.1142/S0217751X1250042X}.
  \href{http://arxiv.org/abs/1104.0532}{\tt arXiv:1104.0532}.
\bibitem[{Kniehl et~al.(2012)Kniehl, Kramer, Schienbein, and
  Spiesberger}]{Kniehl:2012ti}
\bibinfo{author}{B.~A. Kniehl}, \bibinfo{author}{G.~Kramer},
  \bibinfo{author}{I.~Schienbein}, \bibinfo{author}{H.~Spiesberger},
  \bibinfo{journal}{Eur. Phys. J.} \bibinfo{volume}{C72} (\bibinfo{year}{2012})
  \bibinfo{pages}{2082}. \DOIprefix\doi{10.1140/epjc/s10052-012-2082-2}.
  \href{http://arxiv.org/abs/1202.0439}{\tt arXiv:1202.0439}.
\bibitem[{Airapetian et~al.(2014)}]{Airapetian:2013zaw}
\bibinfo{author}{A.~Airapetian}, et~al. (\bibinfo{collaboration}{HERMES}),
  \bibinfo{journal}{Phys. Rev.} \bibinfo{volume}{D89} (\bibinfo{year}{2014})
  \bibinfo{pages}{097101}. \DOIprefix\doi{10.1103/PhysRevD.89.097101}.
  \href{http://arxiv.org/abs/1312.7028}{\tt arXiv:1312.7028}.
\bibitem[{de~Florian et~al.(2007)de~Florian, Sassot, and
  Stratmann}]{deFlorian:2007aj}
\bibinfo{author}{D.~de~Florian}, \bibinfo{author}{R.~Sassot},
  \bibinfo{author}{M.~Stratmann}, \bibinfo{journal}{Phys. Rev.}
  \bibinfo{volume}{D75} (\bibinfo{year}{2007}) \bibinfo{pages}{114010}.
  \DOIprefix\doi{10.1103/PhysRevD.75.114010}.
  \href{http://arxiv.org/abs/hep-ph/0703242}{\tt arXiv:hep-ph/0703242}.
\bibitem[{Lykasov et~al.(2013)Lykasov, Bednyakov, Demichev, and
  Stepanenko}]{Lykasov:2013rva}
\bibinfo{author}{G.~I. Lykasov}, \bibinfo{author}{I.~V. Bednyakov},
  \bibinfo{author}{M.~A. Demichev}, \bibinfo{author}{{\relax Yu}.~{\relax Yu}.
  Stepanenko}, \bibinfo{journal}{Nucl. Phys. Proc. Suppl.}
  \bibinfo{volume}{245} (\bibinfo{year}{2013}) \bibinfo{pages}{215--222}.
  \DOIprefix\doi{10.1016/j.nuclphysbps.2013.10.042}.
  \href{http://arxiv.org/abs/1309.3168}{\tt arXiv:1309.3168}.
\bibitem[{Albino et~al.(2008)Albino, Kniehl, and Kramer}]{Albino:2008fy}
\bibinfo{author}{S.~Albino}, \bibinfo{author}{B.~A. Kniehl},
  \bibinfo{author}{G.~Kramer}, \bibinfo{journal}{Nucl. Phys.}
  \bibinfo{volume}{B803} (\bibinfo{year}{2008}) \bibinfo{pages}{42--104}.
  \DOIprefix\doi{10.1016/j.nuclphysb.2008.05.017}.
  \href{http://arxiv.org/abs/0803.2768}{\tt arXiv:0803.2768}.
\bibitem[{Mangano(2010)}]{Mangano:2010zza}
\bibinfo{author}{M.~L. Mangano}, \bibinfo{journal}{Phys. Usp.}
  \bibinfo{volume}{53} (\bibinfo{year}{2010}) \bibinfo{pages}{109--132}.
  \DOIprefix\doi{10.3367/UFNe.0180.201002a.0113}, \bibinfo{note}{[Usp. Fiz.
  Nauk180,113(2010)]}.
\bibitem[{Rottoli(2016)}]{Rottoli:2016lsg}
\bibinfo{author}{L.~Rottoli}, in: \bibinfo{booktitle}{{Proceedings, 24th
  International Workshop on Deep-Inelastic Scattering and Related Subjects (DIS
  2016)}}. \URLprefix
  \url{https://inspirehep.net/record/1473188/files/arXiv:1606.09289.pdf}.
  \href{http://arxiv.org/abs/1606.09289}{\tt arXiv:1606.09289}.
\bibitem[{Aaij et~al.(2014)}]{Aaij:2014ida}
\bibinfo{author}{R.~Aaij}, et~al. (\bibinfo{collaboration}{LHCb}),
  \bibinfo{journal}{JINST} \bibinfo{volume}{9} (\bibinfo{year}{2014})
  \bibinfo{pages}{P12005}. \DOIprefix\doi{10.1088/1748-0221/9/12/P12005}.
  \href{http://arxiv.org/abs/1410.0149}{\tt arXiv:1410.0149}.
\bibitem[{Campbell and Ellis(2002)}]{Campbell:2002tg}
\bibinfo{author}{J.~M. Campbell}, \bibinfo{author}{R.~K. Ellis},
  \bibinfo{journal}{Phys. Rev.} \bibinfo{volume}{D65} (\bibinfo{year}{2002})
  \bibinfo{pages}{113007}. \DOIprefix\doi{10.1103/PhysRevD.65.113007}.
  \href{http://arxiv.org/abs/hep-ph/0202176}{\tt arXiv:hep-ph/0202176}.
\bibitem[{Levin et~al.(1991)Levin, Ryskin, Shabelski, and
  Shuvaev}]{Levin:1991ry}
\bibinfo{author}{E.~M. Levin}, \bibinfo{author}{M.~G. Ryskin},
  \bibinfo{author}{{\relax Yu}.~M. Shabelski}, \bibinfo{author}{A.~G. Shuvaev},
  \bibinfo{journal}{Sov. J. Nucl. Phys.} \bibinfo{volume}{53}
  (\bibinfo{year}{1991}) \bibinfo{pages}{657}. \bibinfo{note}{[Yad.
  Fiz.53,1059(1991)]}.
\bibitem[{Gribov et~al.(1983)Gribov, Levin, and Ryskin}]{Gribov:1984tu}
\bibinfo{author}{L.~V. Gribov}, \bibinfo{author}{E.~M. Levin},
  \bibinfo{author}{M.~G. Ryskin}, \bibinfo{journal}{Phys. Rept.}
  \bibinfo{volume}{100} (\bibinfo{year}{1983}) \bibinfo{pages}{1--150}.
  \DOIprefix\doi{10.1016/0370-1573(83)90022-4}.
\bibitem[{Kuraev et~al.(1976)Kuraev, Lipatov, and Fadin}]{Kuraev:1976ge}
\bibinfo{author}{E.~A. Kuraev}, \bibinfo{author}{L.~N. Lipatov},
  \bibinfo{author}{V.~S. Fadin}, \bibinfo{journal}{Sov. Phys. JETP}
  \bibinfo{volume}{44} (\bibinfo{year}{1976}) \bibinfo{pages}{443--450}.
  \bibinfo{note}{[Zh. Eksp. Teor. Fiz.71,840(1976)]}.
\bibitem[{Kuraev et~al.(1977)Kuraev, Lipatov, and Fadin}]{Kuraev:1977fs}
\bibinfo{author}{E.~A. Kuraev}, \bibinfo{author}{L.~N. Lipatov},
  \bibinfo{author}{V.~S. Fadin}, \bibinfo{journal}{Sov. Phys. JETP}
  \bibinfo{volume}{45} (\bibinfo{year}{1977}) \bibinfo{pages}{199--204}.
  \bibinfo{note}{[Zh. Eksp. Teor. Fiz.72,377(1977)]}.
\bibitem[{Balitsky and Lipatov(1978)}]{Balitsky:1978ic}
\bibinfo{author}{I.~I. Balitsky}, \bibinfo{author}{L.~N. Lipatov},
  \bibinfo{journal}{Sov. J. Nucl. Phys.} \bibinfo{volume}{28}
  (\bibinfo{year}{1978}) \bibinfo{pages}{822--829}. \bibinfo{note}{[Yad.
  Fiz.28,1597(1978)]}.
\bibitem[{Andersson et~al.(2002)}]{Andersson:2002cf}
\bibinfo{author}{B.~Andersson}, et~al. (\bibinfo{collaboration}{Small x}),
  \bibinfo{journal}{Eur. Phys. J.} \bibinfo{volume}{C25} (\bibinfo{year}{2002})
  \bibinfo{pages}{77--101}. \DOIprefix\doi{10.1007/s10052-002-0998-7}.
  \href{http://arxiv.org/abs/hep-ph/0204115}{\tt arXiv:hep-ph/0204115}.
\bibitem[{Andersen et~al.(2004)}]{Andersen:2003xj}
\bibinfo{author}{J.~R. Andersen}, et~al. (\bibinfo{collaboration}{Small x}),
  \bibinfo{journal}{Eur. Phys. J.} \bibinfo{volume}{C35} (\bibinfo{year}{2004})
  \bibinfo{pages}{67--98}. \DOIprefix\doi{10.1140/epjc/s2004-01775-7}.
  \href{http://arxiv.org/abs/hep-ph/0312333}{\tt arXiv:hep-ph/0312333}.
\bibitem[{Andersen et~al.(2006)}]{Andersen:2006pg}
\bibinfo{author}{J.~R. Andersen}, et~al. (\bibinfo{collaboration}{Small x}),
  \bibinfo{journal}{Eur. Phys. J.} \bibinfo{volume}{C48} (\bibinfo{year}{2006})
  \bibinfo{pages}{53--105}. \DOIprefix\doi{10.1140/epjc/s2006-02615-6}.
  \href{http://arxiv.org/abs/hep-ph/0604189}{\tt arXiv:hep-ph/0604189}.
\bibitem[{Lipatov and Vyazovsky(2001)}]{Lipatov:2000se}
\bibinfo{author}{L.~N. Lipatov}, \bibinfo{author}{M.~I. Vyazovsky},
  \bibinfo{journal}{Nucl. Phys.} \bibinfo{volume}{B597} (\bibinfo{year}{2001})
  \bibinfo{pages}{399--409}. \DOIprefix\doi{10.1016/S0550-3213(00)00709-4}.
  \href{http://arxiv.org/abs/hep-ph/0009340}{\tt arXiv:hep-ph/0009340}.
\bibitem[{Bogdan and Fadin(2006)}]{Bogdan:2006af}
\bibinfo{author}{A.~V. Bogdan}, \bibinfo{author}{V.~S. Fadin},
  \bibinfo{journal}{Nucl. Phys.} \bibinfo{volume}{B740} (\bibinfo{year}{2006})
  \bibinfo{pages}{36--57}. \DOIprefix\doi{10.1016/j.nuclphysb.2006.01.033}.
  \href{http://arxiv.org/abs/hep-ph/0601117}{\tt arXiv:hep-ph/0601117}.
\bibitem[{Hentschinski and Sabio~Vera(2012)}]{Hentschinski:2011tz}
\bibinfo{author}{M.~Hentschinski}, \bibinfo{author}{A.~Sabio~Vera},
  \bibinfo{journal}{Phys. Rev.} \bibinfo{volume}{D85} (\bibinfo{year}{2012})
  \bibinfo{pages}{056006}. \DOIprefix\doi{10.1103/PhysRevD.85.056006}.
  \href{http://arxiv.org/abs/1110.6741}{\tt arXiv:1110.6741}.
\bibitem[{Hentschinski(2012)}]{Hentschinski:2011xg}
\bibinfo{author}{M.~Hentschinski}, \bibinfo{journal}{Nucl. Phys.}
  \bibinfo{volume}{B859} (\bibinfo{year}{2012}) \bibinfo{pages}{129--142}.
  \DOIprefix\doi{10.1016/j.nuclphysb.2012.02.001}.
  \href{http://arxiv.org/abs/1112.4509}{\tt arXiv:1112.4509}.
\bibitem[{Chachamis et~al.(2012)Chachamis, Hentschinski, Madrigal~Martinez, and
  Sabio~Vera}]{Chachamis:2012gh}
\bibinfo{author}{G.~Chachamis}, \bibinfo{author}{M.~Hentschinski},
  \bibinfo{author}{J.~D. Madrigal~Martinez}, \bibinfo{author}{A.~Sabio~Vera},
  \bibinfo{journal}{Nucl. Phys.} \bibinfo{volume}{B861} (\bibinfo{year}{2012})
  \bibinfo{pages}{133--144}. \DOIprefix\doi{10.1016/j.nuclphysb.2012.03.015}.
  \href{http://arxiv.org/abs/1202.0649}{\tt arXiv:1202.0649}.
\bibitem[{Lipatov(1995)}]{Lipatov:1995pn}
\bibinfo{author}{L.~N. Lipatov}, \bibinfo{journal}{Nucl. Phys.}
  \bibinfo{volume}{B452} (\bibinfo{year}{1995}) \bibinfo{pages}{369--400}.
  \DOIprefix\doi{10.1016/0550-3213(95)00390-E}.
  \href{http://arxiv.org/abs/hep-ph/9502308}{\tt arXiv:hep-ph/9502308}.
\bibitem[{Lipatov(1997)}]{Lipatov:1996ts}
\bibinfo{author}{L.~N. Lipatov}, \bibinfo{journal}{Phys. Rept.}
  \bibinfo{volume}{286} (\bibinfo{year}{1997}) \bibinfo{pages}{131--198}.
  \DOIprefix\doi{10.1016/S0370-1573(96)00045-2}.
  \href{http://arxiv.org/abs/hep-ph/9610276}{\tt arXiv:hep-ph/9610276}.
\bibitem[{Lipatov and Malyshev(2016)}]{Lipatov:2016wgr}
\bibinfo{author}{A.~V. Lipatov}, \bibinfo{author}{M.~A. Malyshev}
  (\bibinfo{year}{2016}). \href{http://arxiv.org/abs/1606.02696}{\tt
  arXiv:1606.02696}.
\bibitem[{Martin et~al.(2010)Martin, Ryskin, and Watt}]{Martin:2009ii}
\bibinfo{author}{A.~D. Martin}, \bibinfo{author}{M.~G. Ryskin},
  \bibinfo{author}{G.~Watt}, \bibinfo{journal}{Eur. Phys. J.}
  \bibinfo{volume}{C66} (\bibinfo{year}{2010}) \bibinfo{pages}{163--172}.
  \DOIprefix\doi{10.1140/epjc/s10052-010-1242-5}.
  \href{http://arxiv.org/abs/0909.5529}{\tt arXiv:0909.5529}.
\bibitem[{Olive et~al.(2014)}]{Agashe:2014kda}
\bibinfo{author}{K.~A. Olive}, et~al. (\bibinfo{collaboration}{Particle Data
  Group}), \bibinfo{journal}{Chin. Phys.} \bibinfo{volume}{C38}
  (\bibinfo{year}{2014}) \bibinfo{pages}{090001}.
  \DOIprefix\doi{10.1088/1674-1137/38/9/090001}.

\end{thebibliography}

\end{document}